\documentclass[12pt,openany,twoside]{book}
\usepackage{amsmath}
\usepackage{amssymb}

\usepackage[utf8]{inputenc}
\usepackage[T1]{fontenc}

\def\dju{\mbox{Đurđevich}}

\newtheorem{theorem}{Theorem}[section]

\newtheorem{prop}{Proposition}[section]
\newtheorem{definition}{Definition}[section]
\newtheorem{example}{Example}[section]
\numberwithin{equation}{section}

\begin{document}
\frontmatter

\begin{center}
	{\bf 
    Time Evolution and Probability
	\\
	in Quantum Theory
    \vskip 0.3cm
    The Central Role of Born's Rule}
	\vskip 0.3cm 
	Stephen Bruce Sontz
	\\
	CIMAT
	\\
	Guanajuato, Mexico
\end{center}

\vskip 1cm

\begin{center}
	{\bf Abstract}
\end{center}

\vskip 0.6cm 
\noindent
In this treatise I introduce the time dependent 
{\em Generalized Born's 
Rule} for the probabilities of quantum events,
including conditional and consecutive 
probabilities, 
as the unique fundamental time evolution 
equation of quantum theory. 
Then these probabilities,
computed from {\em states} and {\em events}, 
are to be compared with
relative frequencies of observations.  
Schr\"odinger's equation still is valid in one model 
of the axioms 
of quantum theory, which I call 
the {\em Schr\"odinger model}.
However, the role of Schr\"odinger's equation 
is auxiliary, since 
it serves to help compute the continuous temporal 
evolution of the probabilities given by 
the Generalized Born's Rule. 
In other models, such as the Heisenberg model, 
the auxiliary equations are quite different, 
but the Generalized Born's 
Rule is the same formula ({\em covariance})
and gives the same results 
({\em invariance}). 
Also some aspects of the Schr\"odinger
model are not found in the isomorphic Heisenberg model,
and they therefore do not have any physical significance.
One example of this is the infamous collapse of the 
quantum state. 
Other quantum phenomena, such as entanglement, 
are easy to analyze in terms of the Generalized Born's 
Rule without any reference to the unnecessary 
concept of collapse. 
Finally, this leads to the possibility of quantum 
theory with other sorts 
of auxiliary equations instead of 
Schr\"odinger's equation, and  
examples of this are given. 
Throughout this treatise the {\em leit motif}
is the central importance of quantum probability and
most especially of the simplifying role of the 
time dependent Generalized Born's 
Rule in quantum theory.

\newpage 

\vskip 20cm 
\begin{center}
{\bf A Lilia por tu amor, por tu apoyo, por todo} 
\end{center}

\newpage 

\begin{center}
	{\bf Back Cover Blurb}
\end{center}

\vskip 0.6cm 
\noindent
This treatise presents a careful 
re-organization at 
an advanced level in terms of the 
probability theory 
of basic quantum theory based on 
quantum states and quantum events. 
This leads to the recognition that 
the fundamental time evolution 
equation of quantum theory 
is the {\em Generalized Born's Rule}, 
not Schr\"odinger's equation, which here 
plays a secondary role in one model of 
quantum theory. 
References to classical physics are kept 
to a minimum, since this is meant to be
a stand-alone, axiomatic presentation 
of standard quantum theory. 
This is not a generalization of quantum theory 
nor an interpretation of it. 
The intended audience consists of those 
with a detailed previous knowledge of quantum 
theory, classical 
probability and functional analysis. 
Advanced undergraduate students may find
something of interest, but this is not a 
continuation of an introductory course
in quantum theory.

\chapter{Preface}
\markboth{Preface}{Preface}

\hfill 
What I can not create 

\hfill I do not understand

\vskip 0.2cm
\hfill
Richard Feynman

\vskip 0.4cm
\noindent \indent 
After finishing my recently published introductory 
book \cite{path} 
on quantum theory, I began thinking of 
writing a sequel on the usual advanced topics 
used in applications. 
I am thinking of Hartree-Fock, variational 
methods, atomic and molecular physics, 
solid state physics, 
scattering theory, perturbation theory and so on. 
But these topics are covered in many 
fine texts which get the physics (mostly) right
as well as getting the mathematics (mostly)
right. Some even get everything right! 
There seemed to be no point in writing yet 
another big square book on all that 
well known material. 
And I am tired of the retellings of the story
of the historical 
path that led to the quantum theory. 
Besides history hides the internal logical 
structure of quantum theory. 
So references to history will be few 
and far between. 

I decided to resolve clearly 
at an advanced level, but at the same time in the 
simplest possible way, this  
basic problem of quantum theory: 
its logical structure without any 
reference to classical physics. 
And it seemed to me that the central role of
quantum probability in quantum theory was 
not sufficiently appreciated in the 
scientific community, especially the roles of
conditional and consecutive probability. 
So I aimed at writing an expository article directed at 
a specialized public to explain that.  
I soon realized that I had it wrong. 
Quantum probability is not a central part 
of quantum theory. 
Rather the basics of quantum theory are {\em exactly} 
the same as quantum probability. 
What I mean is that  
the {\em Generalized Born's Rule}, 
and not Schr\"odinger's equation as I previously thought, is the 
fundamental time evolution equation of quantum theory
as well as being the rule for computing all the 
probabilities in quantum theory. 
That is the essence of this treatise. 

My approach is so 
different from the current
trend to give `interpretations' of quantum 
theory that 
I felt obliged to give an apology separately. 

How I happened to arrive at this understanding may be 
of interest. 
I was confused by the equivalence of the 
Schr\"odinger picture and the Heisenberg picture. 
Most particularly, how could the
Schr\"odinger picture have collapse 
of the state as a basic aspect of the theory 
while the equivalent Heisenberg picture
has states which are time independent? 
I then  realized that the equivalence of these pictures is 
a well known statement about time dependent probabilities. 
And therefore, probabilities and how to calculate 
them are the central issue of quantum theory. 
That is exactly what the Generalized Born's Rule
is about. 
That rule is what {\em Quantum 
Probability} means in this treatise. 

I am telling a new story, but all the 
characters in it are familiar. 
So, this is a new organization of standard 
material with Born's rule placed front and 
center. 
This requires introducing a Generalized 
Born's Rule that might be new to you, though 
it is implicit in standard textbook quantum mechanics 
and in many cases is related to L\"uders rule. 
However most importantly, the really new idea in this 
treatise is that the Generalized Born's Rule 
is the one and only fundamental time evolution equation 
in quantum theory. 
Quantum probability is not simply an important 
part of quantum theory, as I have previously
thought. 
Rather it is the central part. 
Schr\"odinger's equation is still present in 
one model of quantum theory and still plays 
a crucial role in understanding quantum systems,
but it is not the fundamental time evolution 
equation.  
It turns out that the model based on 
Schr\"odinger's equation. which I call 
{\em Schr\"odinger's model}, has attributes that 
are not present in other isomorphic models. 
The collapse of the wave function is an 
example of this. 

Another new aspect of this story is that 
there are many non-isomorphic models of the 
basic axioms of quantum theory. 
This seems to fly in the face of trying to find the 
unique and universal laws of nature. 
But the point is that quantum theory is a work in progress 
and in this treatise only the basics are being considered. 
In particular, this is not a complete 
exposition of known quantum theory, since that 
would require many volumes. 
So, many topics are deliberately not included. 
This is not a theory of everything, but a 
theory of something. 
However, I am considering quantum theory consistent with 
Galilean relativity and at the same time 
quantum theory consistent with special relativity.
If things work out that way, 
I am even considering quantum theory 
consistent with general relativity. 
So I hope the reader realizes the importance 
of non-uniqueness. 

This is a story about events 
(as well as states and probabilities)
and could be confused with the consistent histories 
interpretation of quantum theory. 
(See \cite{griffiths}.) 
I deal with this confusion in the chapter 
Interpretations. 
For now let me note that I am not giving any
`interpretation' of quantum theory. 
Nor do I care to, 
since `interpretations' are neither
falsifiable nor  verifiable. 
Rather this is a scientific treatise. 
However, I will touch on some issues that are 
often considered in an `interpretation' of 
quantum theory. 
This is not an attempt to solve 
philosophical problems, but 
instead a way of throwing some light on 
quantum situations by using just the 
basics: events, states and probabilities. 

This story is addressed to a reader with a rather 
substantial background in quantum physics as well 
as in the mathematics used in describing 
that physics. 
This is not an elementary treatise, nor even 
the immediate sequel of such an introduction 
as for example found in my book \cite{path}. 
Advanced undergraduate students in physics 
or mathematics might find this material 
to be quite challenging, though they should
not be discouraged from trying. 
But my intended audience consists of professional
scientists and some graduate students
with lots of math and physics 
knowledge, much more than can be found in 
\cite{path}. 
The prerequisites that the reader should have include,
but are not limited to, an extensive prior 
knowledge of quantum physics and Hilbert space theory. 
I assume knowledge of functional analysis up to at 
least the spectral theory for self-adjoint operators. 
And this will be a stand-alone story. 
References to classical physics will be incidental,
not part of the main story.  
It is not clear whether those with some knowledge of
classical mechanics have an advantage or an
impediment. 
Some easy pedagogical examples, which are somehow
illustrative of the general theory, will also be given. 
But mostly, the presentation will be quite general. 
The experts will note, however, that I limit myself
to examples that are type I von Neumann algebras. 

This treatise is organized as follows. 
The first chapter states and discusses the Axioms. 
Chapter~2 is a long introduction to 
Quantum Probability. 
Here the important topics of conditional and 
consecutive probability are discussed. 
This is not new, neither in theory nor 
experiment, since one has to assign theoretical 
probabilities 
to sequences of events and one measures 
relative frequencies of such sequences. 
In one section 
this theory is compared and contrasted with 
its well known intellectual competitor:   
Classical Probability as a topic 
in measure theory. 
But I do not consider 
comparisons with 
classical physics, Bell's inequalities, hidden 
variables 
and such side issues, since my focus is on 
quantum theory as a new type of
probability theory and nothing else.  
Chapter~3 then applies this theory to Entanglement. 
In Chapters 4 to 7 I discuss various issues that 
have filled volumes for decades. 
These are Schr\"odinger's Cat, the Measurement 
Problem, the EPR paper, 
Determinism and Probability. 
I do not resolve any of these issues, but try to
shed some light on what they are (or should be) about. 

In the rest of the treatise I consider topics that 
strictly speaking have little or nothing to do with 
quantum theory. 
I include them since it is an appropriate moment 
for commenting on them. 
In Chapter~8 I attempt to dismiss Interpretation, in the
contemporary sense of that word, from scientific 
consideration. Chapter~9 
is devoted to the Wave Function about which 
I expect the reader to know a lot already. 
I include it in order to rid quantum theory of 
residual classical language that slips into 
discussions of it. 
Chapter~10 concerns various extensions and alternatives
to quantum theory. 
Some of these are currently well accepted, while 
others are wild speculation at best. 
In the final Chapter~11 it is my turn to tilt at
windmills with an unlikely proposal. 

Readers with a lot of background can get the basic
gist of this treatise by reading the sections on Axiom~5
and Quantum Conditional Probability and then skipping
forward to the Chapter on Entanglement. 
I would like to point out that I originally thought 
that Entanglement could not be explained without
using the collapse of the state associated with a 
measurement. 
So I examined this case carefully in order to 
understand the fundamental role of the collapse 
condition. 
What I found out surprised me; 
an analysis in terms of the conditional
probability of sequences of quantum events suffices. 

There is a multitude of topics that I could have included, 
but I decided to make my point in a slim book. 
I am reminded of the lawyer who was summing up 
on and on the defense of his client to the court. 
His partner seated at his side 
took a legal notepad and wrote on it:
``Sit down! You have made your point!'' 
So, I hope that I have said enough to make my point. 

This treatise was much easier to write than \cite{path},
since here I can rely on the previous knowledge of
my intended reader. 
One day I hope to write a book on basic quantum theory 
in a manner accessible to a general public, since 
I feel that I owe it to my friends who are 
not scientists. 
That will be a much more difficult task. 
In many ways this treatise is prologue. 

Storytelling is a part of human culture and,
by inference, a part of human nature. 
It is an art practiced in all societies
throughout history. 
It pre-exists written language. 
But it is a mistake to think that a 
universal part of the human condition 
is necessarily a talent shared by all
people.
Quite simply, there are many persons 
who can not narrate a story in a fashion
comprehensible to others. 
I am not merely speaking of those born 
with cerebral damage. 
There are also those who do not understand
that a story is for an audience that does 
not yet know completely the topic at hand and, consequently, the storyteller must go into 
the details, step by step, to communicate 
with that audience. 
Young children seem to always be like that;
when they tell a story it is as if they expect 
the listener to already know it and they are 
merely repeating something known by everyone.
One has to learn that the `other' is 
different from the `ego' and so must be
accommodated accordingly. 
But some people never learn this, or learn it
incompletely. 
Such people we can deem as `pre-literate' though
this is meant to be purely
descriptive and not judgmental. 
At best 
they are like children who can repeat fragments 
without realizing what are the processes, 
both mental and observational, by which this
all comes about. 
Such people are not able to organize nor
communicate scientific knowledge.  
They only listen to `authority' and repeat it. 
A fancy way of naming storytelling 
is simply `narrative'. 

Curiosity is also a part of human culture and,
by inference, a part of human nature. 
And similarly, not everyone is curious or, better said, 
not everyone is curious in the same way. 
Moreover, those who are curious may not have much
talent or experience 
for dealing with that curiosity in 
a systematic way. 
Scientific activity is a combination of 
curiosity and narrative. 
While some sort of scientific activity 
in this vague sense is 
found in all cultures, 
the contemporary version of science 
is not a necessary aspect of human culture. 
Many people 
are `pre-scientific' very much in the same way 
as many people are pre-literate.  

The contemporary concept of science is not
an invention of the modern world, even though it
is a human invention of an intellectual sort. 
One can learn how science, basically as 
understood nowadays, was practiced in 
Hellenistic society beginning around 300 B.C. 
from the fascinating book \cite{russo} by
L.~Russo. 
As Russo points out, some `modern' scientists 
do not grasp completely what Archimedes and 
Co.~were doing in Ancient Antiquity. 
One necessary scientific activity, 
then as now, is the 
construction of mathematical models which 
relate to observations. 
It is a two-way street, but logically 
the flow starts on the observational side. 
In this approach the historical process 
of discovery is not of any interest. 
However, I will directly jump over to the theoretical 
side by starting with an axiomatization 
of quantum theory. 
This methodology is what Maxwell's Equations
are all about, though they are not typically 
called Maxwell's Axioms. 
Maybe they should be renamed. 

I thank Micho \dju~and Robert Griffiths 
for graciously 
commenting on a preliminary version of this 
treatise. 
Their comments helped me to revise 
and possibly improve certain things. 

Many people have helped me in 
understanding the matters discussed here. 
Sometimes that help was a detailed explanation, but
often it was a casual question or comment that 
focused my attention on an important matter. 
A very partial list of those whom I have known 
personally must include 
Luigi Accardi, 
David Brydges,  
Jaime Cruz Sampedro, 
Matthew Dawson, 
Micho \dju,  
Jean-Pierre Gazeau, 
Leonard Gross, 
George Hagedorn, 
Brian Hall, 
Ira Herbst, 
Jim Howland, 
Larry Thomas and 
Carlos Villegas. 
And during my undergraduate days 
I was advised  
to read the masters.  
So I am 
indebted also to those whom I have only 
known through their published work. 
Most notable among those are 
Valentine Bargmann, 
Charles Darwin, 
Paul Dirac, Richard Feynman, 
David Hilbert, 
John von Neumann, 
Erwin Schr\"odinger, 
and Eugene Wigner. 
I thank all of those 
mentioned 
as well as many more, too 
numerous to recall and list their names. 
Of course, any errors or shortcomings are 
due to 
my imperfect understanding of
non-intuitive\footnote{Intuition 
	is not an objective criterion. 
	Throughout this treatise intuition refers to
	my own limitations.} matters. 

\vskip 0.5cm 
\noindent 
Stephen Bruce Sontz     
\\
Centro de Investigaci\'on en Matem\'aticas, A.C. 
(CIMAT)
\\
Guanajuato, Mexico 
\\
November 2020   (Preliminary version)
\\
email: sontz@cimat.mx

\chapter{Apology} 
\markboth{Apology}{Apology}

\hfill 
Avant tout il faut savoir poser des probl\`emes.  

\hfill  
Et quoi qu'on dise, dans la vie scientifique, 

\hfill  les probl\`emes 
 ne se posent pas d'eux-m\^emes. 
 
\vskip 0.2cm

\hfill Gaston Bachelard

\vskip 0.6cm

If this treatise receives any reception beyond 
total negligence, it most likely will consist 
mainly of negative criticisms. 
In anticipation of the worst of these, I present 
a defense. 
I only ask those imagined critics to try to 
take what I now say into consideration. 

I will be criticized for not treating some 
favorite topic. 
But this treatise is meant to lay the 
groundwork for building a better understanding 
of quantum theory. 
No claim is made to completeness. 
Nor would this be possible in anything less 
than an encyclopedic set of volumes. 
If some think that it is important that 
other topics be studied from this viewpoint, 
then it behooves 
them to undertake that study. 
I intend to stand not only on the 
shoulders of the giants who preceded me, 
but also on the shoulders of those who follow. 

I will be criticized for not discussing 
classical physics, Bell's inequalities, 
quantization schemes, 
hidden variables or any of a number of 
topics that are not part of quantum theory
itself. 
But my subject matter is quantum theory, 
nothing else. 

I will be criticized for not seeking a 
deterministic, classical explanation for 
quantum situations that have probability $ 1 $, 
despite the fact that probability theory 
always includes the limiting case of probability
$ 1 $ with no need for explanations outside 
of that theory itself. 

I will be criticized for being 
`over-mathematical' when, in fact, I am just 
trying to isolate the purely mathematical 
aspects of quantum theory in order to avoid 
their intrusion into the difficult process 
of creating physical intuition for a probability 
theory. 

I will be criticized for relying on axioms. 
However, this is a time tested methodology for 
organizing an explanatory system of thought
in order to provide objective criteria for
valid advances and to enhance understanding. 
It also serves the important pedagogical 
goal of giving a student a way to learn 
and dominate the subject. 

I will be criticized for giving yet another 
interpretation of quantum theory, no better, 
and quite possibly, worse than others. 
But interpretation is not my concern. 
Interpretation is felt to be 
required when obscurity, 
confusion and contradiction dominate. 
But by isolating and logically 
organizing the basics, I think that 
the need for interpretation recedes. 

I will be criticized because this is mostly 
well known material that has already appeared  
in the scientific literature long ago. 
Well, point partially granted. 
However, I have firstly 
re-organized that 
well known material and secondly put
the Generalized Born's Rule at its head 
as the fundamental time evolution equation 
of quantum theory. 
I have not tried, as many have, to derive 
Born's Rule from more basic principles. 
Rather it is an axiom. 
Moreover, I have emphasized that this is actually
a time evolution equation, even though it is 
{\em not} a differential equation and 
{\em need not} be such. 

I will be criticized  because this is new 
material which rejects standard quantum theory. 
But nothing is rejected, not even the 
collapse condition, although it is language that 
I would prefer to avoid. 
I am only trying to put things in their proper 
perspective given the new viewpoint provided by 
the Generalized Born's Rule. 

I will be criticized for not solving some 
problem or other, whether it be a problem 
of physics, mathematics or epistemology. 
I have posed my own problem and outlined 
my resolution of it. 
Explicitly, the initial problem was how to 
understand the collapse condition as a 
necessary part of the Heisenberg picture 
of the quantum theory of entanglement.  
This is a part of the scientific project, not
all of it. 
I encourage others to work on their problems. 

I will be criticized because I do not have 
the requisite standing in the scientific community 
to opine publicly on these matters. 
This is but a way to avoid confronting the 
treatise on its intrinsic merits and should be
recognized as such. 
And rejected as such.

\newpage

\markboth{Contents}{Contents}
\tableofcontents

\newpage

\markboth{Notation}{Notation}

\begin{center}
	{\bf List of Notations and Abbreviations}
\end{center}

\markboth{Notations}{Notations}

\begin{tabbing}
	
\noindent 
{\bf Symbol} \hskip 1,6cm  \={\bf Meaning}
\\
\\
$ A $ \> Self-adjoint operator
\\
$ B $ \> Borel subset of $ \mathbb{R} $
\\
$ \mathcal{B} $ \>Basis
\\
$ \mathcal{B} (\mathbb{R}) $ 
\>The $ \sigma  $-algebra of Borel subsets of $ \mathbb{R} $  
\\
$\mathrm{Bij} (A)$\> 
The set of bijections of a set $ A $ to itself. 
\\
$ C^{\infty} $ \>Having all derivatives
\\
$ \mathbb{C} $ \>The set of complex numbers
\\
$ \mathbb{C} \mathrm{P} (\mathcal{H}) $ \>
 The complex projective space of $ \mathcal{H} $
\\
$ D $\>Density matrix
\\
$ \dim $ \>Dimension
\\
$ e $\> Base of the natural exponentials and 
natural logarithms
\\
$ E $ \> Quantum event or, equivalently, 
projection
\\
$ E^{c} $ \>$ I - E $, the complementary 
quantum event of $ E $
\\
$ \mathcal{E} $ \>The set of all quantum events
\\
$ E_{t} $ \>One-parameter  
group of bijections of $ \mathcal{E} $
where $ t \in \mathbb{R} $ 
\\
$  \mathcal{E} (X) $ \>Expected value of the 
random variable $ X $
\\
$ \mathcal{E}_{\psi} (A) $ \>Expected value of 
the operator  
$ A $ given the pure state $ \psi $
\\
$ f $ \>A function
\\
$ H $ \>Hamiltonian operator
\\
$ H_{free} $ \>Free Hamiltonian operator
\\
$ H_{int} $ \>Interacting Hamiltonian operator
\\
$ \mathcal{H} $ \>Hilbert space
\\
$ ( \mathcal{H}, E, S )$ \>Model of quantum theory 
\\
$ \hbar $ \>Planck's normalized constant 
\\
$ i $ \> The complex unit $ \sqrt{-1} $ 
\\
$ I $ \>Identity operator 
(on the appropriate vector space)
\\
$ I_{\mathcal{H}} $ \>Identity operator of 
$ \mathcal{H} $
\\
$id$ \>The identity function 
(on the appropriate domain)
\\
$ l $  \>Linear functional 
\\
$ \mathcal{L} (\mathcal{H}) $ \>
The set $
\{ T : \mathcal{H} \to \mathcal{H} \,|\, 
T \mathrm{~is~linear~and~bounded} \} $
\\
$ L^{2} (\mathbb{R}^{3}) $ \>Hilbert space of 
equivalence classes of 
\\
\>square integrable functions 
$ f : \mathbb{R}^{3} \to \mathbb{C} $
\\
$ m_{k} $ \>$ k $th moment of the measure $ \mu $
\\
$ \mathcal{M} (\Omega) $ \>vector space
of all measurable functions 
$ f : \Omega \to \mathbb{C} $
\\
$ \mathbb{N} $ \>The set of 
non-negative integer numbers
\\
$ \mathbb{N}^{+} $ \>The set of 
strictly positive integer numbers
\\
povm \> Positive operator valued measure 
\\
pvm \>Projection valued measure
\\
$ P $ \>Probability; projection valued measure
\\
$ P_{A} $ \>Projection valued measure of 
the self-adjoint operator $ A $
\\
$ P ( E \,|\, \rho ) $ \> 
Quantum probability of the event $ E $ 
\\ \>given the density matrix $ \rho $ 
\\
$ P ( E \,|\, \psi ) $ \> 
Quantum probability of the event $ E $ 
\\ \>given the pure state $ \psi $ 
\\
$ P ( T \in B \,|\, \rho ) $ \> 
Quantum probability
that the observable 
$ T $ lies in the set $ B $,  
\\
\>given 
the density matrix $ \rho $ 
\\
$ P ( T \in B \,|\, \psi ) $ \> 
Quantum probability
that the observable $ T $ lies in the set $ B $, 
\\
\>given 
the pure state $ \psi $ 
\\
$   P ( E_{1} | E_{2} ,\rho ) $ 
\>Conditional quantum 
probability of the event $ E_{1} $, 
\\\>given an earlier event $ E_{2} $ and 
the density matrix $ \rho $
\\
$   P ( E_{1} | E_{2} ,\psi ) $ 
\>Conditional quantum 
probability of the event $ E_{1} $, 
\\\>given an earlier event $ E_{2} $ and 
the state $ \psi $
\\
$ P (E_{1}, E_{2} \,|\, \rho) $ 
\>
Consecutive quantum probability
of the event $ E_{1} $ 
\\
\> and then later 
the event $ E_{2} $, given 
the density matrix $ \rho $
\\
$ P (E_{1}, E_{2} \,|\, \psi) $ 
\>
Consecutive quantum probability
of the event $ E_{1} $ 
\\
\> and then later 
the event $ E_{2} $, given 
the state $ \psi $
\\
$ \mathbb{R} $ \>The set of real numbers
\\
$ \mathrm{Res} (A) $ \>Resolvent set of the operator 
$ A $
\\
$ \mathcal{S} $ \>The set of all quantum states
\\
$ S_{t} $ \> One-parameter 
group of bijections of $ \mathcal{S} $
 where $ t \in \mathbb{R} $ 
\\
$ S_{1}, S_{2}, S_{3} $ \>$ 2 \times 2 $ 
spin matrices
\\
$ S^{2} $ \>$ 4 \times 4 $ total spin matrix
\\
$ \mathrm{Spec} (A)  $ \>The spectrum of the 
operator $ A $
\\
$ SU(2) $ \>The special unitary Lie group of 
$ 2 \times 2 $ matrices 
\\
$ \mathrm{Supp} \, P $ \>Support of a pvm $ P $
\\
$ t $ \>time
\\
$ T $\> Linear operator
\\
$Tr$ \>Trace of an trace class operator
\\
$ U $ \>Unitary operator; open set
\\
$ U (s , t) $  \>Time evolution operator
\\
$ \mathcal{V} $ \>A von Neumann algebra
\\
$  V_{j}  $ \>Spectral subspace
\\
$ X $ \>Classical random variable 
\\
$ \delta_{ij} $ \>Kronecker delta 
\\
$ \varepsilon_{1}, \varepsilon_{2} $ 
\>Standard orthonormal basis of $ \mathbb{C}^{2} $
\\
$ \lambda $ \>Real or complex number
\\
$\{ \lambda \} $ \>The set whose only element is 
$ \lambda $
\\
$ \Lambda $ \>The empty sequence of events 
\\
$ \mu $ \>Measure
\\
$ \mu_{X} $ \>Distribution of the 
random variable $ X $
\\
$ \pi $ \>Quantum probability
\\
$ \rho $\>Density matrix 
\\
$ \sigma $ \>Standard deviation 
\\
$ \sigma_{k} $ \>$ k $th central moment 
of the measure $ \mu $
\\
$ \phi $ \>Unit vector in $ \mathcal{H} $;
pure state
\\
$ \phi_{t} $ \>Pure state that depends  
on time $ t \in \mathbb{R} $
\\
$  | \phi \rangle \langle \phi | $ 
\>Dirac notation for the density matrix 
\\
\>associated with a pure state 
$ \phi $
\\
$ \langle \phi | $ \>Dirac bra notation of the pure 
state $ \phi $ 
\\
$ | \phi \rangle $ \>Dirac ket notation of the pure 
state $ \phi $ 
\\
$ \chi_{S} $ \>The characteristic function of the 
set $ S $
\\
$ \psi $ \>Unit vector in $ \mathcal{H} $;
pure state
\\
$ \psi_{t} $ \>Pure state that depends  
on time $ t \in \mathbb{R} $
\\
$ \Omega $ \>Classical probability space; 
region in Minkowski space-time
\\
$( \Omega, \mathcal{F}, P) $ 
\>Classical probability space, its $ \sigma $-algebra 
\\ \>and its probability function 
\\
$ \emptyset $ \>The empty set
\\
$ \langle \cdot , \cdot \rangle $ 
\>Inner product
\\
$ || \cdot || $ \>Norm
\\
$ || \cdot ||_{op} $ \>Operator norm
\\
$ \cap $ \>Intersection of sets
\\
$ \cup $ \>Union of sets
\\
$ \wedge $ \>Infimum of events or lattice elements
\\
$ \vee $ \>Supremum of events or lattice elements
\\
$ \otimes $ \>Tensor product 
\end{tabbing} 

\mainmatter

\chapter{Axioms}
\markboth{Axioms}{Axioms}
\label{axiom-chapter}

\hfill 
I had been told that Euclid 

\hfill 
proved things and was much 

\hfill 
disappointed that he started 

\hfill 
with axioms.

\vskip 0.2cm
\hfill Bertrand Russell

\vskip 0.4cm

In \cite{path} I put the axioms in the penultimate 
chapter of the book. 
This was a pedagogical choice. 
The intended audience for that book consists 
of people with no prior background in physics, 
and so I wanted to present first the particulars 
and get to the logic behind it all later. 
In fact, since many novices are allergic to 
axioms and logic, I even made that chapter optional. 
But this treatise addresses a much more advanced 
audience. 
I not only expect your interest in these details, 
but also your understanding of their import. 
One point is that there is a way to associate  
certain aspects of this mathematical theory 
with physical phenomena. 
A theory, any theory,  is interesting and important 
if there are a sufficient number 
of physical phenomena 
that are adequately described by it. 
There is no need to claim that this theory, 
or any theory,  
is adequate for describing all possible physical 
phenomena. 
It is in this sense that I use the 
expressions `physical characteristic' or 
`observable quantity', for example. 
These are like `point' and
`line' in geometry; 
they are undefined expressions subject to 
possible correspondence with certain 
physical phenomena. 

These axioms are intended to provide a logical 
starting point for understanding standard 
`textbook' quantum mechanics as is used on 
a daily basis by scientists and engineers. 
If you prefer to start with other axioms that 
have these as logical consequences, then you 
are implicitly accepting the rest of this 
treatise, provided that I have made no mistakes. 
If you prefer to start with other axioms that 
contradict these, then the points of discrepancy 
should be subjected to experimental tests. 
These axioms are not intended to be complete, 
but are meant to give an explicit, 
logical basis for the
rest of the treatise. 
Nor are they intended to be the final, most 
efficient way to do this. 

I will use the spectral theorem for 
self-adjoint operators acting in a Hilbert space. 
I try to use 
standard notations
and conventions. 
Throughout this treatise {\em self-adjoint operators}
are understood to be densely defined. 
These axioms  are not the same as those  
given in \cite{path}. 
Nor is this intended to be a complete list of
axioms. 
The statement of each axiom terminates with 
the symbol 
$ \blacksquare $. 

We let $ \mathbb{C} $ denote the field of complex
numbers and $ i = \sqrt{-1} \in \mathbb{C} $. 
We also define 
$ \mathcal{L} (\mathcal{H}) :=
\{ T : \mathcal{H} \to \mathcal{H} \,|\, 
T \mathrm{~is~linear~and~bounded} \} $, 
where $ \mathcal{H} $ always denotes a complex Hilbert space, usually assumed to be 
separable. 
We note that $ \mathcal{L} (\mathcal{H}) $ 
when equipped with the operator norm, 
denoted by $ || \cdot ||_{op} $, and the 
adjoint operation, denoted by $ T \mapsto T^{*} $, 
is both a $ C^{*} $-algebra as well as a 
von Neumann algebra. 
The inner product on $ \mathcal{H} $, 
denoted as $ \langle \cdot , \cdot \rangle $,  is 
anti-linear in the first entry and linear
in the second. 
The norm on $ \mathcal{H} $ is denoted as 
$ || \cdot || $. 

Here is some Dirac notation we will use. 
Every element in the  dual Hilbert space 
$ \mathcal{H}^{\prime} := 
\{ l : \mathcal{H} \to \mathbb{C} \,|\, l \mathrm{~is~linear~and~bounded} \}$, 
according to the Riesz representation theorem,  
can be written for a unique $ \phi \in \mathcal{H} $
as $ l= \langle \phi | $, where 
the {\em bra} $ \langle \phi | $ is defined by  
$ 
\langle \phi | \, \psi := \langle \phi, 
\psi \rangle  
$ 
for all $ \psi \in \mathcal{H} $. 
Every vector $ \psi \in \mathcal{H}$, 
a Hilbert space,  
can also be denoted as a 
{\em ket} $ | \psi \rangle $. 
Then the definition of bra becomes 
$ 
\langle \phi | \, | \psi\rangle  := \langle \phi,  \psi \rangle  
$. 
For a pair of vectors $ \phi, \psi \in \mathcal{H} $
we define 
\begin{equation*}
 | \psi \rangle \langle \phi |:=  
 | \psi \rangle \otimes \langle \phi |
 \in \mathcal{H} \otimes \mathcal{H}^{\prime}. 
\end{equation*}
This is then identified with the element 
in $ \mathcal{L} (\mathcal{H}) $ defined 
for all $ \alpha \in \mathcal{H} $ by
\begin{equation*}
| \psi \rangle \langle \phi | \, \alpha :=
\langle \phi , \alpha \rangle \psi. 
\end{equation*}
Its operator norm satisfies 
$ || \, | \psi \rangle \langle \phi | \, ||_{op} = 
|| \psi || \, || \phi || $, while its 
operator adjoint satisfies 
$ (  | \psi \rangle \langle \phi | )^{*} =
 | \phi \rangle \langle \psi | $. 
If both $ \psi $ and $ \phi $ are also 
non-zero, then $ | \psi \rangle \langle \phi | $ 
is a rank $ 1 $ operator. 
If $ \phi $ is a unit vector, then
$ | \phi \rangle \langle \phi | $ is both 
a projection and a self-adjoint, positive, 
trace $ 1 $ operator, that is, a 
{\em density matrix}. 

We let $ \hbar $ denote the {\em (normalized) 
Planck's constant}. 
After the Table of Contents 
a list of other standard notations and 
abbreviations from physics and mathematics 
can be found.

\section{\bf Axiom 1 - Kinematics}
\markboth{Axiom 1}{Axiom 1}

{\bf Axiom 1: (Kinematics)}
For every  quantum system there is an 
associated non-commutative von Neumann 
algebra $ \mathcal{V} $ in $ \mathcal{L} (\mathcal{H}) $. 
The projection operators in $ \mathcal{V} $ 
are called {\em (quantum) events} and are associated with  
the physical events of the system. 
There is a non-empty set of self-adjoint 
operators, each of which is associated with a 
physically measurable observable of the system.  
In particular, 
every event is such an observable. 
Also, each such self-adjoint operator $ A $ 
(also called an {\em observable}) is 
associated to $ \mathcal{V} $, which means that its 
projection valued 
measure\footnote{Since projections are quantum 
events, I called this a {\em quantum event valued 
measure} or {\em qevm} in \cite{path}.
This was meant to make physics students feel more 
at home with a quite abstract mathematical structure.} 
(pvm) $ P_{A} $ 
satisfies $ P_{A} (B) \in  \mathcal{V}$ for
all Borel subsets $ B $ of $ \mathbb{R} $.  
$ \blacksquare $

\vskip 0.2cm \noindent 
The first condition in the axiom 
implies that 
$ \dim_{\mathbb{C}} \mathcal{H} \ge 2 $. 
This axiom does not assert that every observable 
quantity of a physical entity has a corresponding 
self-adjoint operator in quantum theory. 
The most important example of such an 
observable is the 
time of a physical event. 
Since the events in $ \mathcal{V} $ are assumed to 
be observables and the events 
($ \equiv $ projections) generate $ \mathcal{V} $, 
it follows that the observables generate $ \mathcal{V} $. 
The criterion of non-commutativity reflects Dirac's 
often expressed opinion that the essential characteristic 
of quantum theory is that the observables do not 
commute. 

Since this is a presentation based on 
events and, in particular events of the 
form $ P_{A} (B) $ (using the notation of the
axiom), it is more convenient, 
almost obligatory, to use 
von Neumann algebras instead of the more 
general structure of $ C^{*} $-algebras. 
The point is that 
if one starts with a $ C^{*} $-algebra 
$ \mathcal{C} $ and 
one has a self-adjoint $ A \in \mathcal{C} $, then 
the associated events $ P_{A} (B) $ do not
necessarily lie in $ \mathcal{C} $, though 
they do lie in the smallest von Neumann 
algebra containing $ \mathcal{C} $. 

The mathematical definition of a quantum event 
as a projection operator must be considered with 
care, since it does not correspond to the word `event'
in common English usage. 
A projection operator $ E \notin \{ 0, I \} $ 
has spectrum $ \{ 0, 1 \} $. 
Sometimes it is said that a quantum event is 
a YES-NO phenomenon. 
In other words, a quantum event has exactly 
two possible eigenvalues. 
But a physical event colloquially means that 
only one thing has occurred. 
For example, suppose there is 
one beta unstable, radioactive nucleus  
in an atomic trap. 
Suppose that it decays in a certain time period. 
Common usage has it that beta decay occurred. 
But if it does not decay in that time period, 
common usage has it that nothing happened, that 
there was no physical event. 
But the second alternative is just the NO or $ 0 $ 
eigenvalue of the event of beta decay in that time 
interval. 
In either case we have the {\em same} quantum event, 
but with two distinct values. 
For example, 
we say that $ P_{A} (B) $ is the quantum 
event that the observable $ A $ takes a value in $ B $. 
But in general this quantum event has {\em two} eigenvalues.  
As with any non-trivial self adjoint operator 
a quantum event can assume more than one value. 
We can also think of the eigenvalue $ 1 $ of 
the event $ P_{A} (B) $ as meaning that 
a measurement associated with $ A $ produced 
a value in $ B $, while the eigenvalue $ 0 $
says that the same measurement produced a 
value in $ \mathbb{R} \setminus B $, the 
set complement of $ B $ in the real line. 
But measurements do not play a distinguished 
role in this treatise; they are just 
a particular type of event. 
A nuclear reaction in a star in a distant galaxy 
is an event and so is an atomic transition  
measured in a laboratory across town. 
And events do play a central role here. 

It is traditional to speak in quantum theory 
in terms of the self-adjoint operators acting
in a given Hilbert space $ \mathcal{H} $. 
This is not logically necessary, since in 
functional analysis one proves that these
mathematical structures are in bijective 
correspondence with two other structures. 
These structures are on the one hand 
projection valued measures (pvm's)
with values in $ \mathcal{L}  (\mathcal{H}) $
defined
on the Borel $ \sigma $-algebra of 
$ \mathbb{R} $ 
and on the other hand 
strongly continuous unitary groups 
acting on $ \mathcal{H} $. 
For example, one can define the commutativity
of two self-adjoint operators $ S $ and $ T $
by one of these equivalent definitions: 
\begin{itemize}
	
	\item 
	Let $ P_{S} $ and $ P_{T} $, denote the
	pvm's of $ S $ and $ T $, 
	respectively. 
	Then we say that {\em $ S $ and $ T $ commute} if
	$ P_{S} (B) $ commutes with 
	$ P_{T} (C) $ for all Borel subsets 
	$ B, C $ of $ \mathbb{R} $. 
	(Notice that each of the families 
	$ \{  P_{S} (B) \,|\, 
	B \in \mathcal{B} (\mathbb{R}) \} $
	and 
	$ \{  P_{T} (C) \,|\, 
	C \in \mathcal{B} (\mathbb{R}) \} $
	is commutative. 
	This definition requires that their union 
	is also a commutative family.) 
	
	\item
	Let $ e^{i r S } $ and $ e^{i s T} $ 
	for $ r,s \in \mathbb{R} $	
	denote the unitary groups of $ S $ and 
	$ T $, respectively. 
	Then we say that {\em $ S $ and $ T $ commute} if 
	$ e^{i r S } $ commutes with $ e^{i s T} $ 
	for all $ r,s \in \mathbb{R} $. 
	(Notice that here as well 
	each of the families of unitary 
	operators $\{ e^{i r S } \,|\, r \in \mathbb{R} \}$
	and $\{ e^{i s T } \,|\, s \in \mathbb{R} \}$ is 
	commutative. 
	This definition requires that their union 
	is also a commutative family.) 
	
\end{itemize}

The next axiom is really 
just a continuation of Axiom~1. 
It is included as a separate axiom because of tradition.

\section{\bf Axiom 2 - States and Events}
\markboth{Axiom 2}{Axiom 2}

{\bf Axiom 2: (States and Events)}
Every quantum system is described by probabilities 
that are computed 
by using its set of events (see Axiom~1)
and the set of states of its von
Neumann algebra $ \mathcal{V} $. 
$ \blacksquare $

\vskip 0.2cm \noindent 
Unfortunately, the terminology `state' is so widely
used that there is no hope of ever replacing it
with a neutral term. 
It is a term that drips with meanings from classical 
physics as well as from everyday life. 
But in quantum theory as presented here it is a
mathematical term with a mathematical definition,
given below, which 
clearly says a state is associated with a quantum system and not with any subsystem or 
particle composing that system. 
We shall see the import of this remark when 
considering entanglement 
in Chapter~\ref{Entanglement-Chapter}. 

In the following we will mainly discuss the 
case when the von Neumann algebra
$ \mathcal{V} = \mathcal{L} (\mathcal{H}) $, 
since many quantum systems are covered by this case. 
In introductory texts it is the only case considered,
since that already puts a lot of mathematical burden 
on beginners. 
I did this myself in \cite{path}.  
While this is  the only case familiar to many physicists,
there are other von Neumann algebras used in 
quantum theory.  
When a physicist speaks of a Hilbert space as being the 
setting for the discussion of a quantum system, 
the underlying, usually implicit assumption 
is that the appropriate von Neumann algebra is 
$ \mathcal{L} (\mathcal{H}) $. 
But that could be an error. 
The problem is to find the 
correct von Neumann algebra,  
and not the correct Hilbert space 
as is often thought.  
This is just a part of the problem of 
{\em quantization}. 
An important point here 
is that the lattice of events is central 
to quantum theory, and its structure depends on the 
von Neumann algebra being used. 
Note that the quantization problem is not a 
problem within quantum theory itself, 
but rather a 
problem of how to arrive at a 
particular quantum theory. 
This non-trivial and important 
problem will not be
addressed in this treatise. 

\begin{definition} 
A {\em state} is 
as a linear functional 
$ l : \mathcal{L} (\mathcal{H}) \to \mathbb{C} $ 
satisfying 
\begin{itemize}
	
	\item 
	(Positivity Preservation) 
	\\
	$ l(T) \ge 0 $ 
	for all $ T \in \mathcal{L} (\mathcal{H}) $ 
	that are {\em positive} (meaning that
	$ T = T^{*} \ge 0$). 
	
	\item	(Normalization) 
	\\
	$ l(I) = 1 $, where
	$ I \in \mathcal{L} (\mathcal{H})$ 
	denotes the identity operator.
		
\end{itemize}
\end{definition}

It turns out that every state $ l $ is a bounded 
linear functional with  norm $ || l || = 1 $. 
So $ l \in ( \mathcal{L} (\mathcal{H}) )^{\prime} $, 
the dual Banach space of the Banach space 
$ \mathcal{L} (\mathcal{H}) $. 
We let $ \mathcal{S} $ denote the set
of all states.  
Then $ \mathcal{S} $ is a convex subset of 
$ ( \mathcal{L} (\mathcal{H}) )^{\prime} $. 

An immediate consequence of this axiom is that every 
{\em unit vector} $ \psi \in \mathcal{H}$,
that is $ || \psi || = 1 $, determines a state
$ l_{\psi} $ defined by
$ l_{\psi} (T) := \langle \psi, T \psi \rangle $ for 
all $ T \in \mathcal{L} (\mathcal{H}) $. 
Moreover, the state $ l_{\psi} $ does {\em not} uniquely
determine a unit vector that defines it,
since $ l_{\psi} = l_{\varphi} $ for all unit 
vectors $ \varphi $ that satisfy $ \varphi = \lambda \psi $ 
for some $ \lambda \in \mathbb{C} $. 
(Necessarily we have $ |\lambda | = 1 $.) 
The easy proof of these facts is left to the reader. 
Such a state $ l_{\psi} $ is called a {\em pure state}.

There are more states. 
Every trace class operator $ D = D^{*} \ge 0$ 
satisfying $ Tr (D) = 1$ is called a {\em density matrix}. 
Any such density matrix $ D $ defines a state 
$ l_{D} $ by $ l_{D} (T) := Tr (D T ) $ for 
all $ T \in \mathcal{L} (\mathcal{H}) $. 
We say that $ l_{D} $ is a {\em mixed state}. 
It is an exercise in functional analysis to show 
that $ l_{D} $ is a state and that this state 
uniquely determines the density matrix $ D $. 
Also, any state $ l_{\psi} $, where $ \psi $ is a 
unit vector, defines a {\em unique} 
density matrix $ D = | \psi \rangle \langle \psi | $ 
(in Dirac notation) 
such that $ l_{D} = l_{\psi} $. 
However, if $ \dim_{\mathbb{C}} \mathcal{H} \ge 2 $,
it is easy to construct 
mixed states which are not pure states. 

I have carefully distinguished the terminology here. 
Often the expressions `density matrix' and
`mixed state' are conflated as if they were 
synonyms. 
But that is not quite so. 
Gleason's Theorem (see \cite{gleason})
speaks to this question. 

In common parlance, especially among physicists, 
the expression mixed state is often
reserved for those states $ l_{D} $ 
which are not pure states. 
This is due in part to the emphasis given to 
pure states in many applications. 
Indeed, 
often one only considers the pure states. 
It is important to note that the set of pure 
states is in bijective correspondence with 
the set of all one-dimensional subspaces of 
the Hilbert space $ \mathcal{H} $. 
The latter set is called the {\em (complex) projective 
space} associated to $ \mathcal{H} $ and 
is denoted as $ \mathbb{C} \mathrm{P} (\mathcal{H}) $. 
This projective space has many interesting 
mathematical properties. 

The set of quantum events 
should be in bijection with 
the physical events of a given quantum system. 
If not, then one has not correctly chosen the
von Neumann algebra for the system. 
The choice of the self-adjoint 
operators and the states, which are physically relevant, 
is not so obvious. 
It is generally accepted that 
the self-adjoint operators should be associated 
with the von Neumann algebra, as specified in 
Axiom~1, though it may be the case that not all 
such self-adjoint operators will have a 
physical significance. 
As for the states, my preference is to include first only 
those states 
$ | \psi \rangle \langle \psi | $ (with $ \psi $ 
a unit vector in $ \mathcal{H} $) 
that are also events in the
von Neumann algebra and then the mixed states 
formed from these, that is, 
their closed convex hull. 
To my way of thinking, this is not an {\em ad hoc} 
super-selection rule, but rather a way of carefully 
choosing a self-consistent model for a quantum system.
Such a choice of model is called a 
{\em quantization}. 
A rule of thumb is that a quantization should be 
based on a von Neumann algebra that is generated 
by the smallest set of events needed to understand 
the quantum system.
However, quantization remains to this day 
as much an art as a science.

\section{\bf Axiom 3 - Spin and Statistics}
\markboth{Axiom 3}{Axiom 3}

{\bf Axiom 3: (Spin and Statistics)}
The most basic quantum entities are either 
{\em bosons}, all of which 
have integer spin, or {\em fermions}, all of 
which have half-integer spin.  
All other quantum entities are composites of these. 
$ \blacksquare $

\vskip 0.2cm 
The Hilbert space $ \mathcal{H} $ 
for a single boson or a single fermion 
carries a {\em unitary representation} 
of the {\em Lie group} $ SU(2) $. 
That representation defines the value of the spin. 
It seems that all `matter' is composed of fermions,
while all `interactions' are mediated by 
bosons. 
But {\em dark matter} could be something else; 
we simply do not know. 
Also, gravitation is thought 
by many physicists 
to be mediated by {\em gravitons}, 
a spin~$ 2 $ boson. 
However, we simply do not know if that is correct. 
So, this axiom may be changed some day. 

The Hilbert space 
of composite quantum entities is given in terms of the 
Hilbert spaces of the constituent 
bosons and fermions by a 
non-intuitive 
mathematical construction. 
As noted in Axiom~2, 
states are associated with the 
total Hilbert space of a quantum system, or 
more specifically with its von Neumann algebra. 
However, the partial trace does define a map 
from the states of the system to the states of any 
subsystem, but as is well known 
 it does not map pure states to pure states. 
So Axiom~3 is the place in this approach where 
partial trace should be introduced. 
Even though 
the partial trace is justifiably considered to be a 
non-commutative version of conditional 
expectation, it is not related to quantum 
conditional probability 
(see Section~\ref{section-conditional-prob}) 
or to quantum integrals 
(see section~\ref{section=quantum-integrals}). 
These remarks become quite relevant when 
considering entanglement 
in Chapter~\ref{Entanglement-Chapter}. 

This axiom is included because of its importance in 
quantum theory. 
However, it is not presented in total detail since
it is not going to play a role in this treatise, 
which focuses on probability and leaves 
spin, statistics and partial trace to a side. 
This axiom is a complicated condition that 
seems to have nothing to do with the other 
axioms, and 
it introduces spin into 
quantum theory in a seemingly {\em ad hoc} manner. 
If you like to think in terms of analogies, 
then the Fifth Axiom of Euclid and 
the Axiom of Choice come to mind. 
While generations have fretted over deeper 
explanations of the origin of probability in 
quantum theory, nary anybody is much concerned with 
an underlying explanation of spin. 
But spin seems to be as much an essential, 
non-classical ingredient of quantum theory as
is probability. 
The reader can refer to the literature 
on this topic, most notably in quantum field
theory. 

Another mathematical structure introduced in this 
axiom is that of a representation. 
There is no denying that 
this is an important mathematical 
concept in quantum physics 
(see \cite{woit}),    
but it will not be 
playing any role in this treatise. 
Perhaps some comments on this omission are in 
order. 
For example, the group symmetries of space and time,
respectively, lead to the laws of conservation of 
linear momentum and energy, respectively. 
Even though 
they are part of relativistic quantum theory, 
these symmetries are not a part of  
basic quantum theory as 
presented in this treatise.  
Actually, the basics presented here are not
explicitly invariant under
representations of the Galilean group nor the 
Poincar\'e group, but rather
consistent with either of these groups. 
Thinking that quantum theory automatically 
carries a representation of the Poincar\'e group 
led to the mistaken conviction that parity is 
conserved in all interactions in quantum theory. 
We also will not be considering quantum 
interactions explicitly, except 
in so far as to say that 
they are a type of quantum event.  

It may seem strange that the non-intuitive 
concept of spin enters an axiom, while the 
intuitive concept of position does not. 
Note that position, together with linear 
momentum, enter quantum theory via a 
representation of the Weyl-Heisenberg group. 
This could be taken as a shortcoming of the approach 
taken in this treatise. 
However, none of these particular observables 
will play a role here, though all of them 
could be present in a more complete 
axiomatization of quantum theory.

\section{\bf Axiom 4 - Time Independent Born's Rule}
\markboth{Axiom 4}{Axiom 4}

{\bf Axiom 4: (Time Independent Born's Rule)}
Let $ T $ be any self-adjoint operator 
(perhaps associated with 
an observable quantity of some physical entity), 
$ \rho $ be a density matrix 
and $ B $ be a Borel subset of $ \mathbb{R} $. 
Then the 
{\em quantum probability} 
that $ T $ has a value 
in the set $ B $ given $ \rho $ is defined to be  
$ 
P ( T \in B \,|\, \rho ) :=
Tr (  P_{T} (B) \rho ). 
$ 
Here $ Tr $ is the trace of a trace class operator, and 
$ P_{T} $ is the pvm associated to 
the self-adjoint operator $ T $. 
$ \blacksquare $

\vskip 0.2cm 
This axiom, or a simple consequence of it about
expected values, is part of standard quantum
theory as found in the textbooks and as practiced 
in the scientific community. 
Only later on will we put time dependence into 
this and, most importantly, 
elevate the resulting formula to be the 
basic time evolution equation of quantum theory.  

For fixed $ T $ and $ \rho $
the assignment 
$ B \mapsto P ( T \in B \,|\, \rho )  \in [0,1]$ 
for $ B $ a Borel subset of $ \mathbb{R} $
is a {\em probability measure} on $ \mathbb{R} $ 
in the sense of measure theory. 
The {\em physical significance} of quantum probability 
is that it corresponds
to the {\em relative frequency} of the 
empirically observed quantities associated to 
the self-adjoint operator $ T $. 

To show the two  normalizations of this 
probability measure, we note first that 
$
P (T \in \emptyset \,|\, \rho) = 
Tr ( P_{T} (\emptyset) \, \rho  ) = 
Tr ( 0 \, \rho) = Tr (0) = 0  
$
and second that 
$
P (T \in \mathbb{R} \,|\, \rho) = 
Tr ( P_{T} ( \mathbb{R} ) \, \rho ) =
Tr ( I \, \rho ) = Tr ( \rho ) =1. 
$

To show $ \sigma $-additivity 
 let $\{ B_{j} \,|\, j \in \mathbb{N} \}$ be
a countable family of disjoint Borel subsets of  
$ \mathbb{R} $. 
We can calculate the trace of a trace class operator 
using any orthonormal basis of $ \mathcal{H} $. 
We choose an orthonormal basis 
$\{ \phi_{k} \}$ which diagonalizes 
the trace class operator $ \rho $. 
Specifically, $ \rho \, \phi_{k} = \lambda_{k} \, \phi_{k}$ 
with all $ \lambda_{k} \ge 0 $ and  
$ \sum_{k} \lambda_{k} = 1 $. 
Then we have 
\begin{align*}
P( T \in \cup_{j} B_{_{j}} \,|\, \rho ) &= 
Tr \big( P_{T} ( \cup_{j} B_{_{j}} ) \, \rho \big)
\\
&= 
\sum_{k} \langle \phi_{k} , 
P_{T} ( \cup_{j} B_{j} ) \, \rho \, \phi_{k} \rangle \\
&= 
\sum_{k} \langle \phi_{k} , 
\sum_{j} P_{T} ( B_{j} ) \, \rho \, \phi_{k} \rangle 
\quad \mathrm{using~strong~operator~topology}
\\
&=
\sum_{k} \langle \phi_{k} , 
\sum_{j} P_{T} ( B_{_{j} } ) 
( \lambda_{k} \, \phi_{k} ) \rangle 
\\
&=
\sum_{k} \sum_{j} \lambda_{k}  \langle \phi_{k} , 
  P_{T} ( B_{_{j}} ) \, \phi_{k} \rangle 
\\
&=
\sum_{j} \sum_{k} \lambda_{k}  \langle \phi_{k} , 
P_{T} ( B_{_{j}} ) \, \phi_{k} \rangle 
\quad \mathrm{using~Fubini's~theorem}
\\
&=
\sum_{j} \sum_{k}  \langle \phi_{k} , 
P_{T} ( B_{_{j}} ) ( \lambda_{k} \, \phi_{k} ) \rangle 
\\
&=
\sum_{j} \sum_{k}  \langle \phi_{k} , 
P_{T} ( B_{_{j}} ) \, \rho \phi_{k}  \rangle 
\\
&=
\sum_{j} Tr \big( P_{T} ( B_{j} ) \, \rho \big)
\\
&=
\sum_{j} P (T \in B_{j} \,|\, \rho). 
\end{align*}
We can apply Fubini's theorem since 
all of the terms in the double sum are non-negative. 

We discuss mainly the case of pure states 
$ \phi \in \mathcal{H} $ where
$ || \phi || =1 $, in which case 
$ \rho =  | \phi \rangle \langle \phi | $. 
We then introduce the notation
$$
    P ( T \in B \,|\, \phi ) := 
    Tr ( P_{T} (B) \, | \phi \rangle \langle \phi | ) 
    =
    \langle \phi, P_{T} (B) \phi \rangle, 
$$
which we read as: 
The probability that $ T $ has a value in $ B $
given $ \phi $. 
The notation as well as the way of reading it
suggests that this is related to the 
concept of conditional probability in classical 
probability. 
We shall come back to this point. 

There are alternative formulas for Born's rule 
for the case of pure states. 
Here are two useful ones:
$$
P ( T \in B \,|\, \phi ) = || P_{T} (B) \phi ||^{2} 
= Tr ( P_{T} (B) E_{\phi} ).  
$$
Here $ E_{\phi} := | \phi \rangle \langle \phi | $ 
in Dirac notation is a rank $ 1 $ projection, 
and $ Tr $ denotes the trace of a trace class operator. 
Note that $ E_{\phi} $ is a both a quantum event and a
density matrix. 
 
Axiom~4 is the simplest form of Born's rule. 
It will be generalized and as such its central 
importance in quantum theory will become apparent. 
Note that {\em quantum probability} in this treatise 
simply means all possible forms of Born's rule.

\section{\bf Axiom 5 - Dynamics}
\markboth{Axiom 5}{Axiom 5}

{\bf Axiom 5: (Dynamics)}
Every physical system has two associated actions of 
one-parameter groups. 
These are denoted as $ S_{t} $ and $ E_{t} $, 
where $ t \in \mathbb{R} $ 
is considered as a 
parameter in the theory which corresponds to 
time. 
(By {\em one-parameter group} we mean that $ S_{a} S_{b} = S_{a+b} $
for all $ a,b \in \mathbb{R} $ and that 
$ S_{0} = id $, the appropriate identity 
map. 
Similar formulas hold for $ E_{t} $.) 
 
The one-parameter 
group $ S_{t} $ maps the convex set $ \mathcal{S} $ of
{\em all} states (including mixed states) to itself, while
the one-parameter group $ E_{t} $ maps the set 
$ \mathcal{E} $ of all quantum events to itself. 

The {\em dynamics (or time evolution)} 
of the physical system 
itself 
for an initial 
observable represented by 
a self-adjoint operator 
$ T = T^{*}$ and an initial density matrix 
$ \rho $ is given by 
the {\em time dependent Born's rule} 
\begin{equation}
\label{time-depend-borns-rule}
t \mapsto 
Tr ( E_{t} ( P_{T} (B) ) \, S_{t} \rho ) \quad 
\mathrm{for~} t \in \mathbb{R}, 
\end{equation}
provided that  {\em conservation of probability} holds,
which by definition means that for fixed $ \rho $ 
and $ T = T^{*} $  the mapping 
$ B \mapsto Tr ( E_{t} ( P_{T} (B) ) \, S_{t} \rho ) $ 
is a probability measure for every $ t \in \mathbb{R} $. 
Here {\em initial} means at time $ t = 0 $. 
$ \blacksquare $

\vskip 0.2cm 
The reason for having two one-parameter groups
in \eqref{time-depend-borns-rule} will become 
apparent in the next section. 

Following tradition 
I have stated these as five separate axioms. 
Nonetheless, logically speaking 
Axioms~1 and 2 form one statement
on kinematics, while Axioms~4 and 5 are one 
statement on dynamics. 
Axiom~3 as noted above stands out as being quiet 
different, although in a technical sense it is also
a statement on kinematics. 
A long, convoluted statement. 

A special case of Axiom~5 
is when $ S_{t} $ maps the 
set of pure states to itself. 
Then we have that 
\begin{equation}
\label{St-of-pure-state}
    S_{t} | \phi \rangle \langle \phi | =
    | \phi_{t} \rangle \langle \phi_{t} | 
\end{equation}
for some unit vector $ \phi_{t} \in \mathcal{H} $  
which is not uniquely determined, but is unique modulo 
a phase factor. 
In this case the dynamics is given by another 
version of Born's rule:
\begin{equation}
\label{common-borns-rule}
t \mapsto 
Tr ( E_{t} ( P_{T} (B) ) \, | 
\phi_{t} \rangle \langle \phi_{t} | ) =
\langle \phi_{t}, E_{t} ( P_{T} (B) ) \phi_{t} \rangle 
= || E_{t} ( P_{T} (B) ) \phi_{t} ||^{2}\!\!, 
\end{equation}
provided again that 
conservation of probability holds.

\section{\bf Models of the Axioms}
\markboth{Models of Axioms}{Models of Axioms}

The time evolution of neither the events 
(given by $ E_{t} $)
nor of the states (given by $ S_{t} $) 
is fundamental.  
What is fundamental in the sense that it 
corresponds to observations is the combination of 
these two one-parameter groups in the  above expressions 
for the time dependent probability 
given by Born's rule. 
Neither of these two 
one-parameter groups is uniquely 
determined by the axioms. 
There are various {\em models}
of these axioms for which $ S_{t} $ and 
$ E_{t} $ are quite different. 
Some of these models are typically called 
{\em pictures} in the literature. 
Axiom~5 was written to include the three most 
commonly used
models, which we present next. 
But other models are possible. 

I am not the first to 
say that quantum theory is 
a way for calculating probabilities, and 
nothing else. 
But I am saying that different models have 
different ways for doing those calculations. 
Moreover, the fundamental, final formula 
for arriving at those calculations
is the same in all models (namely, 
the Generalized Born's Rule),
which is {\em covariance}, 
and that the resulting number is the same in all models, 
which is {\em invariance}. 

The most commonly used model is called the
{\em Schr\"odinger picture}, but I prefer to 
call it the {\em Schr\"odinger model}. 
This model has  
the property that $ E_{t} = id $ for
all $ t \in \mathbb{R} $, that is,
the events have trivial time evolution. 
One also says that the observables, 
which are events since they are pvm's, 
are {\em time independent} in this model. 
In this model 
the time evolution maps pure states 
to pure states and 
is given by 
{\em Schr\"odinger's equation}. 
The solution of this equation with an initial 
condition $ \phi $ at time $ t = 0 $ is 
given by functional analysis 
as $ \phi_{t} =  e^{- i t H / \hbar} \, \phi $, 
where $ H = H^{*} $ is the Hamiltonian in 
Schr\"odinger's equation. 
(The minus sign in the exponent of $ e^{- i t H / \hbar} $ 
is a purely conventional.    
It has no physical significance;
it does not need to be `explained'.) 
Taking the initial condition $ \phi $ 
to be a unit vector, it follows by the unitarity of 
$ e^{- i t H / \hbar} $ that $ \phi_{t} $ is 
a unit vector for all $ t \in \mathbb{R} $. 
Then $ S_{t} $ in the Schr\"odinger model 
is defined on pure states by the 
formula \eqref{St-of-pure-state},  
and the time dependent Born's rule 
\eqref{common-borns-rule} 
gives the  
probability that the observable $ T = T^{*} $
is in the Borel subset $ B $ of $ \mathbb{R} $
as a function of
time $ t \in \mathbb{R} $ as
\begin{equation}
\label{schrodinger-model-borns-rule}
  t \mapsto \langle \phi_{t}, P_{T} (B) \phi_{t} \rangle.  
\end{equation}
Since $ \phi_{t} $ is a unit vector, for each fixed time
$ t $ this is a probability measure 
as we have seen earlier.  
That is to say,
conservation of probability holds in 
the Schr\"odinger model.

Notice that Schr\"odinger's equation 
is pushed into the background even in the 
Schr\"odinger model. 
All that is important is the Hamiltonian $ H $
which in and of itself specifies the flow 
$ t \mapsto \phi_{t} =  e^{- i t H / \hbar} \, \phi $ 
in the Hilbert space~$ \mathcal{H} $. 
However, 
there is a preference for thinking that differential 
equations are basic and that their solutions,
in this case the flow, are secondary. 
Of course, to understand this flow in specific 
cases it often is a good idea to solve 
Schr\"odinger's equation.  
In introductory texts it is usually considered to be 
pedagogically advantageous to give 
Schr\"odinger's equation a central role. 
And I do this in \cite{path} for example. 
But this should be taken by those with more
knowledge with a ton, not a grain, of salt, 
since \eqref{schrodinger-model-borns-rule} is 
the fundamental time evolution equation 
for one observable in 
the Schr\"odinger model for pure states. 
For those who prefer differential equations, 
the time derivative of 
\eqref{schrodinger-model-borns-rule} 
formally gives for any Borel subset 
$ B $ of $ \mathbb{R} $ that 
\begin{equation}
\label{ode-for-borns-rule}
\dfrac{d}{d t}
 \langle \psi, P_{T} (B) \psi \rangle = 
 \langle \psi , 
 \dfrac{i}{\hbar} [ H, P_{T} (B) ] \, \psi \rangle. 
\end{equation}
where $ \psi = \phi_{t} $ and 
$ [ R, S]:= R S = S R $ is the {\em commutator} 
of the two operators $ R , S $. 
This is an easy non-rigorous exercise, provided 
domain considerations are ignored. 
As is typical, the differential equation 
\eqref{ode-for-borns-rule},
which in no way do I  wish to 
consider to be fundamental, 
is a more singular object than its `solution' 
\eqref{schrodinger-model-borns-rule}, which 
I do consider to be fundamental.  
The mapping 
$ B \mapsto \frac{i}{\hbar} [ H, P_{T} (B) ] $
maps the empty set $ \emptyset $ to $ 0 $ and
is formally $ \sigma$-additive . 
But it also maps $ \mathbb{R} $ to $ 0 $, and  
so it is not a reasonable 
operator valued measure, even 
if $ H $ is a bounded operator. 
On the other hand,  
\eqref{ode-for-borns-rule} is 
formally valid in the Heisenberg and interaction 
models, as described below. 
It seems that for Born's rule 
there is no `nice' differential 
equation that holds (even formally) in all models. 

Next, we extend $ S_{t} $  
to act on density matrices $ \rho $ as 
follows. 
First, by the spectral theorem there exists 
an orthonormal basis $\{ \phi_{k} \}$ 
of $ \mathcal{H} $
and real numbers $ \lambda_{k} \ge 0 $
such that $ \sum_{k} \lambda_{k} = 1$ and
$
  \rho = \sum_{k} \lambda_{k} 
  | \phi_{k} \rangle \langle \phi_{k} | 
$.
We also define $ U_{t} := e^{- i t H / \hbar}  $, the
unitary time evolution operator, 
for all $ t \in \mathbb{R} $. 
The time evolution of the ket is 
$| \phi_{k} \rangle \mapsto U_{t} | \phi_{k} \rangle $, while that of its dual bra is 
$ \langle \phi_{k} | \mapsto \langle \phi_{k} | U_{t}^{*}$. 
So, 
$  | \phi_{k} \rangle \langle \phi_{k} | 
\mapsto U_{t} | \phi_{k} \rangle \langle \phi_{k} | U_{t}^{*}$. 
Putting this together, the time evolution of $ \rho $ is
defined by 
$$
\rho \mapsto \sum_{k} \lambda_{k} 
\big( 
U_{t}| \phi_{k} \rangle \langle \phi_{k} | U_{t}^{*} \big) 
= 
U_{t}
   \Big( 
      \sum _{k}| \lambda_{k} \phi_{k} \rangle \langle \phi_{k} | 
   \Big) 
U_{t}^{*}
=  
U_{t} \, \rho \, U_{t}^{*}. 
$$
This model is so widely used that properties
specific to it are often thought to be 
general properties of quantum theory.  
I will discuss this misunderstanding 
in more detail later on.

Another model is called the 
{\em Heisenberg picture}, which I prefer
to call the {\em Heisenberg model.}
In this model one has $ S_{t} = id $ for
all $ t \in \mathbb{R} $. 
One says that the states are {\em time independent}
in this model. 
In particular, $ S_{t} $ maps pure states to pure states. 
The time evolution of a quantum event $ P $ is given 
in this model by the {\em Heisenberg equation} 
\begin{equation}
\label{Heisenberg-eqn}
P \mapsto E_{t} \, P := U_{t}^{*} P U_{t}, 
\end{equation}
where $ U_{t} $ for $ t \in \mathbb{R} $ 
is a strongly continuous  
unitary group acting on $ \mathcal{H} $. 
It is important to note that this flow is basic; 
it is not arrived at as 
the solution of a differential equation. 
Next it follows that  
$ U_{t} := e^{- i t H / \hbar} $ with $ H = H^{*} $ by 
Stone's theorem. 
(The minus sign in the exponent of $ e^{- i t H / \hbar} $ 
is again purely conventional.  
It has no physical significance;
it does not need to be `explained'.) 
If $ P_{A} $ is the pvm of a self-adjoint operator $ A $,
then the same time evolution applies to it:
$$
P_{A} \mapsto E_{t} \, P_{A} := U_{t}^{*} P_{A} U_{t}, 
$$
where 
$ U_{t}^{*} P_{A} U_{t} (B) := U_{t}^{*} P_{A} (B) U_{t} $ 
for all Borel subsets $ B $ of $ \mathbb{R} $. 
It turns out from functional analysis that this time 
evolution maps a pvm, which is a family of projections,
to another family of projections, which turns out itself
to be a pvm. 
It follows that the pvm $ P_{A} $ of a self-adjoint operator 
maps to the pvm of another self-adjoint operator. 
Explicitly, one can show that 
$$
E_{t} \, P_{A} = 
U_{t}^{*} P_{A} U_{t} =  P_{ U_{t}^{*}A U_{t} }. 
$$

Again, the Heisenberg 
model has certain specific properties 
which are not general properties 
of quantum theory. 
Such a particular property of the Heisenberg model is 
that $ E_{t} $ extends naturally to a time evolution of 
$ \mathcal{L} (\mathcal{H})$ 
defined by $ E_{t} \, T := U_{t}^{*} \, T \, U_{t} $ 
for all  $T \in \mathcal{L} (\mathcal{H})$. 
The group $ E_{t} $ in any of these manifestations 
is difficult to accommodate with ordinary intuition. 
This says that observables are time dependent, which 
might seem sensible enough if one is speaking of position 
or angular momentum. 
However, events are self-adjoint operators; 
so they are observables too. 
So, in the Heisenberg model events are time dependent! 
On the other hand states, which intuitively tell us 
everything about a system at any moment, 
are time independent.
This is backwards from the common intuition of what
`events' are and what `states' are. 
In part this is due to a poor choice in 
terminology. 
A quantum event $ E \notin \{ 0, I \} $ 
is an observable that 
has spectrum $ \{ 0, 1\} $. 
This corresponds to two possible observed values. 
Sometimes an event is called a Yes-No 
experiment. 
We tend to think that the value $ 1 $ (Yes) 
means that the event occurred, while 
the value $ 0 $ (No) means that the event did not occur. 
But this is misleading; 
an event is an observable that can give 
either of these two values. 
In either case the event has been observed. 
In any event (in another sense yet of that word) 
I find 
quantum theory to be non-intuitive. 

Also notice that 
the time evolution of events in the Schr\"odinger 
model extends naturally to 
$ \mathcal{L} (\mathcal{H})$ by setting 
$ E_{t} := I_{\mathcal{L} (\mathcal{H})} $ 
for all $ t \in \mathbb{R} $. 
This particular extension property is not 
a necessary aspect of quantum theory. 

To show that 
conservation of probability holds in 
the Heisenberg model 
let $ \rho $ be a density matrix and
$ T = T^{*} $ be a self-adjoint operator. 
Then 
\begin{align}
Tr ( E_{t} (  P_{T}  (B) ) \, \rho ) &= 
Tr ( U_{t}^{*} P_{T}  (B) U_{t} \rho ) 
=
Tr ( P_{T}  (B) \, U_{t} \rho  U_{t}^{*} ) 
\nonumber 
\\
&=
Tr ( P_{T}  (B) \rho_{t} ), 
\label{hei-to-sch} 
\end{align}
where $ \rho_{t} := U_{t} \rho U_{t}^{*} $,
the time evolved density matrix 
for every $ t \in \mathbb{R} $, is also a 
density matrix. 
But we have already proved in our discussion of the
Schr\"odinger model that the last 
expression gives a probability measure on $ \mathbb{R} $. 

The same Born's rule \eqref{common-borns-rule}
is used in the Schr\"odinger model and the 
Heisenberg model. 
What that means is that the 
same {\em formula} is used in both. 
This is {\em covariance}.  
But more is true. 
The two models give the same numerical probability using 
\eqref{common-borns-rule} provided that the same 
self-adjoint operator $ H $ is used in both models. 
This is {\em invariance}.
(In other words, {\em compatible} sign conventions 
are being used
in the two models. 
This does have a significance, since it makes 
the following proof work.) 

Here is the proof of this well-known, yet 
important result. 
We assume that in both models 
at time $ t = 0 $ the state is
$ \phi $  and that the pvm is $ P_{T} $. 
Of course, in the Schr\"odinger model only 
the state can be time dependent, while in the Heisenberg 
model only the pvm can be time dependent. 
Be aware that all of the following 
expressions are time dependent. 
Then for all $ t \in \mathbb{R} $ we have 
in an obvious notation that 
\begin{align*}
  P^{Sch} ( T \in B \,| \, \phi_{t}) &=  
 ||  P_{T} (B)  \, \phi_{t} ||^{2}
 \\
 &=
 ||  P_{T} (B) \, e^{-i t H / \hbar} \, \phi ||^{2} 
 \\
 &=
 ||  e^{i t H / \hbar} P_{T} (B)  \, e^{-i t H / \hbar} \, \phi ||^{2} 
 \\
 &=
 || U_{t}^{*}  P_{T} (B) U_{t} \, \phi ||^{2}
 \\
 &=
 || E_{t} ( P_{T} (B) ) \phi ||^{2} 
 = P^{Heis} ( E_{t} T \in B \,|\, \phi ). 
\end{align*}

This is, of course, a special case of the following 
calculation.
Let $ \rho $ be a density matrix. 
Then we see that 
\begin{align}
P^{Sch} ( T \in B \,|\, \rho_{t} ) &= 
Tr ( P_{T} (B) \, \rho_{t} ) 
= 
Tr ( P_{T} (B) \, U_{t} \rho U_{t}^{*} ) 
\nonumber 
\\
&=  
Tr ( U_{t}^{*} P_{T} (B) \, U_{t} \rho  ) 
=  
Tr ( E_{t} P_{T} (B) \,  \rho  ) 
\nonumber 
\\
&=  
\label{sch-equals-hei}
P^{Heis}  ( E_{t} T \in B \,|\, \rho  ). 
\end{align}
This calculation should be compared with 
\eqref{hei-to-sch}. 

But still other models are used for which both 
$ S_{t} $ and $ E_{t} $ are non-trivial. 
The {\em interaction pictures} are such models. 
There are a multitude of such models. 
One starts by taking 
any one-parameter {\em family} of unitary 
operators $ U_{t} $ for $ t \in \mathbb{R} $ acting on the 
Hilbert space $ \mathcal{H} $. 
Next one changes the Schr\"odinger model 
to become the new {\em interaction model} 
by transforming the pure states 
by $ \psi \mapsto U_{t} \psi =: \psi^{\prime} $ 
and mapping the pvm's by 
$ P_{S} \mapsto U_{t} P_{S} U_{t}^{*} =:  P_{S^{\prime}} $,
where $ S $ is a self-adjoint operator and
$ U_{t} S U_{t}^{*} =:  S^{\prime} $. 
One then sees that 
$$
P( S \in B \,|\, \psi )  = 
\langle \psi , P_{S} (B) \psi \rangle =
\langle U_{t} \psi , U_{t} P_{S} (B) U_{t}^{*}
U_{t} \psi \rangle = 
P ( S^{\prime} \in B \,|\, \psi^{\prime} ). 
$$
This shows that the time dependent probability
as calculated in each model is 
exactly the same even though the time dependence of
both the states and the pvm's has been changed.  
The transformation from the Schr\"odinger model 
to the Heisenberg model is a special case of this. 
Another very special case is to take $ U_{t}:= U $
for all $ t \in \mathbb{R} $ where $ U $ is a fixed 
unitary transformation. 

Note that in the interaction model 
$ U_{t} $ need not be a unitary group nor 
does $ t \mapsto U_{t} $ need to have any sort of 
continuity.
The lack of continuity, for example, 
may seem to you  to be a mathematical trick 
with no physical intuition behind it. 
If so, you are right. 
The interaction model is just used as a convenient 
mathematical technique to help one 
calculate probabilities, and its intermediary steps 
have no physical significance.

All of these models are {\em isomorphic} 
(defined below) to the 
Schr\"odinger model, but there are other 
models which are not. 
The point of the axioms is to capture certain
features, which are to be considered as basic 
to quantum theory. 
They are not meant to be {\em categorical}, in the 
same sense as, for example,  
the axioms of Euclidean plane geometry are meant to 
describe completely their topic. 
Rather the axioms for quantum theory 
are meant to be like the axioms in mathematics of 
a vector space, of which there are many non-isomorphic objects. 
Similarly, as we shall discuss in detail later, 
there are many non-isomorphic quantum theories. 
Here is that important definition,  
where $\mathrm{Bij} (A)$ denotes 
the set of bijections of a set $ A $ to itself. 

\begin{definition}
A {\rm model of quantum theory} is a 
triple $ ( \mathcal{H}, E , S )$ 
where $ \mathcal{H} $ is a 
Hilbert space and
$ E: \mathbb{R} \to \mathrm{Bij} (\mathcal{E})  $ 
is a one-parameter group of bijections of the 
set $ \mathcal{E} $ of events in 
$ \mathcal{L} ( \mathcal{H})$ to itself, and 
$ S : \mathbb{R} \to \mathrm{Bij} (\mathcal{S}) $ 
is a one-parameter group of bijections of the 
set $ \mathcal{S} $ of states in 
$ \mathcal{L} ( \mathcal{H})^{\prime}$ to itself
such that for every $ t \in \mathbb{R} $
the {\rm time dependent Born's rule}
\begin{equation*}
t \mapsto 
Tr ( E_{t} ( P_{T} (B) ) \, S_{t} \rho ) 
\end{equation*}
is a probability measure on $ \mathbb{R} $ 
for every density matrix $ \rho  $, for all 
self-adjoint operators $ T $ and all Borel 
subsets $ B $ of $ \mathbb{R} $.  
As noted earlier this last condition is called 
{\rm conservation of probability}. 

An {\rm isomorphism} of the quantum theories 
$ ( \mathcal{H}, E, S )$ and 
$ ( \mathcal{H}^{\prime}, E^{\prime}, 
S^{\prime} )$ 
is a {\rm one-parameter family} of unitary transformations 
$ U_{t} : \mathcal{H} \to \mathcal{H}^{\prime}$ which are 
onto and satisfy condition 
(\ref{isomorphism-condition}) below. 
Here the parameter is $ t \in \mathbb{R} $. 
In the model $ ( \mathcal{H}, E, S )$ 
we let $ T = T^{*} $ be a self-adjoint operator and 
$ \rho $ be a density matrix. 
These correspond in the usual way in the model 
$ ( \mathcal{H}^{\prime}, E^{\prime}, 
S^{\prime} )$
to the self-adjoint operator  
$ T^{\prime} := U_{t} T U_{t}^{*} $ 
and to the density matrix 
$ \rho^{\prime} (A) := \rho ( U_{t}^{*} A U_{t} )  $ 
for all $ A \in \mathcal{L} (\mathcal{H}^{\prime} ) $. 
Then for all times $ t \in \mathbb{R} $ 
and for all Borel subsets $ B $ 
of $ \mathbb{R} $ we require that   
\begin{equation}
\label{isomorphism-condition}
Tr ( E_{t} ( P_{T} (B) ) \, S_{t} \rho ) = 
Tr ( E^{\prime}_{t} ( P_{T^{\prime}} (B) ) 
\, S^{\prime}_{t} \rho^{\prime} ) 
\end{equation}
which is called {\em preservation of probability}.
(This condition is based on
the equality of probabilities \eqref{sch-equals-hei} 
for the Schr\"odinger and Heisenberg models.) 

Finally, we say that two models of quantum theory 
are {\rm isomorphic} if there exists an 
isomorphism between them. 
\end{definition}

Please be careful to note the difference between 
{\em conservation of probability}
and
{\em preservation of probability}.
The definition of isomorphism of models must be 
modified in an obvious 
way to include preservation of probability 
for consecutive and
conditional probabilities after these are defined 
later on. 

There will surely be be those who wish to maintain 
that the time changing 
state and the Schr\"odinger model 
are correct while the corresponding 
theory in the Heisenberg model is not correct. 
This just  amounts to rejecting the importance of the 
previous definition. 
In that case the test is to design an experiment 
whose results would be different in the two models. 
And then do the experiment to see which model is wrong. 
(Maybe both will be wrong!) 
But the Schr\"odinger model as presented here would 
have to be augmented with another measurable 
property besides the Generalized Born's Rule, 
which is the same in the two models. 
In other words one would have to propose 
a property that only holds in the Schr\"odinger 
model.

\section{\bf How Many Time Evolutions?}
\markboth{How Many Time Evolutions?}{How Many Time Evolutions?}

A standard criticism of standard quantum theory 
is that it has two time evolutions:
one from Schr\"odinger's equation and the other 
from the collapse of the state. 
It is widely held that neither of these can be 
the consequence of the other. 
After all, Schr\"odinger's equation gives 
a continuous time evolving state while the 
collapse is discontinuous.
  
Of course, one could try to explain collapse 
as a continuous change that is so quick as to 
appear discontinuous. 
Or, on the other hand, one might try to explain 
continuous time evolution as a rapid succession 
of discontinuous jumps, much as a motion picture 
is not a picture of continuous motion but rather 
many static pictures in succession giving the 
impression of motion. 
While these remain as logical possibilities, 
neither is readily implementable. 

So, leaving these possibilities to a side, 
these two time evolutions 
are not only incompatible, but also
neither can be {\em the} basic description of time
evolution in quantum theory. 
There are logically then two alternatives. 
The first is to accept that nature is described by 
two independent time evolutions, and that's the 
end of the matter.  
The second is that there is one time evolution 
which is 
somehow more basic than these two. 
I am advocating for the second option in a rather
specific way.

\section{\bf Relation to Observation}
\markboth{Relation to Observation}{Relation to Observation}

It is crucial to understand that 
the relation of quantum theory to physical 
observations is based only on those 
statements in the theory
that are model independent within an 
isomorphism class of models. 
Unitary transformations are the 
isomorphisms of Hilbert spaces. 
Physically, this means that the formulation of a 
theory in the context of one specific Hilbert space can 
always be translated into the context of a 
different isomorphic Hilbert space via the 
application of a unitary transformation. 
Such a unitary transformation changes the model 
being used to understand the physics. 
Only properties that hold in all isomorphic 
models can possibly be  
relevant to understanding physical phenomena. 
A property that holds in one model, but not 
in all other isomorphic models, 
must absolutely never be considered as having any
physical significance. 
Such properties can be useful as intermediate steps
in the analysis of a physical system, but nothing more
than that. 
Such model dependent properties do not need 
to be `interpreted' nor `explained' {\em vis a vis} 
their physical significance. 
They simply have no physical significance. 

As an example, a Schr\"odinger operator
acting in $ L^{2} (\mathbb{R}^{3}) $ could 
have an eigenstate 
which is represented by a $ C^{\infty} $  function. 
That function corresponds to a unit vector 
in any isomorphic Hilbert space, whose elements need
not be represented by functions 
on a differential manifold.  
So the concept of $ C^{\infty} $ does  
not apply in that isomorphic model. 
Nonetheless, the eigenvalue 
of the corresponding self-adjoint operator for the 
corresponding  
eigenstate is the {\em same} real number for all isomorphic 
models and has a physical significance, 
provided that the original Schr\"odinger operator 
represents a physical system. 
For example, the {\em Segal-Bargmann space}
on $ \mathbb{C}^{3} $ 
(see \cite{bargmann}) can be used instead of
$ L^{2} (\mathbb{R}^{3}) $. 
(A unitary transformation effecting this change
of model is the {\em Segal-Bargmann transform}.)
In the Segal-Bargmann
model all pure states are $ C^{\infty} $ functions. 
So in that model the property of a pure state being 
$ C^{\infty} $ is unremarkable. 
As we shall see later, the 
failure to recognize this elementary aspect of 
quantum theory leads 
to pointless discussions about the `meaning' 
or `interpretation'
of model dependent properties.

\section{\bf Omissions from the Axioms}
\markboth{Omissions}{Omissions}

Many important properties of quantum theory 
have been omitted from the axioms. 
This is not because they lack importance.  
Rather I think they are not basic, but 
instead are secondary. 
These include, but are not limited to, 
uncertainty principles, complementarity, 
decoherence, 
symmetries (including broken symmetries) 
and their related conservation laws, 
the existence and properties of particles 
(e.g. photons) and of 
pseudo-particles (e.g. phonons).

\chapter{Quantum Probability}
\label{qprob-chapter}

\hfill 
le hasard . . .

\hfill
qui est en r\'ealit\'e

\hfill 
l'ordre de la nature

\vskip 0.2cm

\hfill 
Anatole France

\vskip 0.4cm 
\section{\bf Introduction}
\markboth{Introduction}{Introduction}

This chapter is meant to 
motivate, define and study 
the fundamentals of quantum probability. 
I start by explaining  
how certain classical probability measures 
(in the sense of the formulation of 
Kolmogorov in his book \cite{kolmogorov}) 
arise as a consequence of various
hypotheses, which are accepted generally 
by physicists who work in quantum theory. 
The idea that probability is an added-on
`interpretation' of quantum theory
is a misconception, that leads many
to think that quantum probability can 
be replaced by using some alternative
`interpretation'. 
While this has been known since at least
the publication of
von~Neumann's book \cite{von-neumann} in 1932, 
it seems not to be well known, neither 
in the mathematics nor the physics 
communities. 
Actually, just about everything in this chapter 
can be found in the literature, with the unique 
exception of the Generalized Born's Rule as the 
fundamental time evolution equation of 
quantum theory. 
I have tried to re-organize this known, but not 
well known, material which has too often been 
overlooked  
by people who were, I am afraid, more interested 
in `interpreting' quantum theory than in understanding it

Before going into details it is necessary to 
understand what is meant by a {\em probability 
theory} in the most general terms. 
But this is rather clear. 
It is any theory which contains formulas whose values
are real numbers 
lying in the closed interval $ [0,1] $. 
The application of probability theory to 
experimental science is that these 
real numbers, called {\em probabilities}, have the
significance that they describe {\em relative 
frequencies} of certain observations. 
There are many more details on both the theoretical 
and experimental sides, but this is the basic idea. 
One of the more important achievements of contemporary 
mathematics is the introduction of many new 
probability theories, such as {\em free probability}. 
The approach, known as {\em classical 
probability theory}, given by Kolmogorov 
in \cite{kolmogorov} is in this view just one 
of many possible probability theories. 
We will now see its relation to Quantum Probability. 

I understand {\em Quantum Probability} to mean that 
part of {\em Quantum Theory} which gives rise to 
probabilities and the rules that apply to them. 
As such it is not an independent theory, since
it is essentially 
linked to the rest of Quantum Theory. 
In other words, Quantum Probability 
is a part of physics. 
It also is an example, (the first 
historically, I guess), of a Non-commutative 
Probability Theory, which is an ongoing 
research area in basic mathematics. 
But I shall not deal with this more 
general mathematical topic, except to note 
that it (typically?) does not come as part 
and parcel of a theory with time evolution 
nor is it intended to be a 
physical theory. 
For a glimpse into some beautiful advanced topics
in quantum probability beyond the 
Generaiized Born's Rule see \cite{accardi}, 
\cite{parthasarathy}, \cite{sinha} and 
references therein.

\section{\bf The Physical Assumptions}
\markboth{Physical Assumptions}{Physical Assumptions}

A given quantum system is described by a formulation
based on a specific complex, non-zero  
Hilbert space $ \mathcal{H} $ 
which is 
often, but not necessarily, infinite dimensional. 
This is implicit in Dirac's notation, 
though it is usually accepted explicitly
by those in the quantum 
physics community. 
Note that the case when $ \mathcal{H} $ has
dimension $ 1 $ is included, even though the 
resulting quantum physics is trivial.
But we do exclude the case $ \mathcal{H} = 0 $, 
since that leads to no physics whatsoever, since 
there are no states and no events. 
So, $ \mathcal{H} \ne 0 $ in this treatise. 

Within such a mathematical model quantum 
physics is assumed by most in the physics community 
to satisfy these 
properties among others: 
\begin{itemize}
	
	\item 
	The state of a quantum system
	is described by vectors $ \psi \in \mathcal{H} $ 
	with $ || \psi || = 1  $ with two such unit
	vectors $ \psi_{1}, \psi_{2} $ describing the same state provided that there exists 
	$ \lambda \in \mathbb{C} $ such that
	$ \psi_{1} = \lambda \psi_{2} $. 
	(Necessarily $ |\lambda| =1 $.)
	
	\item
	Most physical measurements (with the remarkable 
	exception of time measurements) 
	are represented by
	self-adjoint, densely defined linear 
	operators acting on a linear subspace 
	of $ \mathcal{H} $. 
	
	\item 
	The same physical measurement 
	performed on an ensemble
	of quantum systems, which are all
	described by the same state, 
    does not in general yield the same 
    measured value, but rather values 
    with some relative frequencies 
    of numbers in some 
    subset of $ \mathbb{R} $ 
    with {\em more} than one number.  
    
\end{itemize}

Various comments are in order. 
The first property only describes 
{\em pure states} and not the more 
general {\em mixed states}, which also
enter into quantum theory.  
However, what we will be saying is easily 
generalized to the setting with all states,
including the mixed states, under consideration. 

The second property may have a converse, namely
that every self-adjoint, densely defined linear
operator corresponds to a physical 
measurement. 
This converse is denied by theories with 
{\em superselection rules}. 
But this possibility does not 
concern us, since we will not
be using the converse statement. 

The third property concerns what one observes 
in experiment. 
The only way to deny it is to re-define the 
meaning of the word `same'. 
But we will not use this property is the 
following argument, but rather offer a 
explanation of it using 
classical probability theory.

\section{\bf The Mathematical Model}
\markboth{Mathematical Model}{Mathematical Model}

The central mathematical fact is that there is 
a complete description of densely defined, 
self-adjoint operators which act in a Hilbert 
space. 
This is the content of
the {\em Spectral Theorem.} 
This description is given in terms of 
{\em projection-valued measures}, which we will define 
momentarily. 
However, this section is not a complete 
presentation of the appropriate functional analysis, 
which I expect the reader already to know. 
Mainly it serves to establish notation 
and as a quick review. 

First, we recall some basics 
from Hilbert space theory. 
We let $ \langle \cdot, \cdot \rangle $ denote 
the inner product in $ \mathcal{H}  $, 
a complex Hilbert space, and 
$ || \cdot || $ its associated norm. 
One has 
$ || \psi || := \langle \psi, \psi \rangle^{1/2} $ 
for all $ \psi \in \mathcal{H} $. 
A function $ T: \mathcal{H} \to \mathcal{H}  $
is said to be {\em linear} if, just as in linear 
algebra, we have 
$$ 
 T (\alpha_{1}
\psi_{1} + \alpha_{2} \psi_{2}) 
= 
 \alpha_{1}
 T (\psi_{1}) + \alpha_{2} T (  \psi_{2}) 
$$
for all $ \psi_{1}, \psi_{2} \in \mathcal{H} $
and all $ \alpha_{1}, \alpha_{2} \in \mathbb{C} $. 
Furthermore, we say that such a function is 
a {\em (linear) operator}. 
If such an operator $ T $ has the property that 
there exists some real number $ M \ge 0 $ such that   
$ || T \psi || \le M || \psi || $ holds 
for all $ \psi \in \mathcal{H} $, then we say 
that the operator $ T $ is {\em bounded}.

A {\em projection}
$ E : \mathcal{H} \to \mathcal{H} $ is by 
definition a linear operator such that
$ E $ is idempotent (meaning that $ E^{2} = E$) 
and $ E $ is self-adjoint 
(meaning that $ E^{*} = E $). 
Projections are bounded operators. 
It is important to note that $ 0 $, 
the zero operator (i.e., the operator which 
maps every $ \psi \in \mathcal{H} $ 
to $ 0 \in \mathcal{H} $) is a projection. 
Also, $ I_{\mathcal{H}} $, 
the identity operator acting on $ \mathcal{H} $
is a projection, where by definition
$ I_{\mathcal{H}} \psi := \psi$ for all 
$ \psi \in \mathcal{H} $. 
Also, we define 
$$
\mathcal{L} (\mathcal{H}) :=
\{ T: \mathcal{H} \to \mathcal{H} ~|~
T \mathrm{~is~linear~and~bounded}\}. 
$$
For each 
$ T \in \mathcal{L} (\mathcal{H}) $ we 
define its {\em operator norm} as 
$$
   || T ||_{op} := 
   \sup \, \{ || T \psi || \,\big|\, 
   \psi \in \mathcal{H} \mathrm{~and~}
   ||\psi|| \le 1  \}. 
$$
With this norm $ \mathcal{L} (\mathcal{H}) $ 
becomes a complete, normed space, which 
is to say that it is a {\em Banach space}. 
(Complete means that every Cauchy sequence 
with respect to the metric associated to the norm 
is convergent 
to an element in $ \mathcal{L} (\mathcal{H}) $. 
The {\em associated metric} is defined by 
$ d_{op}(T_{1}), T_{2} := 
|| T_{1}- T_{2} ||_{op}  $ for all 
$ T_{1}, T_{2} \in \mathcal{L} (\mathcal{H})  $.) 

As examples we mention that for any projection
$ E \ne 0 $ we have that $ || E ||_{op} = 1 $. 
Also, $ || I_{\mathcal{H}} ||_{op} =1 $ since 
$ \mathcal{H} \ne 0 $. 

We recall from standard measure theory 
that the smallest $ \sigma  $-algebra, denoted as
$ \mathcal{B} (\mathbb{R}) $, of 
subsets of $ \mathbb{R} $ that contains all
finite open interval $ (a,b) $ (with 
$ a,b \in \mathbb{R} $ and $ a < b $) 
is called the {\em Borel $ \sigma  $-algebra} of 
$ \mathbb{R} $. 
Also, a set in 
$ \mathcal{B} (\mathbb{R}) $ is called a
{\em Borel set}.
Notice that the empty set $ \emptyset $ 
and the whole real line $ \mathbb{R} $ are
Borel sets. 

We now have enough to give this definition.
\begin{definition}
A {\em projection-valued measure} 
(or simply {\em pvm}) 
associated to a Hilbert space $ \mathcal{H} $
is a function 
$ P : \mathcal{B} (\mathbb{R}) \to 
\mathcal{L} (\mathcal{H})  $ 
satisfying these properties:
\begin{itemize}
	
	\item 
	$ P(B) \in \mathcal{L} (\mathcal{H}) $ 
	is a projection for every
	Borel set $ B \in \mathcal{B} (\mathbb{R}) $. 
	
	\item 
	$ P(\emptyset) =0 $, the zero operator,
	and $ P ( \mathbb{R} ) = I_{\mathcal{H}} $, 
	the identity operator acting on $ \mathcal{H} $. 
	
	\item 
	Let $\{ B_{j} \,|\, j \in J \}$ be a finite or 
	countably infinite family of
	Borel sets in $ \mathcal{B} (\mathbb{R}) $, 
	which is disjoint 
	(meaning that $ B_{j} \cap B_{k} = \emptyset $ 
	whenever $ j \ne k $). 
	Then we have the  
	{\rm $ \sigma $-additivity condition}
	$$
	P \big( \cup_{j \in J} B_{j} \big)= 
	\sum_{j \in J} P (B_{j}), 
	$$
	where the possibly infinite	sum 
	on the right side has the meaning that 
	$$
	P \big( \cup_{j \in J} B_{j} \big) \psi= 
	\sum_{j \in J} P (B_{j}) \psi 
	$$
	holds for every $ \psi \in \mathcal{H} $. 
	Here the possibly infinite sum 
	on the right side is understood to mean  
	the convergence with respect to the metric 
	topology on $ \mathcal{H} $ induced by 
	its norm. 
	That metric, denoted as $ d $, 
	is defined by 
	$ d (\psi_{1}, \psi_{2}) := 
	|| \psi_{1} - \psi_{2} ||$ 
	for all $ \psi_{1}, \psi_{2} \in \mathcal{H} $. 
    Also notice that $  \cup_{j \in J} B_{j} $ is 
    a Borel set, 
    by definition of $ \sigma $-algebra, 	
	and so the left side is also well defined.

\end{itemize}

\end{definition}

A curious consequence of this definition is that
for every pvm we have  
$ P ( B \cap C) = P (B) P (C) $ for all 
$ B,C \in \mathcal{B} (\mathbb{R}) $. 
And this in turn implies that the family of operators 
$ \{ P(B) \,|\, B \in \mathcal{B} (\mathbb{R}) \} $
is commutative. 

The point of measure theory is to use measures 
in order to develop a well behaved theory of
integrals. 
In courses this is typically done with measures
with non-negative real values (such as {\em Lebesgue 
measure}), though also the 
generalization to measures with all real values 
or with complex values is sometimes presented. 
These generalizations are essentially the same 
as the theory with non-negative valued measures. 
Actually, this also works out for 
a measure $ \mu : \mathcal{B} (\mathbb{R}) \to X $, 
where $ X $ is any Hausdorff topological vector space. 
The topology gives a meaning to  
convergent infinite sums in $ X $.  
In particular, for any Borel measurable, 
function $ f : \mathbb{R} \to \mathbb{C} $
we can consider whether the 
integral 
$ \int_{\mathbb{R}} f (\lambda) \, d \mu (\lambda) $
exists as a uniquely defined element in $ X $. 
The usual technical, measure-theoretic details apply. 
In the case of interest here the vector space 
is the complex vector space 
$ \mathcal{L} (\mathcal{H}) $, but 
not with its norm topology. 
Instead, we use the {\em strong operator topology}
on $ \mathcal{L} (\mathcal{H}) $. 
Without going into all the technical details, 
let's simply note that a {\em sequence}
of operators  $A_{n} \in  \mathcal{L} (\mathcal{H}) $
converges to an operator 
$ A \in \mathcal{L} (\mathcal{H}) $ in the 
strong operator topology 
if and only if for every $ \phi \in \mathcal{H} $ 
the sequence $ A_{n} \phi \in \mathcal{H} $ 
converges to $ A \phi \in \mathcal{H}$ in the 
norm topology of $ \mathcal{H} $.  
The strong operator topology is the same as the operator 
norm topology on $ \mathcal{L} (\mathcal{H}) $
if and only if the dimension of
$ \mathcal{H} $ is finite. 
Using the norm topology in the infinite dimensional 
case gives an integral inadequate for 
{\em spectral theory} and
hence inadequate for quantum theory. 

A pvm $ P $ is a measure taking 
values in $ \mathcal{L} (\mathcal{H}) $,
a Hausdorff topological vector space
space when endowed with the strong operator topology. 
And so we have a well defined 
(and well behaved in the sense of measure theory) integral  
$ \int_{\mathbb{R}} d P (\lambda) \, f (\lambda)
\in \mathcal{L} (\mathcal{H}) $ 
for any {\em essentially  bounded}, Borel function 
$ f : \mathbb{R} \to \mathbb{C} $.
Also, an integral can even be defined for some 
unbounded Borel functions, but that leads to 
ever more technical details, which the reader 
can find in any good functional analysis text. 
Unfortunately, in the statement of the Spectral 
Theorem, which we state next, the integrand 
is an unbounded Borel function in the important
case when the operator is not bounded. 

\begin{theorem}[Spectral Theorem] 
Suppose $ A $ that is a densely defined self-adjoint 
operator acting in a Hilbert space 
$ \mathcal{H} $. 
Then there exists a unique pvm 
$ P : \mathcal{B} (\mathbb{R}) \to 
\mathcal{L} (\mathcal{H})  $  
such that 
\begin{equation}
\label{spectral-theorem}
   A = \int_{\mathbb{R}} \lambda \, dP (\lambda). 
\end{equation}
\end{theorem}

Many comments are in order, including why 
this theorem is named so. 
As a start we have the annoying fact that the
function that we are integrating  
in \eqref{spectral-theorem} is the 
function $ f(\lambda) = \lambda $ for 
all $ \lambda \in \mathbb{R} $, and this 
is not a bounded function, though it is a
Borel function. 
Here measure theory helps, if the pvm has 
bounded support. 
The definition of the support of a pvm $ P $ is
much the same as in regular measure theory:
$$
\mathrm{Supp} \, P := 
\mathbb{R} \setminus 
\cup \{  U \subset \mathbb{R} \,|\, P(U) = 0
\mathrm{~and~} U \mathrm{~is~open} \}. 
$$
The only, quite minor difference is that $ 0 $
in the condition $  P(U) = 0 $
refers to the projection operator $ 0 $. 
By definition, $ \mathrm{Supp} \, P $ is a
closed subset of $ \mathbb{R} $, being the 
complement of an open subset. 
It also can not be empty, since if it were
we would have that 
$I_{\mathcal{H}} = P(\mathbb{R}) = 0 $, 
which is impossible since $ \mathcal{H} \ne 0 $. 

By measure theory we always have
$$
A = \int_{\mathrm{Supp} \, P} \lambda \, dP (\lambda).
$$
So, if $ \mathrm{Supp} \, P $ is a bounded 
subset of $ \mathbb{R} $, then this integral 
exists, since the integrand {\em on this subset} is
a bounded Borel function. 

Here is where we see spectral theory entering 
the theory. 
First, we review some material from the
functional analysis of Hilbert spaces. 
\begin{definition}
Suppose that $ A : \mathcal{D} \to \mathcal{H} $ 
is a linear operator defined on the dense
subspace $ \mathcal{D} $ of 
a Hilbert space $ \mathcal{H} $. 
Then we define the {\rm resolvent set} of $ A $ to be
the following subset of the complex numbers:
$$
\mathrm{Res} (A) := \{ \lambda \in \mathbb{C}
\,\big|\, \exists T_{\lambda} \in \mathcal{L} (\mathcal{H}) \mathrm{~s.t.~}
(\lambda I - A) T_{\lambda} = I_{\mathcal{H}}
\mathrm{~and~} 
T_{\lambda} (\lambda I - A) = id_{\mathcal{D}} \}. 
$$
One says $ \lambda $ is in the resolvent set 
provided 
that the (densely defined) operator
$ \lambda I - A $ has a globally defined, bounded 
inverse. 
Then the spectrum is defined as the complementary
subset of $ \mathbb{C} $, that is 
$$
\mathrm{Spec} (A) := 
\mathbb{C} \setminus \mathrm{Res} (A). 
$$
\end{definition}

In functional analysis one proves that 
for self-adjoint operators $ A $ we have that 
$ \mathrm{Spec} (A) $ is a closed, non-empty 
subset of the real line $ \mathbb{R} $ and, 
moreover, that $ \mathrm{Spec} (A) $ is a 
bounded subset if and only if $ A $ is a 
bounded operator. 
(The diligent reader will have noticed that 
we have not defined what it means for
a densely defined operator to be bounded. 
Even worse, we have not defined what it means for
a densely defined operator to be self-adjoint. 
See your favorite functional analysis text 
for these details.)

Now we come to a relation between the pvm 
and spectral theory. 

\begin{theorem}
Let $ A $ be a densely defined 
self-adjoint operator, and  
let $ P_{A} $ be the unique pvm associated 
to $ A $ by the spectral theorem.
Then
$$
\mathrm{Supp} \, P_{A} = \mathrm{Spec} (A). 
$$
\end{theorem}

Putting these results together we see that 
for bounded self-adjoint operators the integral 
in \eqref{spectral-theorem} is well-defined. 
Of course, it remains to prove the 
equation \eqref{spectral-theorem}. 
As is typical in functional analysis there are 
a lot of technical details to deal with in order 
to give meaning to the integral in 
\eqref{spectral-theorem} 
for an unbounded self-adjoint operator $ A $
and,
having done that, to prove 
the equation \eqref{spectral-theorem} holds.

\section{\bf The Finite Dimensional Case}
\markboth{Finite Dimensional Case}{Finite Dimensional Case}

Leaving those technical details to the 
texts, let us understand in detail what the Spectral 
Theorem says in the case when $ \mathcal{H} $ 
is finite dimensional, since this is often not
presented in a course of functional analysis. 
In that case we have a situation in linear algebra. 
The self-adjoint operator is then identified 
with a Hermitian $ n \times n $ matrix $ A $
with $ n \ge 1 $. 
Since only the whole space $ \mathcal{H} $ is
a dense subspace (because $ \dim \mathcal{H} $ 
is finite), the matrix 
$ A $ maps all of $ \mathcal{H} $ to itself. 
In linear algebra one proves that the set of 
eigenvalues is a finite, non-empty subset 
of $ \mathbb{R} $. 
Let $ \{ \lambda_{1}, \cdots ,\lambda_{k} \} $ 
where $ 1 \le k \le n $, 
denote that subset of eigenvalues, 
that is these are the 
$ k $ {\em distinct} eigenvalues of $ A $. 
The diagonalization theorem for $ A $ says 
that there is a basis of $ A $ with 
respect to which $ A $ 
has diagonal form with the eigenvalues 
appearing along the diagonal, each with a 
number of times equal to its multiplicity.  
Let's put this statement into another 
equivalent notation.
For each $ 1 \le j \le k $ we define the
spectral subspace associated with the
eigenvalue $ \lambda_{j} $ by
$$
V_{j} :=\{ \psi \in \mathcal{H} \,|\, 
A \psi = \lambda_{j} \psi \}. 
$$
Since $ \lambda_{j} $ is an eigenvalue, 
there is an eigenvector (non-zero, by definition) 
in $ V_{j} $. 
In short, $ V_{j} \ne 0 $. 
One proves readily that $ V_{i} $ and $ V_{j} $
are orthogonal subspaces, since they correspond 
to eigenvalues $ \lambda_{i} \ne \lambda_{j} $. 
Then the diagonalization of $ A $ is realized 
by choosing a basis $ \mathcal{B}_{j} $ of 
$ V_{j} $ for each $ 1 \le j \le k $ and 
proving that the union 
$\mathcal{B}:= \cup_{j} \mathcal{B}_{j} $ 
is a basis of $ \mathcal{H} $. 
Since $ A $ restricted to $ V_{j} $ is simply 
multiplication by $ \lambda_{j} $, the matrix
of $ A $ in the basis $ \mathcal{B} $ is the sought 
for diagonal matrix representation of $ A $. 
What this means is that we have an orthogonal 
decomposition
\begin{equation}
\label{decomp-of-H}
   \mathcal{H} = V_{1} \oplus \cdots \oplus V_{k}
\end{equation}
such that $ A $ acts as multiplication by a scalar
on each summand. 

We now put this into the language of projections. 
For each spectral subspace $ V_{j} $ there is a unique 
projection such that 
$ \mathrm{Ran} \, P_{j} = V_{j} $.
The very fact that these are projections says
that $ P_{j}^{*} = P_{j}  $ and $ P_{j}^{2} = P_{j} $.
Moreover, the fact that $ V_{i} $ is orthogonal 
to $ V_{j} $ for all $ i \ne j $ implies that 
$ P_{i} P_{j} = 0$ for $ i \ne j $. 
Note that this property is not shared by 
numbers (say, real or complex numbers as you wish)
 since in that case the product $ a b =0 $ 
implies either $ a =0 $ or $ b = 0 $. 
But these non-zero projections satisfy 
$ P_{i} P_{j} = 0$ for $ i \ne j $. 
In any event we use this convenient condensed 
notation: $ P_{i} P_{j} = \delta_{ij} P_{i}$ 
for all $ 1 \le i,j \le k $, where 
$ \delta_{ij} $ is the Kronecker delta. 
And we can write \eqref{decomp-of-H} in this 
notation as:
$$
I_{\mathcal{H}} = P_{1} + \cdots + P_{k},
$$
which is called a {\em resolution of the identity}.
This is not too exciting, since there are many, 
many resolutions of the identity.
What makes this resolution of the identity 
interesting is that it comes from the 
diagonalization of $ A $. 
In fact, since $ A $ acts as multiplication 
by the eigenvalue $ \lambda_{j} $
on the subspace $ V_{j} $, it follows 
quickly that 
\begin{equation}
\label{spectral-rep-of-A}
A = \lambda_{1} P_{1} + \cdots + \lambda_{k} P_{k},
\end{equation}
which is called the spectral representation 
of the Hermitian matrix $ A $. 
This is completely equivalent to the 
diagonalization of $ A $. 
We next re-write the 
finite sum on the right side of 
\eqref{spectral-rep-of-A} as an
integral with respect to a pvm $ P $.
We clearly want $ P $ to satisfy 
$ P (\{\lambda_{j}\}) = P_{j} $ for 
each $ 1 \le j \le k $. 
We also want 
$ P ( \mathbb{R} \setminus \mathrm{Spec} A) = 0 $, 
the zero operator. 
If we can do that, then by standard identities 
of measure theory (only now applied to a pvm), 
we have
\begin{align*}
\int_{\mathbb{R}} \lambda \, dP (\lambda) 
&= \int_{\mathbb{R} \setminus \mathrm{Spec} A} 
\lambda \, dP (\lambda) 
+ \int_{\mathrm{Spec} A} \lambda \, dP (\lambda) 
\\
&= 0 + 
\int_{ \cup_{j=1}^{k} \{ \lambda_{j} \}} \lambda \, dP (\lambda) 
\\
&= \sum_{j=1}^{k} 
\int_{\{ \lambda_{j} \}} \lambda \, dP (\lambda)
\\
&= \lambda_{1} P_{1} + \cdots + \lambda_{k} P_{k} 
\end{align*}
as desired. 
So it only remains to define the pvm with 
the required properties. 
But for every Borel subset $ B \subset \mathbb{R} $ 
we simply define 
$$
P (B) :=  \sum_{j=1}^{k} 
\chi_{B} \, (\lambda^{j}) \, P_{j},
$$
where the characteristic function 
$ \chi_{S} $
of any set 
$ S $ is defined as
$$
    \chi_{S} (\lambda) := 
    \left\lbrace 
    \begin{array}{cc}
    1 & \mathrm{if~} \lambda \in S,
    \\
    0 & \mathrm{if~} \lambda \notin S. 
    \end{array}
    \right. 
$$
An alternative way to write this is as
$$
P (B) =  \!\!\!\!
\sum_{ \{ j  \,|\, \lambda_{j} \in B \}  }
\!\!\!\! P_{j},
$$
that is, add up all the projections 
associated to the eigenvalues that lie in $ B $.

One readily checks that $ P $ so defined 
is a pvm and that it satisfies the 
desired properties. 
The moral of this story is that 
the diagonalization of a Hermitian 
matrix is exactly the finite-dimensional special 
case of the Spectral Theorem. 
Another way to phrase this is to say 
that the Spectral Theorem generalizes 
the diagonalization of a Hermitian 
matrix to the infinite dimensional case.
Notice that in this generalization a finite
sum has been replaced by an integral, 
which in certain cases will be a finite sum
and in other cases will be an infinite sum.
However, the complete generalization to 
the infinite dimensional case requires 
integrals and the corresponding 
measure theory with pvm's.

\section{\bf Quantum Probability: The First Steps}
\markboth{Quantum Probability: The First Steps}{Quantum Probability: The First Steps}

With all this mathematical background we are ready to 
see probability theory in the context 
of quantum theory. 
This way of introducing probability into
quantum physics is what is called Quantum Probability. 
 
The first thing to note is that 
a pvm is quite similar, though not 
identical, to a probability measure. 
This is seen for example in the normalization
conditions. 
But even more impressive is that there is a 
partial order on self-adjoint operators, 
which we can restrict to the special case
of projection operators (which are, as you will 
recall, self-adjoint). 
It turns out that for any projection $ P $
we have $ 0 \le P \le I_{\mathcal{H}}  $, which are 
inequalities of self-adjoint operators.  
So projections lie between the two extreme 
projections possible. 
This is analogous to classical probability 
theory, where any probability $ p $ lies
between the two extreme probabilities, 
namely $ 0 \le p \le 1 $, which are inequalities
of numbers. 

To make the relation tighter with classical 
probability, we note that 
there is a converse
of the Spectral Theorem. 
\begin{theorem}[Converse to Spectral Theorem]
Suppose that $ \mathcal{H} $ is a Hilbert space and
$ P : \mathcal{B} (\mathbb{R}) \to 
\mathcal{L} (\mathcal{H})  $ be a pvm. 
Then 
\begin{equation}
\label{converse}
A := \int_{\mathbb{R}} \lambda \, dP (\lambda)
\end{equation}
defines a self-adjoint operator, densely defined
in $ \mathcal{H} $. 
Moreover, the pvm $ P_{A} $ associated to $ A $ 
satisfies $ P_{A} = A $. 
\end{theorem}

Moreover, and more importantly, this result 
together with the Spectral Theorem give a
bijection between the set of all 
self-adjoint operators acting in 
$ \mathcal{H} $ and the set of 
all pvm's taking values in 
$ \mathcal{L} (\mathcal{H}) $. 
One way the bijection sends the self-adjoint 
operator $ A $ to its pvm $ P_{A} $. 
The other way a pvm $ P $ is sent to the self-adjoint 
operator given in \eqref{converse}. 
This is an amazing result, since it sets up 
a dictionary between objects of analysis
(self-adjoint operators) and objects of measure
theory (pvm's). 

Therefore the second basic assumption of quantum theory 
which says that physical measurements are 
represented by self-adjoint operators translates 
to saying that physical measurements are 
represented by pvm's. 
We are getting very close to classical 
probability theory using just a 
basic assumption of quantum theory. 
To take the next step we consider a self-adjoint 
operator~$ A $ (or equivalently, its pvm $ P_{A} $) and 
a state $ \psi \in \mathcal{H} $. 
(Recall that $ || \psi || = 1 $.)
Here we are using the first basic assumption of quantum theory. 
From these mathematical objects, given to 
us by quantum physics, we define for every
Borel subset $ B \subset \mathbb{R} $ and unit 
vector $ \psi \in \mathcal{H} $ the expression
on the left side here: 
\begin{equation}
\label{borns-rule}
P ( A \in B \,|\, \psi) := 
\langle \psi , P_{A}(B) \, \psi \rangle.  
\end{equation}

We could 
read the left side in full detail 
as follows: 
The probability that the measurement of $ A $ 
gives a value in $ B $ given that the quantum 
system is `described' by the state $ \psi $. 
A more colloquial reading without referencing 
any measurement is: 
The probability that $A$ is in $B$ given $ \psi $. 
Of course, this terminology should be justified. 
The first result to prove is that the right side 
satisfies 
$0 \le \langle \psi , P_{A}(B) \, \psi \rangle \le 1 $. 
The next thing to prove is that for the state 
$ \psi $ fixed and for the 
self-adjoint operator $ A $ fixed, 
the function defined for every Borel subset
$ B $ of $ \mathbb{R} $ by 
$ B \mapsto \langle \psi , P_{A}(B) \, \psi \rangle $
is a classical probability measure on  
the real line $ \mathbb{R} $, that is, 
it is $ \sigma $-additive and satisfies the 
normalization of a classical probability measure: 
$$
\langle \psi , P_{\emptyset}(B) \, \psi \rangle = 0
\quad \mathrm{and} \quad 
\langle \psi , P_{A}(\mathbb{R}) \, \psi \rangle = 1.
$$
All of this follows trivially from the definitions
and standard properties of Hilbert space. 
I hope the reader appreciates that the tricky 
bit in this theory is getting the definitions right. 
In functional analysis one calls the probability measures in 
\eqref{borns-rule} the {\em spectral measures}.
But in the physics community, 
\eqref{borns-rule} is known as 
{\em Born's rule}, 
and although Max Born stated it differently 
it was worthy of his receiving the Nobel prize 
mainly for this achievement in quantum physics.  
It is no exaggeration to say that Born's rule is
Quantum Probability.  

Let us pause for an interlude to remark that there are some
equivalent formulas for Born's rule \eqref{borns-rule}. 
Continuing with the notation established above 
and using standard results in Hilbert space theory 
we have 
\begin{align*}
P ( A \in B \,|\, \psi) &= 
\langle \psi , P_{A}(B) \, \psi \rangle =
||  P_{A}(B) \, \psi ||^{2} 
\\
& =Tr ( P_{A}(B) \, E_{\psi} )
= Tr ( E_{\psi} \, P_{A}(B)   )
\end{align*}
where $ E_{\psi} := | \psi \rangle \langle \psi | $ 
is a rank $ 1 $ quantum event (using Dirac notation) and 
$ Tr $ denotes the trace of a trace class operator. 
The last two formulas have 
the advantage of expressing the probability
of Born's rule in terms of the trace. 

This is also a good time to note that there is
something more general going on here. 
The event $ P_{A} (B) $ is the formulas above 
can be replaced by any event $ E $ to give this
definition of the 
{\em probability of a quantum event, given 
a pure state $ \psi $}, as
\begin{equation}
\label{quantum-prob-of-event}
 P ( E \,|\, \psi ) :=  \langle \psi, E \psi \rangle = 
||  E \psi ||^{2} = Tr ( E_{\psi} \, E ). 
\end{equation}
The reader should note that this is a new meaning for
the word ``probability''.  
Quantum events do not form 
a $ \sigma $-algebra if $ \dim \mathcal{H} \ge 2 $, 
However, there are enough properties to justify this 
terminology. 
First, $ 0 \le  P ( E \,|\, \psi ) \le 1 $ by basic 
Hilbert space theory. 
So we get a real number that can be 
compared with a relative frequency. 
Next, it satisfies the normalizations 
$$
P ( I \,|\, \psi) = 1 \quad \mathrm{and} \quad 
P ( 0 \,|\, \psi) = 0,
$$ 
where the $ 0 $ on the left side of the last  
equation is the zero operator. 
This topic will be presented in more detail 
in Section~\ref{section-two-events}. 

And definition \eqref{quantum-prob-of-event} 
in turn is a particular case of the 
{\em expected value} of an observable $ A = A^{*} $ 
with respect to $ \psi $, which is defined as
\begin{equation}
\label{expectf-value-of-A}
   \mathcal{E}_{\psi} (A) := \langle \psi, A \psi \rangle . 
\end{equation}
And one more generalization is obtained by using 
the definition 
\eqref{expectf-value-of-A} for all 
$ A \in \mathcal{L} (\mathcal{H}) $. 
The linear map 
$ \mathcal{E}_{\psi} : 
\mathcal{L} (\mathcal{H}) \to \mathbb{C}$ so obtained 
gives an example of a {\em non-commutative integral}. 
The elegant theory of non-commutative integration 
is found in many places in the
literature. 

Let's return to the main argument. 
So by applying two basic properties of quantum theory
(which really are widely accepted) we have 
arrived at an infinite number of classical 
probability measures for a given observable. 
As I like to say: This is a feature, not a bug. 
It only remains to understand what, if any, role these
classical probability measures play in quantum physics. 
It is difficult to argue that they have no 
relevance or meaning in quantum physics, since they 
arise from 
two basic properties of quantum theory plus some
(well, admittedly a lot of) mathematics. 
It is difficult to argue that there they are, 
but who cares? 
The standard way of dealing with this situation 
is to use these probability measures 
to save (`explain' if you wish) the phenomena  of 
the relative frequencies 
that occur in the third basic property 
of quantum theory. 
However, do notice that our argument for arriving at these
classical probability measures did not use this 
third basic property as an assumption. 

It might appear that I have proved Axiom~4 in Chapter~1 
from other axioms of quantum theory. 
That is not correct, since 
Axiom~4 is only a mathematical definition within the theory. 
However, it is a definition with a lot of physical significance 
since it is related to observations. 
Using the probability measures in \eqref{borns-rule} 
to understand measured relative frequencies 
is the physical significance of Axiom~4 
as well as of the time dependent 
Born's rule \eqref{time-depend-borns-rule} 
in Axiom~5. 

However, two generally accepted first principles 
of quantum theory have led us to the doorstep of
Axiom~4. 
We just need an extra shove to cross the threshold 
to arrive at Axiom~4. 
This is quite different from Schr\"odinger's equation, 
though some authors put it on an equal footing 
with Born's Rule. 
But mathematical considerations alone do not get 
one close to Schr\"odinger's equation. 
One could just as well arrive at the Klein-Gordon 
equation or the Dirac equation. 
I consider Born's rule as very nearly inevitable, 
but not 
Schr\"odinger's equation. 
I have argued earlier that it is reasonable to 
replace Schr\"odinger's equation 
with Born's Rule as the
fundamental time evolution equation of 
quantum theory. 
But we have yet to see the {\em Generalized 
Born's Rule}. 
The next section is a step in that direction.

\section{\bf Quantum Conditional Probability}
\markboth{Quantum Conditional Probability}{Quantum Conditional Probability}
\label{section-conditional-prob}

The probability of an event occurring, given that
another event has already occurred, is a 
standard topic in classical probability. 
This corresponds to the relative frequencies 
measured for such a pair of events. 
This is known as {\em conditional probability.}
So we now consider the quantum analogue 
of conditional probability. 
Recall that if $( \Omega, \mathcal{F}, P) $
is a classical probability space 
we define the conditional probability of 
an event $ E_{1} \in \mathcal{F} $ given 
the event $ E_{2} \in \mathcal{F} $ 
as
\begin{equation}
     P ( E_{1} | E_{2}) := 
     \left\lbrace 
     \begin{array}{cc}
     \dfrac{P( E_{1} \cap E_{2} )}{P(E_{2})}
     \quad &\mathrm{if~} P(E_{2}) \ne 0
     \\
     \\
     0 \quad &\mathrm{otherwise}.
     \end{array}
     \right. 
\end{equation}
We immediately have 
$  P ( E_{1} | E_{2}) P(E_{2}) = P( E_{1} \cap E_{2} ) $ 
for all $ E_{1}, E_{2} \in \mathcal{F} $. 
From this we get Bayes Theorem
$$
 P ( E_{1} | E_{2}) P(E_{2}) = P( E_{1} \cap E_{2} ) =
  P( E_{2} \cap E_{1} ) = 
 P ( E_{2} | E_{1}) P(E_{1}).  
$$

We shall see that these simple formulas do not 
carry over to quantum probability. 
So this classical probability theory serves as 
an analogy, nothing more. 

Now suppose that $ S,T $ are self-adjoint operators
acting in $ \mathcal{H} $ and that $ B,C $
are Borel subsets of $ \mathbb{R} $. 
The probability that $ S$ has a value in $ B $ 
for a state $ \psi $ is given, as we have seen,  
by {\em Born's rule:}
\begin{equation}
\label{borns-rule-reprise}
    P ( S \in B \,|\, \psi) = 
    \langle \psi , P_{S} (B) \psi \rangle =
    ||   P_{S} (B) \psi  ||^{2}. 
\end{equation}
We wish to define the conditional probability
that $ T \in C $ given that $ S \in B $  for a 
state $ \psi $. 
This will be denoted as
$ P ( T  \in C \,|\, S \in B, \psi )$. 
We assume that this has the form 
$ P ( T \in C \,|\, \psi_{1} ) $ for some
state $ \psi_{1} $, that is to say,
it will be a quantum probability of a certain form. 
The choice of $ \psi_{1} $ depends on the first 
event $ S \in B $ and the state $ \psi $. 
The usual definition is 
\begin{equation}
\label{collapse}
   \psi_{1} := \dfrac{P_{S} (B) \psi}{|| P_{S} (B) \psi||} 
\end{equation}
provided $ P_{S} (B) \psi \ne 0 $. 
This definition has been given several unfortunate 
names.  
I will comment on those names and the significance 
of this definition after we have seen some of this
material developed further. 
But first here is 
the important definition that this 
leads up to. 

\begin{definition}
Let $ S,T $ be self-adjoint operators, 
$ B,C $ Borel subsets of $ \mathbb{R} $ and 
$ \psi $ a unit vector. $  $
Then the {\em (quantum) conditional probability} 
that $ T $ has a value in $ C $, given both that  
a state $ \psi $ is given and that  
$ S $ has a given value in B,  
is defined by 
\begin{equation}
\label{define-quantum-cond=prob}
P ( T  \in C \,|\, S \in B, \psi ) :=
P ( T \in C \,|\, \psi_{1} ) = 
P \Big( T \in C \,\Big|\, 
\dfrac{P_{S} (B) \psi}{|| P_{S} (B) \psi||} \Big)
\end{equation}
if $ P_{S} (B) \psi \ne 0 $. 
Also, if $ P_{S} (B) \psi = 0 $ we 
define
$ P ( T  \in C \,|\, S \in B, \psi ) := 0 $. 
\end{definition}

Some might prefer to leave the expression 
$ P ( T  \in C \,|\, S \in B, \psi )  $
undefined in case $ P_{S} (B) \psi = 0$. 
Taking $ S,T, B, \psi  $ fixed and 
if also $ P_{S} (B) \psi \ne 0$, then 
the formula 
$C \mapsto  P ( T  \in C \,|\, S \in B, \psi ) =
P ( T \in C \,|\, \psi_{1} ) $ shows that 
this expression defines a probability measure
on $ \mathbb{R} $. 
By using the formula in the 
next proposition one can see that the expression 
$  P ( T  \in C \,|\, S \in B, \psi ) $ is not 
even finitely additive in $ B $. 

\begin{prop}
With the notation of this definition  we have that 
\begin{equation}
\label{quantum-cond-prob}
P ( T  \in C \,|\, S \in B, \psi ) =
\dfrac{  || P_{T} (C_{}) P_{S} (B) \psi ||^{2}  }
{P ( S \in B \,|\, \psi)}
\end{equation}
provided that $ P ( S \in B \,|\, \psi) \ne 0 $. 
\end{prop}
\noindent 
{\bf Proof:}
Expanding this definition out we see 
that 
the quantum conditional probability satisfies 
\begin{align*}
&P ( T  \in C \,|\, S \in B, \psi ) = 
P \Big( T \in C \,\Big|\, 
\dfrac{P_{S} (B) \psi}{|| P_{S} (B) \psi||} \Big)
\\
&= 
\Big\langle 
\dfrac{P_{S} (B) \psi}{|| P_{S} (B) \psi||} , 
P_{T} (C) 
\dfrac{P_{S} (B) \psi}{|| P_{S} (B) \psi||} 
\Big\rangle 
=
\dfrac{1}{ || P_{S} (B) \psi||^{2} }
\langle 
 P_{S} (B) \psi , P_{T} (C_{}) P_{S} (B) \psi
\rangle 
\\
&= 
\dfrac{  || P_{T} (C_{}) P_{S} (B) \psi ||^{2}  }
{P ( S \in B \,|\, \psi)}. \qquad \blacksquare 
\end{align*}

\vskip 0.2cm 
For doing computations 
the expression \eqref{quantum-cond-prob} is much more
useful than 
Definition~\ref{define-quantum-cond=prob} 
and, in fact, is a preferable 
definition of quantum conditional probability, thereby avoiding the intermediate 
step \eqref{collapse} used to define $ \psi_{1} $. 
Also, it immediately follows 
for {\em any} value of $ P_{S} (B) \psi $ that 
$$
P ( T  \in C \,|\, S \in B, \psi ) \,
P ( S \in B \,|\, \psi) = 
 || P_{T} (C_{}) P_{S} (B) \psi ||^{2}. 
$$
This definition of conditional probability 
immediately allows us to define independence. 
This topic is discussed in the next section 
in the more general context of quantum events. 

Another immediate consequence of the definition is 
that for any $ T = T^{*} $ and Borel subset 
$ C $ of $ \mathbb{R} $ we have 
$ P ( T \in C \,|\, T \in C, \psi) = 1 $ 
for  all $ \psi $ satisfying 
$ P (T \in C \,|\, \psi) \ne 0 $.
This should be compared with the 
equally trivial identity 
that holds in classical probability theory: 
$ P (E | E) = 1 $ for every event $ E $, 
provided that $ P (E) \ne 0 $. 

After seeing all this the reader should realize that 
the definition \eqref{collapse} plays an
intermediate, minor role 
and could have been incorporated directly into 
the definition \eqref{quantum-cond-prob}
without further ado.   
But much ado has been spent on this so-called 
{\em collapse condition} \eqref{collapse}. 
It has also been dubbed the {\em quantum jump} and  
the {\em projection postulate}. 
So much ink has been spilled to `explain' or
`justify' this formula (often under the 
term `interpretation') that I am obliged to make 
some comments about it. 
The points of contact with observational 
data are the relevant, essential 
features of this, or any, theory. 
And for quantum theory those are its
quantum probabilities. 
This is the central importance of Born's rule 
\eqref{borns-rule}, which on the theoretical side 
is simply a definition. 
Similarly, the Generalized Born's Rule 
\eqref{generalized-borns-rule} below is simply 
a definition in the theory. 
Its importance comes from its relevance 
for understanding experimentally 
measured relative frequencies. 
When looking at different models of
quantum theory, it is perfectly acceptable to 
have discrepancies over matters that do not have
anything to do with observations. 
What is relevant are the probabilities, not the 
specific manners for calculating them in a 
particular model. 

For example, in the Schr\"odinger model the 
states vary in time while the pvm's do not. 
On the contrary, in the Heisenberg model 
the opposite obtains: The pvm's vary in time 
while the states do not. 
What is the same in both are the probabilities 
for the same events. 
This is what makes these two models isomorphic. 
It is important to note that Born's rule 
and its generalization are exactly 
the same formulas in both of these models. 
This is {\em covariance}.
When the time dependence is introduced into these 
equations (but in different manners depending on the model 
chosen), the calculated time dependent probabilities 
are identically equal in these two models. 
This is {\em invariance}. 
In the Schr\"odinger model 
the collapse formula  
\eqref{collapse} does have 
a mathematical sense to it, since 
states are allowed to change in that model.  
However, it is devoid of any 
physical significance since it is 
not part of other equivalent models, such as 
the Heisenberg model in which states do not 
change. 
Do be careful!
The physical significance of the theory 
comes only later in  
the formula for the quantum 
conditional probability 
\eqref{quantum-cond-prob} and 
the application of it 
to quantum events.

\section{\bf Quantum Consecutive Probability}
\markboth{Quantum Consecutive Probability}{Quantum Consecutive Probability}

Classical probability is a guiding analogy 
in the next definition. 
But both experiment and theory indicate that 
a time ordered sequence of two specific 
observations yielding certain values should 
have an associated probability. 
After all the relative frequencies of such 
sequences are often measured. 

\begin{definition}
Let $ S,T $ be self-adjoint operators, $ B,C $ 
Borel subsets of $ \mathbb{R} $ and $ \psi $ a unit
vector. 
Then the 
{\em (quantum) consecutive probability} is 
defined as 
\begin{equation}
\label{a-more-general-borns-rule}
P (S \in B, T \in C \,|\, \psi) := 
|| P_{T} (C) P_{S} (B) \psi ||^{2}. 
\end{equation}
The left side of this should be read as follows:
The probability that $ S $ takes a value in $ B $ 
and {\em subsequently} $ T $ takes a value in $ C $, 
given the state $ \psi $.
\end{definition}
Of course, one motivation of this definition is that
it allows us to write 
\eqref{quantum-cond-prob} in this form 
reminiscent of classical conditional probability: 
\begin{equation}
\label{cond-prob-TC-SB-bis}
P ( T  \in C \,|\, S \in B, \psi ) =
\dfrac{ P (S \in B, T \in C \,|\, \psi) }
{P ( S \in B \,|\, \psi)}
\end{equation}
provided that $ P ( S \in B \,|\, \psi) \ne 0 $. 

Some authors have the previous definition 
only for the case of commuting operators. 
Some say it is not meaningful 
for non-commutating operators, while others
react adversely to 
the fact that $  P_{T} (C) P_{S} (B) $ 
is not necessarily an event and so should 
not have an associated probability.  
However, I have found this definition in the 
general non-commutative case in the literature 
though I think it has been  unduly neglected. 
But that literature is vast, and my 
reading of it is necessarily limited. 
 
The expressions \eqref{quantum-cond-prob} and
\eqref{a-more-general-borns-rule} are 
generalization of Born's rule for calculating 
quantum probabilities. 
They are special cases of 
the {\em Generalized Born's Rule}, which will be 
presented later. 

We previously showed that Born's rule 
\eqref{borns-rule-reprise} is invariant if we 
transform the states by 
$ \psi \mapsto U_{t} \psi =: \psi^{\prime} $ and 
the pvm's by 
$ P_{S} \mapsto U_{t} P_{S} U_{t}^{*} =: P_{S^{\prime}} $,
where $ U_{t} $ is unitary for all $ t \in \mathbb{R} $. 
We will now show that the conditional probability 
\eqref{quantum-cond-prob} 
and the generalized Born's rule 
\eqref{a-more-general-borns-rule} are also 
invariant under this transformation. 
Of course, we also transform 
$ P_{T} \mapsto U_{t} P_{T} U_{t}^{*} =: P_{T^{\prime}} $. 
Given this notation 
we now have for 
$ P( S^{\prime} \in B ) \,|\, \psi^{\prime}) =
P( S \in B ) \,|\, \psi) \ne 0$ that 
the conditional probability \eqref{quantum-cond-prob} 
satisfies 
\begin{align*}
P ( T^{\prime} \in C \,|\, S^{\prime}\in B, \psi^{\prime} )
&=
\dfrac
{|| P_{T^{\prime}} (C)  P_{S^{\prime}} (B) \psi^{\prime}||^{2}}
{P( S^{\prime} \in B ) \,|\, \psi^{\prime})}
\\
&=
\dfrac{||U_{t}P_{T}(C) U_{t}^{*} U_{t} P_{S}(B) U_{t}^{*}
U_{t }\psi  ||^{2}}
{P( S \in B ) \,|\, \psi)}
\\
&=
\dfrac{||P_{T}(C) P_{S}(B) \psi  ||^{2}}
{P( S \in B ) \,|\, \psi)}
\\
&=
P ( T \in C \,|\, S \in B, \psi ). 
\end{align*}
On the other hand, if  
$ P( S^{\prime} \in B ) \,|\, \psi^{\prime}) =
P( S \in B ) \,|\, \psi) = 0$, then
$$
P ( T \in C \,|\, S \in B, \psi ) = 0 =
P ( T^{\prime} \in C \,|\, S^{\prime} \in B, \psi^{\prime}). 
$$
This proves that the quantum conditional probability is 
invariant under this transformation. 
The proof that the generalized Born's rule is 
also invariant is much the same. 
The importance 
of these invariances is that these quantum 
probabilities give the same results 
in all isomorphic quantum theories. 

Note that the self-adjoint 
operators $ S $ and $ T $ commute 
(by definition!) if and only if every projection
$ P_{S} (B)$ commutes with every projection $ P_{T} (C) $. 
So for commuting $ S $ and $ T $ we have by 
\eqref{a-more-general-borns-rule} that  
\begin{equation}
\label{combined-prob}
P (S \in B, T \in C \,|\, \psi) =
P (T \in C, S \in B \,|\, \psi). 
\end{equation}
This says that {\em in this case}
changing the temporal order of 
the two events does not change the combined probability. 
For such events simultaneity of the two events makes 
sense and, again, gives the same combined probability
\eqref{combined-prob}. 

However, it is not too difficult to find examples 
(thinking of spin $ 1/2 $, say)  
of non-commuting self-adjoint operators $ S $ and $ T $
for which there exist Borel sets $ B,C $ and a 
state $ \psi $ satisfying 
\begin{equation}
\label{and-does-not-exist}
P (S \in B, T \in C \,|\, \psi) \ne 
P (T \in C, S \in B \,|\, \psi). 
\end{equation}
This implies that for such $ S $ and $ T $ the
time order of the two events must not include 
simultaneity. 
This contrasts with the case 
of commuting $ S $ and $ T $ where
simultaneity is allowed. 

Perhaps a word or two is in order about 
the notion of simultaneity. 
I have not assumed Galilean invariance of this 
theory. 
Nor have I assumed Lorentz invariance. 
Either one of these invariances 
can be incorporated 
into quantum theory. 
Of course, these lead to non-isomorphic models 
of quantum theory. 
But in either case there is a well defined notion
of simultaneity of events with respect to an 
inertial frame of reference. 
That is the notion of simultaneity that should 
be applied here. 
I suppose that this is how simultaneity should 
be understood in a quantum theory of gravitation, 
even though such a theory does not exist at the 
time of writing this.  
This also relates to the question of the time ordering 
of events in general. 
In a Galilean theory this order is absolute, but 
in a Lorentzian theory it is only absolute for 
events that are light-like or time-like with 
respect to each other.
(I am excluding time reversal as a symmetry, of course.)
This motivates the extra hypothesis for 
Lorentz invariant quantum theories that two events that 
are space-like should commute.

This discussion has relevance to the 
well known--{\bf and false}--assertion that
``the electron can be at two places at the 
same time''. 
Let $ Q $ represent the position operator
for the electron. 
It is clear that $ Q $ commutes with itself. 
So for any state $ \psi $
\begin{equation}
\label{Q-in-B-and-Q-in-C}
    P ( Q \in B, Q \in C \,|\, \psi) =
    P ( Q \in C, Q \in B \,|\, \psi)
\end{equation}
for any pair of Borel subset $ B,C $ of $ \mathbb{R} $. 
This is important, since it includes simultaneous events, 
which is the case we want to consider. 
It also says that the probability of these simultaneous 
events makes sense in quantum theory. 
It is now just a matter of computing the probability
in \eqref{Q-in-B-and-Q-in-C}. 
So we start that calculation as follows:
\begin{equation*}
    P ( Q \in C, Q \in B \,|\, \psi) = 
    || P_{Q} (B) P_{Q} (C) \psi ||^{2}
    = || P_{Q} (B \cap C) \psi ||^{2}. 
\end{equation*}
The last equality is a property of pvm's that 
is proved in functional analysis. 
The expression `two places' I interpret as 
meaning that the sets $ B $ and $ C $ are disjoint,
that is, $ B \cap C = \emptyset$, the empty set. 
But, by the definition of pvm we have that 
$ P_{Q} (\emptyset) =0 $. 
We conclude that the probability in \eqref{Q-in-B-and-Q-in-C}
is $ 0 $ for disjoint Borel sets $ B,C $ and for {\em any} 
state $ \psi $.  
So much for the theory. 
What about observations? 
Well, I claim that there has never been an observation 
of an electron in two places at the same time. 
Of course, if such an observation were ever 
made, this aspect of quantum theory would have to be 
modified. 
But the principle of conservation of electric charge 
would also have to be modified. 
So, if there is anything `strange' or `mysterious' 
to understand here, 
I do not know what that might be. 
I also will not discuss the relation of this result 
with classical physics, because I do not wish to 
discuss classical physics at all in this treatise. 
As a final point let me note that this argument is 
quite abstract; it applies to any self-adjoint 
operator and any pair of disjoint Borel subsets 
of $ \mathbb{R} $.

The inequality \eqref{and-does-not-exist} is 
equivalent to  
$$ 
P ( T  \in C \,|\, S \in B, \psi ) \,
P ( S \in B \,|\, \psi) \ne 
P ( S  \in B \,|\, T \in C, \psi ) \,
P ( T \in C \,|\, \psi)
$$
which says that Bayes Theorem does not 
generalize to quantum probability.
Notice that the inequality \eqref{and-does-not-exist}
also contrasts with the classical case where 
$ P (E_{1} \cap E_{2}) = P (E_{2} \cap E_{1}) $. 
Moreover, \eqref{and-does-not-exist} precludes defining
a classical probability measure $ \pi $ 
on $ \mathbb{R}^{2} $
such that $ S $ takes a value in $ B $ and 
$ T $ takes a value in $ C $ in any order is equal 
to $ \pi ( B \times C ) $. 
Nonetheless, \eqref{quantum-cond-prob} has 
reasonable `marginals', namely
$$
P (S \in \mathbb{R}, T \in C \,|\, \psi) = 
|| P_{T} (C) P_{S} (\mathbb{R}) \psi ||^{2} 
= || P_{T} (C) \psi ||^{2} =
P (T \in C \,|\, \psi ),  
$$
since $ P_{S} (\mathbb{R}) = I $, the identity map. 
Similarly,
$$
P (S \in B, T \in \mathbb{R} \,|\, \psi) := 
|| P_{T} (\mathbb{R}) P_{S} (B) \psi ||^{2} 
= || P_{S} (B) \psi ||^{2} =
P ( S \in B \,|\, \psi). 
$$

On the other hand, taking $ S $, $ T $, $ B $ and $ \psi  $ 
fixed, the map that sends the Borel subset $ C $ 
of $ \mathbb{R} $ to $ P (  T \in C \,|\, S \in B, \psi) $ 
is a classical 
probability measure on $ \mathbb{R} $ as already 
remarked before.

\section{\bf Quantum Probability of Two Events}
\markboth{Quantum Probability of Two Events}{Quantum Probability of Two Events}
\label{section-two-events}

The results of the previous section have been presented 
in terms of events associated to pvm's. 
So we have considered events such as 
$ S \in B = P_{S} (B) $ and so forth. 
But it is useful to express these results more 
abstractly in terms of arbitrary events. 
Here are the definitions.
The first part of the next 
definition is Born's rule for one event, 
given a pure state. 
The second part  is Born's rule for one event, 
given a density matrix; 
it is a generalization of the first part. 

\begin{definition}
	Let $ E $ be an event, $ \psi $ be a unit vector 
	and $ \rho $ be a density matrix.
	Then the probability of $ E $ given $ \psi $ 
	is defined by Born's rule 
	\begin{equation}
	\label{borns-rule-one-event}
	P (E \,|\, \psi) := \langle \psi, E \psi \rangle 
	= || E \psi ||^{2}.
	\end{equation}
	Moreover, the probability of $ E $ given $ \rho $ 
	is defined by 
	\begin{equation}
	\label{rho-rule}
	P (E \,|\, \rho) := Tr ( E \rho). 
	\end{equation}
\end{definition}
In the sequel \eqref{rho-rule} 
will be regarded as 
a special case of the Generalized Born's Rule. 

Let's recall the standard justification of 
\eqref{rho-rule} in terms
of \eqref{borns-rule-one-event}. 
So we take an orthonormal basis $ \phi_{k} $
which diagonalizes the density matrix $ \rho $, 
say 
$ \rho = \sum_{k} \lambda_{k} 
|\phi_{k}\rangle \langle \phi_{k} |$ 
where
$ 0 \le \lambda_{k} \le 1 $ and $ \sum_{k} \lambda_{k} =1 $. 
Then we compute 
\begin{align*}
    P (E \,|\, \rho) &= Tr ( E \rho) 
    \\
    &= 
    \sum_{k} 
    \langle 
        \phi_{k} , E \, \rho \, \phi_{k} 
    \rangle 
\\
    &= 
   \sum_{k} 
   \langle 
      \phi_{k} , E \, \lambda_{k} \, \phi_{k} 
   \rangle 
\\
   &= 
   \sum_{k} 
   \lambda_{k} \,
   \langle 
       \phi_{k} , E \,  \phi_{k} 
   \rangle 
\\
   &= 
   \sum_{k} 
   \lambda_{k} \, P ( E \,|\, \phi_{k} ). 
\end{align*}
In short, 
$ P \big( E \,|\, 
\sum_{k} \lambda_{k} |\phi_{k}\rangle \langle \phi_{k} | 
\big) = 
\sum_{k} 
\lambda_{k} \, P ( E \,|\, \phi_{k} )$. 
This exhibits $ P (E \,|\, \rho) $ as an infinite 
combination of the component probabilities 
$ P ( E \,|\, \phi_{k} ) $, each with its 
corresponding weight factor $ \lambda_{k} $. 
An immediate consequence of the previous formula 
is $ 0 \le P (E \,|\, \rho) \le 1$. 
Also the following normalizations are easily shown 
and provide more justification for \eqref{rho-rule}: 
$$
   P ( I \,|\, \rho ) = 1 \quad \mathrm{and} 
   \quad  P ( 0 \,|\, \rho ) = 0. 
$$
Finally, there is a form of $ \sigma $-additivity 
which says that for any countable family 
$\{ E_{j} \,|\, j \in \mathbb{N} \} $ 
of orthogonal events, meaning that 
$ E_{j} \wedge E_{k} = 0$ for all $ j \ne k $, 
we have by a standard argument that 
\begin{equation}
\label{event-sigma-additivity}
   P \big( \textsf{V}_{j} E_{j} \,|\, \rho \big) =
   \sum_{j} P ( E_{j} \,|\, \rho ). 
\end{equation}
While this looks like the 
$ \sigma $-countable condition 
for classical measures, it is 
quite different. 
Notice that a necessary, but not sufficient, 
condition for the family
$\{ E_{j} \}$ to be orthogonal is that
$\{ E_{j} \}$ is a commutative family, 
a very restrictive condition. 
These facts motivate the next definition. 

\begin{definition}
	\label{define-qprob-function}
A {\em (quantum) probability} on the set $ \mathcal{E} $ 
of quantum events is  a function
$ \pi :  \mathcal{E} \to [0,1]$ such that 
\begin{itemize}
	
	\item 
	Normalization:
    $ \pi (0) = 0 $ and $ \pi (I) = 1 $.
    
    \item 
    $ \sigma $-additivity; 
     For any countable family 
     $\{ E_{j} \,|\, j \in \mathbb{N} \} $ 
     of orthogonal events in $ \mathcal{E} $ we have 
     $$
      \pi \big( \textsf{V}_{j} E_{j} \big) =
      \sum_{j} \pi ( E_{j}  ).
     $$
	
\end{itemize}
\end{definition}
It follows from the discussion above 
that for any density matrix $ \rho $ 
the function 
$ E \mapsto P ( E \,\, \rho )$ is a probability
on $ \mathcal{E} $. 
The question of the converse arises, that is, whether
all probabilities on 
$ \mathcal{E} \subset \mathcal{L} (\mathcal{H}) $   
have this form. 
The answer is yes, if $ \dim \, \mathcal{H} \ge 3$ 
by Gleason's theorem in \cite{gleason}.

The next definition gives two more versions of 
Born's rule, but now for {\em two time ordered} events. 
It is based on \eqref{quantum-cond-prob} 
and \eqref{a-more-general-borns-rule} of the previous section. 
\begin{definition}
	Let $ E_{1}, E_{2} $ be quantum events and
	$ \psi $ be a unit vector.  
	Then the {\em (quantum) consecutive probability} 
	of the event $ E_{1} $ and then later 
	the event $ E_{2} $, given 
	the state $ \psi $, is
	$$
	P (E_{1}, E_{2} \,|\, \psi) := ||  E_{2} E_{1} \psi  ||^{2}.
	$$
	(Notice that on the right side 
	the earlier event $ E_{1} $ 
	goes on the right.) 

	The {\em (quantum) condition probability} of 
	$ E_{2} $, given that $ E_{1} $ has occurred 
	and given a state $ \psi $, 
	is defined provided that $ E_{1} \psi \ne 0 $ by
	$$
	P ( E_{2} \,|\, E_{1}, \psi ) := 
	\dfrac{	P (E_{1}, E_{2} \,|\, \psi)}
	{P (E_{1} \,|\, \psi)}
	=
	\dfrac{ || E_{2} E_{1} \psi ||^{2}}{|| E_{1} \psi ||^{2}}.  
	$$
	If $ E_{1} \psi = 0 $, then we define 
	$ P ( E_{2} \,|\, E_{1}, \psi ) := 0 $. 

    We say that the {\em event $ E_{2} $ is 
    (quantum) independent
    of the event $ E_{1} $ with respect to 
    a state $ \psi $} if 
    $ E_{1} \psi \ne 0 $ and 
   \begin{equation}
    \label{E2-indept-of-E1}
    P ( E_{2} \,|\, E_{1}, \psi ) = 
    P ( E_{2} \,|\, \psi ). 
    \end{equation}

If $ S $ and $T $ are self-adjoint operators, then the 
{\em ordered pair} $ (S,T) $ is said to be 
{\em independent with respect to $ \psi $}
if for all Borel subsets $ B,C $ of $ \mathbb{R} $ 
the event $ S \in B $ is independent of the event 
$ T \in C $ with respect to $ \psi $, that is
$$
P ( S \in B, T \in C \,|\, \psi) = 
P ( S \in B \,|\, \psi ) \, P ( T \in C \,|\, \psi ). 
$$
(The motivation for this equation is given later.) 
\end{definition}

As we shall see in 
Chapter~\ref{Entanglement-Chapter}, 
entanglement is a special case 
of lack of independence. 
Due to the non-commutativity of quantum theory
quantum independence is not necessarily a symmetric relation. 
More specifically, $ E_{2} $ is independent of 
$ E_{1} $ given $ \psi $  if and only if 
$ E_{1} \psi \ne 0 $ and
$$
 || E_{2} E_{1} \psi ||^{2} = 
 || E_{1} \psi ||^{2} \, || E_{2} \psi ||^{2}.
$$
This equation trivially holds if $ E_{1} \psi = 0 $.
So the condition $ E_{1} \psi \ne 0 $ need not be 
imposed here and will be dropped when  we use 
this formula. 
Therefore the last equation is equivalent to 
$$
P ( E_{1}, E_{2} \,|\, \psi) = 
P ( E_{1} \,|\, \psi ) \, P ( E_{2} \,|\, \psi ),
$$
which gives the promised 
motivation of the last part of 
the previous definition.
Using \eqref{E2-indept-of-E1} we see that 
$ E_{1} $ is independent of 
$ E_{2} $ given $ \psi $  if and only if 
$$
|| E_{1} E_{2} \psi ||^{2} = 
|| E_{1} \psi ||^{2} \, || E_{2} \psi ||^{2} 
$$
or equivalently
$$
P ( E_{2}, E_{1} \,|\, \psi) = 
P ( E_{1} \,|\, \psi ) \, P ( E_{2} \,|\, \psi ). 
$$ 

Clearly if $ E_{1} $ and $ E_{2} $ commute, then this 
is a symmetric relation, that is,
$ E_{1} $ is independent from $ E_{2} $, given $ \psi $,
if and only if 
$ E_{2} $ is independent from $ E_{1} $, given $ \psi $. 
Also 
$ P ( E_{1}, E_{1} \,|\, \psi) = 
P ( E_{1} \,|\, \psi) $ trivially holds. 

This definition is taking us ever deeper 
into non-commutative territory. 
We assume $ \dim \mathcal{H} \ge 2 $, 
since that guarantees that $ \mathcal{L}( \mathcal{H} ) $ 
is non-commutative. 
For each state 
we already have a probability 
(cp. Definition~\ref{define-qprob-function}) 
defined on the 
set of quantum events $ \mathcal{E} $
in $ \mathcal{L}( \mathcal{H} ) $,  
and $ \mathcal{E} $ is not a $ \sigma $-algebra. 
But now we have, given a state, a probability 
defined on {\em ordered pairs} of events. 
(Neither $ E_{1} $ nor $ E_{2} $ 
is associated to a specific time. 
We only require the time order that $ E_{1} $ 
occurs first and then later $ E_{2} $.)  
This is yet another step beyond the
probability theory of  Kolmogorov  
based on $ \sigma $-algebras. 
The question arises as to what are the 
properties of quantum consecutive probability. 

Throughout the following we let $ \psi $ be any 
unit vector. 
First notice that 
$0 \le P (E_{1}, E_{2} \,|\, \psi)   \le 1  $, 
since $ ||  E_{2} E_{1}|| \le 1 $. 
Also, we have for any event $ E $
 the intuitively transparent formulas for the marginals 
$$
P (E, I \,|\, \psi) = P (E  \,|\, \psi) 
\quad \mathrm{and} \quad 
 P (I, E \,|\, \psi) = P ( E \,|\, \psi). 
$$
So, any event $ E $ and $ I $ are independent, 
and in both orders, since $ P ( I \,|\, \psi ) =1 $. 
Moreover, at the other extreme 
we have the normalizations 
$$
P (E, 0 \,|\, \psi) = 0
\quad \mathrm{and} \quad 
P (0, E \,|\, \psi) = 0. 
$$
And so any event $ E $ and the `never-YES'
event $ 0 $ are
independent (and again in both orders), 
since $ P ( 0 \,|\, \psi ) =0 $. 

Also, we have a form of $ \sigma $-additivity: 
Let the event $ F = \mathsf{V}_{j} F_{j} $, where 
$\{ F_{j} \}$ is a countable family of pairwise 
orthogonal events. 
Then for any event $ E $ by a standard argument we have 
for the consecutive probability that 
$$
P ( E, F \,|\, \psi  ) = \sum_{j} P ( E, F_{j} \,|\, \psi ). 
$$
This is $ \sigma $-additivity in the second event 
of the ordered pair of events. 

However, $ \sigma $-additivity fails in general 
in the first event 
for consecutive probability. 
We continue with the above notation 
and compute 
\begin{align*}
P ( F, E \,|\, \psi ) &= ||  E F  \, \psi ||^{2}
= || E \, (\mathsf{V}_{j} F_{j}  \, \psi ) ||^{2} 
\\
&=
\langle  E \, (\mathsf{V}_{j} F_{j} \, \psi ) , 
 E \, (\mathsf{V}_{k} F_{k} \, \psi ) \rangle 
\\
&= \sum_{j,k} 
\langle  E F_{j} \, \psi  , 
E \, F_{k} \, \psi \rangle 
\\
&= \sum_{j} 
\langle  E F_{j} \, \psi   , 
E  F_{j} \, \psi  \rangle 
+
\sum_{j \ne k} 
\langle  E F_{j} \, \psi , 
E F_{k} \, \psi  \rangle 
\\
&=
\sum_{j} 
P ( F_{j}, E \,|\, \psi ) 
+
\sum_{j \ne k} 
\langle \psi , 
F_{j} E F_{k} \, \psi  \rangle.  
\end{align*} 
The first summation is what one expects 
from $ \sigma $-additivity. 
But the second summation is a typical quantum term,
which already  
in the case $ F = F_{1} \vee F_{2} $ is  
producing interference with the first term. 
That is, the second sum, 
which is called an {\em interference term}, 
can either increase or decrease the first term. 
However, if $ E $ commutes with all of the 
events $ F_{j} $ (which already commute among
themselves by orthogonality), then we have 
for each pair $ j \ne k $ that 
$$
\langle \psi , F_{j} E F_{k} \, \psi  \rangle = 
\langle \psi , E F_{j} F_{k} \, \psi  \rangle = 0
$$
by the orthogonality condition $ F_{j} F_{k} = 0$. 
Therefore, in the commutative case this characteristic 
quantum interference term vanishes identically. 

Also it is important to remark that this interference term 
arises from the rules for computing quantum 
probabilities
and from nothing else. 
Of course, those rules are based on 
Hilbert space properties, particularly on the fact 
that the set of events is not a $ \sigma $-algebra. 
Note, for example, that 
there is no particle/wave duality being 
invoked here. 
In fact, there is no wave equation. 
There is no Superposition Principle 
for solutions or for states.  
There is no mention of the Uncertainty Principle, 
of Complementarity  
or of the Measurement Problem. 
There is no so-called `self-interference' of a
particle with itself. 
Even Schr\"odinger's equation is absent 
from the derivation of this result. 
If this interference term is not intuitive for 
you, it means that quantum probability is 
not intuitive for you.

The phrase `consecutive probability' 
is never even defined in classical probability 
theory for two good reasons. 
First, the order of events is not important. 
Second, the conjunction of two events 
$ A_{1}, A_{2} $ in a $ \sigma $-algebra
is their intersection $ A_{1} \cap A_{2} $, which is 
again in the $ \sigma $-algebra, that is, it is also
an event. 
So, in classical probability the probability of
two (and by induction any finite sequence of) 
events is itself the probability 
of just one event. 
This is not so in quantum probability. 
It is easy to construct events $ E_{1} $ and $ E_{2} $
in $ \mathcal{L} (\mathbb{C}^{2})  $ such that
$ 0 \ne E_{1} E_{2} \ne E_{2} E_{1} \ne 0$ and yet 
$ E_{1} \wedge E_{2} = 0 $, the `never-YES' event. 
In this case neither 
$ E_{1} E_{2} $ nor $ E_{2} E_{1} $ is an event. 
Nonetheless, the probabilities of sequences 
$ P ( E_{1}, E_{2} \,|\, \psi ) $ and 
$ P ( E_{2}, E_{1} \,|\, \psi ) $ make good sense 
and are not equal in general.  

\begin{example}
\label{example-independence} 
Here is a general example. 
Let $ \mathcal{H}_{1} $ and  $ \mathcal{H}_{2} $
be Hilbert spaces. 
Then define 
$\mathcal{H} \!:=\! \mathcal{H}_{1} \otimes\mathcal{H}_{2}$, 
the Hilbert space tensor product. 
Let $ F_{1} $ (resp.,~$ F_{2} $) be an event acting on
$ \mathcal{H}_{1} $ (resp., $ \mathcal{H}_{2} $).
Put $ E_{1} := F_{1} \otimes I_{2} $, 
$\, E_{2} := I_{1} \otimes F_{2} $, where
$ I_{j} $ is the identity map of 
$ \mathcal{H}_{j} $ for $ j=1,2 $. 
Then $ E_{1} $ and $ E_{2} $ are commuting events, 
but that 
is not enough to have independence. 
We also have to choose appropriately 
a unit vector $  \psi \in \mathcal{H} $. 
We choose $ \psi = \psi_{1} \otimes \psi_{2} $, 
where $ \psi_{1} $ (resp.,~$ \psi_{2} $)
is a unit vector in 
$ \mathcal{H}_{1} $ (resp., $ \mathcal{H}_{2} $). 
Then it is an easy exercise to show that 
$ E_{1} $ is independent of $ E_{2} $ with respect 
to $ \psi $. 
Of course, this example gets very interesting if 
we use a unit vector that does not factorize, 
namely an {\rm entangled state}. 
We will come back to this in 
Chapter~\ref{Entanglement-Chapter}. 
\end{example}

To generalize the last definition for the case 
of a density matrix, consider this expression:
\begin{align*}
|| E_{2} E_{1} \psi ||^{2} &= 
\langle E_{2} E_{1} \psi, E_{2} E_{1} \psi \rangle 
= 
\langle \psi, (E_{2} E_{1})^{*} E_{2} E_{1} \psi \rangle 
\\ 
&=
\langle 
\psi, (E_{2} E_{1})^{*} E_{2} E_{1} E_{\psi} \psi \rangle 
\\ 
&=
Tr ( (E_{2} E_{1})^{*} E_{2} E_{1} E_{\psi}  ). 
\end{align*}
where $ E_{\psi} = | \psi \rangle \langle \psi | $ is 
a rank $ 1 $ projection operator. 
So, using the same notation, 
the corresponding  definitions for a density 
matrix $ \rho $ are 
\begin{equation}
\label{conjunctive-prob-E1-E2}
P (E_{1}, E_{2} \,|\, \rho ) := 
Tr((E_{2} E_{1})^{*} E_{2} E_{1} \rho), 
\end{equation}
the {\em consecutive conjunctive probability} 
of first $ E_{1} $ and 
then later $ E_{2} $, given $ \rho $. 
Notice that $ (E_{2} E_{1})^{*} E_{2} E_{1} \rho $ is
a trace class operator, since $ \rho $ is. 
So its trace in \eqref{conjunctive-prob-E1-E2} 
is well defined. 
However, its trace norm 
need not be equal to its trace. 

Also the {\em conditional probability} of $ E_{2} $, 
given $ E_{1} $ and $ \rho $, is defined by
\begin{equation}
\label{cond-prob-2-events}
P( E_{2} \,|\, E_{1}, \rho ) := 
\dfrac{P (E_{1}, E_{2} \,|\, \rho )}
{P (E_{1} \,|\, \rho )}
=
\dfrac{Tr((E_{2} E_{1})^{*} E_{2} E_{1} \rho)}
{Tr ( E_{1} \rho )},
\end{equation}
provided that $ P (E_{1} \,|\, \rho ) \ne 0 $. 
The formula \eqref{cond-prob-2-events} is usually 
derived from L\"uders rule. 
(See \cite{luders}.) 
Here I have by-passed L\"uders rule and given 
\eqref{cond-prob-2-events} directly 
as a definition. 
One can easily manipulate \eqref{cond-prob-2-events} 
to arrive at the formula usually found in the literature
for this conditional probability. 
For example see \cite{cassinelli}, where 
a uniqueness result for this formula is also found.  
Since I arrived at \eqref{cond-prob-2-events} without 
being aware of the literature
(such as \cite{cassinelli}), my notation 
is not standard. 

Since these formulas may be unfamiliar in the
non-commutative context, let's see a justification
of \eqref{conjunctive-prob-E1-E2}. 
As before, 
we let $ \phi_{k} $ be an orthonormal basis 
that diagonalizes $ \rho $, that is, 
$ \rho = \sum_{k} \lambda_{k} |\phi_{k}\rangle \langle \phi_{k} |$ 
where
$ 0 \le \lambda_{k} \le1 $ and $ \sum_{k} \lambda_{k} =1 $. 
Then we see that 
\begin{align*}
P (E_{1}, E_{2} \,|\, \rho ) &= 
Tr((E_{2} E_{1})^{*} E_{2} E_{1} \rho) 
= 
\sum_{k} 
\langle 
\phi_{k} , (E_{2} E_{1})^{*} E_{2} E_{1} \rho \, \phi_{k}
\rangle 
\\
&= 
\sum_{k} 
\langle 
 E_{2} E_{1} \phi_{k} , E_{2} E_{1} \lambda_{k} \, \phi_{k}
\rangle 
= 
\sum_{k} \lambda_{k} \, || E_{2} E_{1} \phi_{k} ||^{2}
\\
&=
\sum_{k} \lambda_{k} \, P ( E_{1}, E_{2} \,|\, \phi_{k}). 
\end{align*}
So 
$ P \left(  E_{1}, E_{2} \,|\, \sum_{k} \lambda_{k} |\phi_{k}\rangle \langle \phi_{k} |  \right)  = 
\sum_{k} 
\lambda_{k} \, P ( E_{1}, E_{2} \,|\, \phi_{k} )$, 
showing $ P (E_{1}, E_{2} \,|\, \rho) $ as an
infinite 
combination of the component probabilities 
$ P ( E_{1}, E_{2} \,|\, \phi_{k} ) $, each with its 
corresponding weight factor $ \lambda_{k} $. 
As before, these probabilities are easily shown to be 
real numbers in the interval $ [0 , 1] $. 

As in the case of pure states, we define 
the {\em event $ E_{2} $ to be independent of the 
event $ E_{1} $ given a density matrix $ \rho $}, 
if $ P (E_{1}  \,|\, \rho ) \ne 0$ and 
$$ 
P (E_{2}  \,|\, E_{1}, \rho) = P (E_{2}  \,|\, \rho). 
$$
This is equivalent to
$$
P ( E_{1}, E_{2}  \,|\, \rho) = 
P (E_{1}  \,|\, \rho) 
P (E_{2}  \,|\, \rho).
$$
In general this is not a symmetric relation, 
though it is if $ E_{1} $ and $ E_{2} $ commute.

\begin{example}
	
The events $ E_{1} = F_{1} \otimes I $ and 
$ E_{2} = I \otimes F_{2} $ of Example 
\ref{example-independence}  
are independent in either order 
with respect to any density matrix 
$ \rho = \rho_{1} \otimes \rho_{2}  $, where 
$ \rho_{1} $ (resp., $ \rho_{2} $) is a density matrix 
acting on 
$ \mathcal{H}_{1} $ (resp., $ \mathcal{H}_{2} $), 
provided that $ P (F_{1} \,|\, \rho_{1} ) \ne 0 $
and $ P (F_{2} \,|\, \rho_{2} ) \ne 0 $. 
The details are left to the reader. 
\end{example}

\section{\bf Generalized Born's Rule with a State}
\markboth{Generalized Born's Rule with a State}{Generalized Born's Rule with a State}

Rather than spell out more details
of the case of two events, we continue with 
the generalization to a finite sequence 
of time ordered events,  
which is now readily at hand. 
This is the central definition of this treatise. 
\begin{definition}
\label{def-gen-borns-rule}
{\bf Generalized Born's Rule  with a State. }
Suppose that $ E_{1}, E_{2}, \dots , E_{n} $
for an integer $ n \ge 1 $ is an ordered sequence 
of events (possibly with repetitions) 
and let $ \psi $ be a unit vector.  
Then the {\em consecutive (conjunctive) 
probability} that first $ E_{1} $ occurs and then 
$ E_{2} $ occurs and so on continuing until  
$ E_{n} $ occurs, given $ \psi $, is defined as 
\begin{equation}
\label{n-event-Borns-rule}
P( E_{1}, E_{2}, \dots , E_{n} \,|\, \psi ):= 
|| E_{n} \cdots E_{2} E_{1} \psi ||^{2}. 
\end{equation}
Let $ \Lambda $ denote the {\em empty sequence} of events,
not to be confused with $ \emptyset $, the empty set. 
Note that $ \Lambda $ is vacuously ordered.  
Then we define $ P ( \Lambda  \,|\, \psi) :=1$. 

The {\em conditional probability} that the sequence of events
$ E_{1}, \dots , E_{n} $ occurs in that time order 
given that the sequence of events $ F_{1}, \dots, F_{k} $
has already occurred in that time order, 
given $ \psi $, is defined as  
\begin{align*}
P ( E_{1}, \dots , E_{n} \,|\, F_{1}, \dots, F_{k}, \psi )
&:= \dfrac
{P( F_{1}, \dots , F_{k}, E_{1}, \dots , E_{n}  
	\,|\, \psi )}
{P( F_{1} , \dots ,  F_{k} \,|\, \psi )} 
\\
&= 
\dfrac{|| E_{n} \cdots E_{1} F_{k} \cdots F_{1} \psi ||^{2}}{||  F_{k} \cdots F_{1} \psi ||^{2}} 
\end{align*}
provided the denominator is not zero. 
(The definition on the first line is for all integers 
$ k \ge 0 $, while the second line only holds 
for $ k \ge 1 $.) 
Otherwise we define 
$$
P ( E_{1}, \dots , E_{n} \,|\, F_{1}, \dots, F_{k}, \psi ) 
:= 0.
$$
These definitions are special cases of the following 
corresponding definitions given the same 
sequences of events and a density matrix $ \rho $. 
 
The {\em consecutive probability} 
of the ordered sequence of events 
$  E_{1},  \dots , E_{n} $
for $ n \ge 1 $ is defined as
$$
P( E_{1}, \dots , E_{n} \,|\, \rho ):= 
Tr \big((E_{n}\cdots E_{1} )^{*}E_{n}\cdots E_{1}\rho \big). 
$$
Also, we define $ P ( \Lambda \,|\, \rho) := 1 $. 

We say that a family of events 
$\{ E_{\alpha} \,|\, \alpha \in A \}$ which has an order 
induced from a linear order on $ A $ is {\em independent} 
with respect to $ \rho $ 
if for every finite ordered subset 
$  E_{\alpha_{1}}, E_{\alpha_{2}}, \dots , E_{\alpha_{n}} $
for $ n \ge 1 $ with 
$ \alpha_{1} <  \alpha_{2} < \cdots < \alpha_{n}$
we have 
$$
P(  E_{\alpha_{1}}, E_{\alpha_{2}}, \dots , E_{\alpha_{n}} \,|\, \rho ) =
P( E_{\alpha_{1}} \,|\, \rho ) \, 
P( E_{\alpha_{2}} \,|\, \rho )
\cdots 
P( E_{\alpha_{n}} \,|\, \rho ). 
$$
The ordered sequence $ T_{1}, \dots , T_{n} $ 
of self-adjoint operators is {\em independent} 
with respect to $ \rho $ if for every 
sequence of Borel subsets $ B_{1}, \dots , B_{n} $ 
of $ \mathbb{R} $ the ordered sequence of 
events 
$T_{1} \in B_{1}, \dots , T_{n} \in B_{n}$ 
is independent with respect to $ \rho $. 
(This can be defined for arbitrary families as well.) 

With the same notation as above 
the {\em conditional probability} is defined as
\begin{equation}
\label{generalized-borns-rule} 
P ( E_{1}, \dots , E_{n} \,|\, F_{1}, \dots, F_{k}, \rho )
:= \dfrac
{P( F_{1}, \dots , F_{k}, E_{1}, \dots , E_{n}  \,|\, \rho )}
{P( F_{1} , \dots ,  F_{k} \,|\, \rho )} 
\end{equation}
provided $ P( F_{1}, \dots , F_{k} \,|\, \rho ) \ne 0 $; 
otherwise it is defined to be $ 0 $. 

All of the previous definitions of quantum probability 
are special cases of 
\eqref{generalized-borns-rule}, which is called 
the {\em Generalized Born's Rule}. 

\end{definition} 

All of these probabilities are real numbers in 
the interval $ [0,1] $, and they have other obvious 
properties. 
However, the language of $ \sigma $-algebras has 
been left far behind. 

The ordering here of events reflects the time order 
of their occurrences, but does not associate  
them to specific times. 
However, since events are self-adjoint operators, 
they can have a time dependence in some 
models (such as the Heisenberg model). 
But this dependence does not impact the time
order of these events. 

Consider the union  
$ \{ F_{1}, \dots , F_{k}, E_{1}, \dots , E_{n} \} $  
of the ordered sequences 
as a new ordered sequence of events.  
If this new ordered sequence is independent, 
then {\em necessarily} we have that  
$$
P ( E_{1}, \dots ,E_{n} \,|\, F_{1},\dots,F_{k}, \rho ) = 
P ( E_{1}, \dots ,E_{n} \,|\, \rho ) \,
P( F_{1}, \dots , F_{k} \,|\, \rho ).   
$$
But this is not a {\em sufficient} condition for 
independence of the new ordered sequence, 
except in the case when $ n = k = 1 $. 

Having defined independence of an ordered sequence of 
observables (which are quantum random variables), 
it is desirable to define identically distributed 
observables as well. 
Then we will be able to speak of an ordered sequence of 
independent, identically distributed 
(iid) observables. 
\begin{definition}
	Let $ S $ and $ T $ be self-adjoint operators. 
	Then we say that $ S $ and $ T $ are {\em identically 
	distributed} with respect to a density matrix $ \rho $ 
	if
	$
	P (S \in B \,|\, \rho) = 	P (T \in B \,|\, \rho)
	$
	for all Borel subsets $ B $ of $ \mathbb{R} $. 
\end{definition}
Notice that this definition does not depend on the order of
$ S $ and $ T $. 
Clearly, this is a symmetric relation.
In fact, it is an equivalence relation 
on the set of self-adjoint operators. 
The construction of a finite sequence of iid 
observables can be done easily using tensor products. 

Everything in the formulas of these definitions 
is well known to anyone working in 
classical probability where each sequence of events 
becomes just one event. 
What is really 
new is not so much generalizing to sequences of 
non-commuting events, but rather identifying 
all of this as the 
Generalized Born's Rule 
\eqref{generalized-borns-rule} of quantum theory. 
Moreover, \eqref{generalized-borns-rule} is 
the fundamental time 
evolution equation of quantum theory, provided that 
the secondary, model dependent 
time evolutions are given for
the set of events and for the set of states. 

The normalization conditions  
$ P ( \Lambda  \,|\, \psi) =1$
and $ P ( \Lambda \,|\, \rho) := 1 $
could seem non-intuitive. 
In their defense  
they make the conditional probabilities work out 
when $ k = 0 $. 
But they seem to say that the probability that nothing
happens is $ 1 $. 
I suspect that this is a trap of language. 
Think of ever longer sequences of events. 
Easily such very long sequences can have probability $ 0 $ 
or very near $ 0 $. 
But then thinking of starting with such a long sequence 
of events with small 
probability and then considering shorter and shorter 
sub-sequences of it. 
The intuition is that the probability increases. 
And when one arrives at the empty 
sequence of events, one then  has the most probable 
situation. 
In more mundane terms we can say that events impose 
restrictions on probability and that, by removing all
such restrictions, one gets the most 
probable outcome, i.e., probability~$ 1 $. 

Continuing with the intuition in the previous 
paragraph, we can define the probability of an 
infinite sequence of events.
\begin{definition}
Let $\{ E_{j} \,|\, j \in \mathbb{N} \}$ be an infinite sequence
of linearly ordered events with $ E_{i} $ preceding 
$ E_{j} $ if and only if $ i < j $.  
Let $ \rho $ be a density matrix. 
Then we define the probability of the ordered 
sequence, given $ \rho $, as
\begin{align*}
   P ( \{ E_{j} \} \,|\, \rho ) &:= 
   \lim_{n \to \infty} P ( E_{1}, \dots , E_{n} \,|\, \rho) 
   \\
   &= 
    \lim_{n \to \infty} 
    Tr \big((E_{n}\cdots E_{1} )^{*}E_{n}\cdots E_{1}\rho \big)
    \\
    &=
    \inf_{n} \, Tr \big((E_{n}\cdots E_{1} )^{*}E_{n}\cdots E_{1}\rho \big). 
\end{align*}
\end{definition}

If this limit exists, then clearly 
$ 0 \le P ( \{ E_{j} \} \,|\, \rho )  \le 1 $. 
It also satisfies obvious normalization condition if 
all $ E_{j} = I $ or if any $ E_{j} = 0. $

The limit in this definition exists since the sequence 
is both bounded below by $ 0 $ and non-increasing. 
To prove the latter statement 
we compare the $ n+1$--st term with 
the $ n $--th term. 
Write the density matrix as 
 $ \rho = \sum_{k} \lambda_{k} |\phi_{k}\rangle \langle \phi_{k} |$ where
 $ 0 \le \lambda_{k} \le1 $, $ \sum_{k} \lambda_{k} =1 $ 
 and $\{ \phi_{k}  \}$ is an orthonormal basis 
 of $ \mathcal{H} $ which 
 diagonalizes $ \rho $. 
 The result is trivially true if $ E_{n+1} =0 $. 
 So we assume that $ E_{n+1} \ne 0 $ which 
 implies that $ || E_{n+1} || = 1 $. 
Then we have
\begin{align*}
 Tr 
 &\big( (E_{n+1} E_{n}\cdots E_{1} )^{*}
 E_{n+1} E_{n}\cdots E_{1}\rho \big) = 
 \\
 &= 
 \sum_{k} 
 \big\langle
 \phi_{k}, 
 (E_{n+1} E_{n}\cdots E_{1} )^{*}
 E_{n+1} E_{n}\cdots E_{1}\rho
 \phi_{k} 
 \big\rangle 
 \\
 &= 
 \sum_{k} 
 \big\langle
 E_{n+1} E_{n}\cdots E_{1} \phi_{k}, 
 E_{n+1} E_{n}\cdots E_{1}\rho
 \phi_{k} 
 \big\rangle 
\\
 &= 
 \sum_{k} 
 \lambda_{k} \big\langle
 E_{n+1} E_{n}\cdots E_{1} \phi_{k}, 
 E_{n+1} E_{n}\cdots E_{1}
 \phi_{k} 
 \big\rangle 
\\
 &= 
 \sum_{k} 
 \lambda_{k} 
 || E_{n+1} E_{n}\cdots E_{1} \phi_{k} ||^{2}
\\
 &\le \sum_{k} 
 \lambda_{k} 
 || E_{n+1} ||^{2} \, ||E_{n}\cdots E_{1} \phi_{k} ||^{2}
 \\
 &=
 \sum_{k} 
 \lambda_{k} 
 ||E_{n}\cdots E_{1} \phi_{k} ||^{2}
 \\
 &= 
 \sum_{k} 
 \lambda_{k} 
 \big\langle 
 E_{n}\cdots E_{1} \phi_{k}  , E_{n}\cdots E_{1} \phi_{k} 
 \big\rangle 
 \\
 &= 
 \sum_{k} 
 \big\langle 
 E_{n}\cdots E_{1} \phi_{k}  , E_{n}\cdots E_{1}
 \lambda_{k} \phi_{k} 
 \big\rangle 
 \\
 &= 
 \sum_{k} 
 \big\langle 
 \phi_{k}  , ( E_{n}\cdots E_{1})^{*} E_{n}\cdots E_{1}
 \rho \phi_{k} 
 \big\rangle 
 \\
 &=
 Tr 
 \big(  
 (E_{n}\cdots E_{1})^{*} E_{n}\cdots E_{1} \rho 
 \big).  
\end{align*}
This proves that the sequence is non-increasing. 
And this fact is behind the assertion that 
the conditional probabilities are $ \le 1 $. 
However, this result does {\em not} mean that more events
lowers the probability.
It only says that adding more events {\em after} 
a given sequence
of events lowers the probability. 
It is well known one can find events $ E_{1} $ and 
$ E_{2} $ acting on $ \mathbb{C}^{2} $ such that 
$ E_{2} E_{1} = 0 $ and so, in particular,
$ P ( E_{2} E_{1} \,|\, \psi ) = 0$, but that there 
exists an event $ F $ satisfying 
$ P ( E_{2} F E_{1} \,|\, \psi ) > 0$. 
(Think about light polarizing filters.)

Since $ E_{1} $ is the first event and 
$ E_{n} $ is the last event, 
the expression $ E_{n} \cdots E_{1} $ 
is well-known in quantum field theory. 
It is called a {\em time-ordered product}.
Clearly, the actual calculation of these probabilities 
can be rather challenging in practice. 
Such probabilities could be difficult 
to check in the laboratory as well. 
One typically prefers experimental 
situations with few events in play. 
However, nature does not always smile favorably 
on the experimental scientist. 
Even if one wished to study just two consecutive 
events, there may be other uncontrolled 
intermediate events so that one is studying 
a situation with many events instead of just two. 
Such undesired intermediate events are pejoratively 
dubbed as {\em noise} (as if they were not physical phenomena
which one could study) and the experimenter then 
works hard to eliminate them or, at least, to minimize 
their collective impact. 
Neither at an experimental level nor a theoretical 
level is there any `mystery' about such noise 
that needs special explanation.
It is simply very annoying. 
But to their merit some physicists do try to study 
this noise which they rename as {\em decoherence}. 

It is important to remark, and trivial to verify, 
that the probabilities in 
Definition~\ref{def-gen-borns-rule} 
are invariant  under isomorphisms 
between models. 
So it makes sense to suppose these probabilities  
could have physical significance. 
The Generalized Born's Rule clearly applies to the 
special case when all of the events have the 
form $ P (T \in B) $, where $ T $ is a self-adjoint 
operator and $ B $ is a Borel subset of $ \mathbb{R} $. 

Also, these quantum probabilities can be defined for events 
and states associated to any von Neumann algebra.

\section{\bf Generalized Born's Rule with no State}
\markboth{Generalized Born's Rule with no State}{Generalized Born's Rule with no State}

It makes mathematical sense to drop the state 
from the formulas and arrive at a definition of
probability for sequences of events with no 
mention of a state. 
This also has some physical motivation behind it 
though, as we shall see, it clashes with ideas from
classical probability. 
But these counter-intuitive definitions 
will be used to discuss entanglement 
in Chapter~\ref{Entanglement-Chapter}. 
\begin{definition}
Suppose that $ E_{1}, E_{2}, \dots , E_{n} $
for an integer $ n \ge 1 $ is an ordered sequence 
of events. 
Then the {\em consecutive (conjunctive) 
	probability} that first $ E_{1} $ occurs and then 
$ E_{2} $ occurs and so on continuing until  
$ E_{n} $ occurs is defined as 
\begin{equation}
\label{n-event-Borns-rule-no-state}
P( E_{1}, E_{2}, \dots , E_{n}  ):= 
|| E_{n} \cdots E_{2} E_{1} ||^{2}. 
\end{equation}
Let $ \Lambda $ denote the {\em empty sequence} of events. 
Then we define $ P ( \Lambda ) :=1$. 

The {\em conditional probability} that the sequence of events
$ E_{1}, \dots , E_{n} $ occurs in that time order 
given that the sequence of events $ F_{1}, \dots, F_{k} $
has already occurred in that time order 
is defined as  
\begin{align*}
P ( E_{1}, \dots , E_{n} \,|\, F_{1}, \dots, F_{k} )
&:= \dfrac
{P( F_{1}, \dots , F_{k}, E_{1}, \dots , E_{n} )}
{P( F_{1} , \dots ,  F_{k} )} 
\\
&= 
\dfrac{|| E_{n} \cdots E_{1} F_{k} \cdots F_{1} ||^{2}}{||  F_{k} \cdots F_{1} ||^{2}} 
\end{align*}
provided the denominator is not zero. 
(The definition on the first line is for all integers 
$ k \ge 0 $, while the second line only holds 
for $ k \ge 1 $.) 
Otherwise we define 
$$
P ( E_{1}, \dots , E_{n} \,|\, F_{1}, \dots, F_{k}) 
:= 0.
$$
\end{definition}

This definition will be used in the Chapter on 
Entanglement. 
So it does have relevance to physics. 
However, it is time independent in the
Heisenberg model. 
(Note that the Schr\"odinger model 
does not apply here.)  
Thus, it is only applicable in time 
independent situations, that is, with 
dynamics given by the Hamiltonian $ H = 0 $. 
And it is non-intuitive, at least for me. 
Here is a puzzling particular case. 
If $ E \ne 0 $ is an event, then 
$$
   P (E) = || E ||^{2} = 1. 
$$

This contradicts the intuition that 
non-zero events should have non-trivial 
probabilities. 
In particular, the map $ P : \mathcal{E} \to [0,1] $ is not
$ \sigma $-additive, 
where $  \mathcal{E} $ is the set of all 
events in $  \mathcal{H} $. 
Therefore, $ P $ is not a pvm. 
Also, when restricted to  a subset 
of $ \mathcal{E} $ which is a $ \sigma $-algebra
(the commutative, classical case), 
the range of $ P $ is 
just $ \{ 0,1\} $. 
So in Chapter~\ref{Entanglement-Chapter} 
on Entanglement at first 
only the conditional 
probability with a state will be used although 
a similar analysis with the less intuitive 
conditional probability with no state 
will be discussed as well. 

This probability with no state 
contrasts sharply with the probability of
an event, given a state $ \psi $, for which we
have that 
$$
   P (E \,|\, \psi) = || E \psi ||^{2}
$$
can be any number in the interval $ [ 0,1 ] $
for events $ 0 \ne E \ne I $. 
This accords with the classical idea that 
the probability of an event reflects to some degree 
the state in which the system `finds' itself. 
But $ P (E) = 1 $ for all non-zero events 
seems to mean something decidedly weaker. 
Since $ P (0) = 0 $ is the only exceptional 
case and $ 0 $ is the event that has
only the value NO, one is led to think that 
$ P (E) = 1 $ says that YES is a possible 
value of $ E $. 
Again, this is a very weak condition.  

Here is a situation where the 
probability with no state
makes more sense. 
This is an identity 
that we will use later.
We let $ \psi \in \mathcal{H} $ be a 
unit vector and let $ E $ be an event or a 
product of events. 
Then we have
\begin{align}
P ( E  \,| \psi \rangle \langle \psi | ) 
\nonumber 
&= 
|| \, E  \,| \psi \rangle \langle \psi | \,
||^{2}
\\
&= \left( 
\sup_{||\phi||=1} 
|| E  \,| \psi \rangle \langle \psi | \, \phi ||
\right)^{2} 
\nonumber 
\\
&= \sup_{||\phi||=1} 
|\langle \psi , \phi \rangle |^{2} 
|| E \psi ||^{2}
\nonumber 
\\
&= || E \psi ||^{2}
\nonumber
\\
&= P (E \,|\, \psi ).
\label{E-psi-psi-is-E-psi}
\end{align}

Let's see how this works in a familiar 
example. 
Let $ Q $ denote the position operator 
defined in the 
Hilbert space $ L^{2} ( \mathbb{R} )$. 
Let $ a,b \in  \mathbb{R} $ satisfy $ a < b $. 
Then the quantum event 
$$ 
Q \in [a,b]= P_{Q} \big( [a,b] \big) \ne 0 
$$
and so $ P (  Q \in [a,b] ) = 1$. 
On the other hand 
$$
  P ( Q \in [a,b] \,|\, \psi) = 
  || \big( Q \in [a,b] \big)  \psi  ||^{2}
$$
can assume any value in $ [0,1] $, depending 
on the value of the unit vector $ \psi $. 
Since there exist some states such that 
the event $  Q \in [a,b]  $ gives the value 
YES, the probability of the event itself 
is $ 1 $. 
However, the quantum event 
$$ 
Q \in [a]= P_{Q} \big( [a] \big) = 0 
$$
and so $ P (  Q \in [a] ) = 0$. 
This is a way of saying in the 
language of quantum probability 
that the position observable can not give sharp 
numerical values, but can give values in 
non-trivial intervals.

\section{\bf Probability Amplitudes}
\markboth{Probability Amplitudes}{Probability Amplitudes}

In many formulations of quantum theory 
it is emphasized that the quantum probabilities 
are calculated as the absolute value squared 
of a probability amplitude. 
That is implicit in this approach in the 
appropriate case when we have two unit 
vectors $ \psi, \phi \in \mathcal{H}$, 
in which case we define their 
probability amplitude to be 
$ A ( \phi , \psi ) := \langle \phi, \psi \rangle $. 
We use the notation 
$ E_{\phi} = | \phi \rangle \langle \phi |$, 
the event that `$ \phi $ occurs'. 
Then the probability of $ E_{\phi} $, given 
$ \psi $, is 
$$
   P ( E_{\phi} \,|\, \psi ) = 
   ||  E_{\phi} \, \psi ||^{2} =
   || \langle \phi , \psi \rangle \phi ||^{2} 
   = |  A ( \phi , \psi ) |^{2}. 
$$
For fixed $ \phi $ the map 
$ \psi \mapsto A ( \phi, \psi ) $ 
is linear 
provided that $ \psi $ is allowed to be {\em any} 
vector in $ \mathcal{H} $, but that 
$ \psi \mapsto | A ( \phi, \psi ) |^{2} $
is not linear. 
Of course, the more relevant concept here 
is convexity. 
So, suppose that 
$ \psi = \sum_{k} \lambda_{k} \phi_{k}$, where 
$\{ \phi_{k} \} $ is an orthonormal basis and
the complex numbers $ \lambda_{k} $ satisfy 
$ \sum_{k} | \lambda_{k} |^{2} = 1 $. 
Then we have a form of $ \sigma $-additivity 
of amplitudes: 
$$
A ( \phi, \psi ) = 
\sum_{k} \lambda_{k} \langle \phi, \phi_{k} \rangle. 
$$
But in general 
$$
|A ( \phi, \psi )|^{2} \quad \mathrm{and} \quad  
\sum_{k} | \lambda_{k} 
\langle \phi, \phi_{k} \rangle|^{2} 
= \sum_{k} | \lambda_{k} |^{2} \, 
| A ( \phi, \phi_{k} ) |^{2} 
$$
are not equal due to the well known 
interference terms. 

Probability amplitudes can also be defined for all 
the other examples of quantum probability, but
they are elements in the Hilbert space instead 
of being complex numbers. 
For example, for a quantum event $ E $ and a unit 
vector $ \psi $ we have 
$ P ( E \,|\, \psi ) = || A ( E, \psi ) ||^{2} $, 
where we define the probability amplitude as 
$ A ( E, \psi ) := E \, \psi \in \mathcal{H} $. 

Essentially, probability amplitudes add more 
notation to the theory but without shedding much more
light on it. 
However, that language is available if you wish 
to use it.

\section{\bf Quantum Integrals}
\markboth{Quantum Integrals}{Quantum Integrals}
\label{section=quantum-integrals}

Having defined and studied quantum 
probability, 
it is now straightforward to 
define and study 
the quantum theory of integration as 
was mentioned earlier as
$ \mathcal{E} (A \,|\, \rho) = Tr ( A \,  \rho ) $
for any $ A \in \mathcal{L} ( \mathcal{H} )$ and
density matrix $ \rho  $. 
We can say that this is the {\em (non-commutative) integral 
of $ A $ with respect to the state $ \rho $}.
Using terminology from classical probability theory we 
can call this the {\em expectation of $ A $ 
with respect to $ \rho $}. 
In quantum physics one says that this is the {\em expected 
value of $ A $ in the state $ \rho $}; 
this is well known since the early days 
of quantum theory, although only after Born's seminal paper 
appeared. 
So the next definition seems to be natural, though 
its importance is not clear. 
\begin{definition}
	Let $ A_{1}, A_{2}, \dots , A_{n} $ be an 
	ordered sequence 
	in $ \mathcal{L} ( \mathcal{H} ) $ and 
	let $ \rho $ be a density matrix. 
	We define the {\em time-ordered integral} of 
	this sequence of operators with respect to $ \rho $ 
	to be 
	\begin{equation}
	\label{non-commutative-n-integral}
	\mathcal{E} ( A_{1}, A_{2}, \dots , A_{n} \,|\, \rho ) :=
	Tr ( 
	A_{n} \cdots A_{2}  A_{1} \, \rho ). 
	\end{equation}
	This is also called the 
	{\em expectation of the ordered sequence
	$ A_{1}, A_{2}, \dots , A_{n} $~with 
	respect to $ \rho $}. 
\end{definition}

Of course, the time ordered integral of an ordered 
sequence of operators is
equal to the integral of a single operator, since
$
\mathcal{E} ( A_{1}, A_{2}, \dots , A_{n} \,|\, \rho ) =
\mathcal{E} (  
 A_{n} \cdots A_{2} A_{1} \,|\, \rho ).
$
But the point is that non-commutativity 
of $ \mathcal{L} ( \mathcal{H} ) $ makes the order
of the operators important. 
And that is what is underlining this definition. 
The corresponding definition in usual measure 
theory, which is a commutative integration theory, 
would not have such importance. 

There is a temptation to say that 
\eqref{non-commutative-n-integral}
is a 'state' that is associated to the 
probability for ordered sequences of events. 
However, it has some properties that argue 
against being so named. 
For example, for an ordered sequence 
$  E_{1}, E_{2}, \dots , E_{n}  $ of events 
with $ n \ge 2 $
we have in general that 
$$
\mathcal{E} ( E_{1}, E_{2}, \dots , E_{n} \,|\, \rho )
\ne P ( E_{1}, E_{2}, \dots , E_{n} \,|\, \rho ). 
$$
So with this definition the expectation does 
not extend the probability. 
Also, there is no apparent positivity property. 
One `nice' property that it does have is the normalization 
$\mathcal{E} ( I, I, \dots , I \,|\, \rho ) = 
Tr \, \rho = 1 $. 
It also has reasonable marginals such as
$ \mathcal{E} ( I, E_{2}, \dots , E_{n} \,|\, \rho ) = 
\mathcal{E} ( E_{2}, \dots , E_{n} \,|\, \rho )$ 
and so forth. 

But now we also have quantum conditional probability 
at our disposal, and so we should have 
quantum conditional expectation as well. 
\begin{definition}
Let $ E_{1}, \dots , E_{k} $ be an 
ordered sequence of events 
in $ \mathcal{L} ( \mathcal{H} ) $ 
and $ A_{1}, \dots , A_{n} $ be an 
ordered sequence of operators 
in $ \mathcal{L} ( \mathcal{H} ) $. 
Also let $ \rho $ be a density matrix. 
Then the 
{\em (quantum) conditional expectation} is defined as
\begin{equation}
\label{quantum-conditional-expectation} 
\mathcal{E} ( A_{1}, \dots , A_{n} \,|\, 
E_{1},\dots, E_{k}, \rho ):= 
\dfrac
{\mathcal{E}( A_{1}, \dots A_{n}, 
E_{1} ,\dots, E_{k} \,|\, \rho )}
{\mathcal{E}( E_{1}, \dots , E_{k} \,|\, \rho )} 
\end{equation}
provided 
$\mathcal{E}( E_{1}, \dots , E_{k} \,|\, \rho ) \ne 0$. 
(Note that the denominator here is not 
$ P ( E_{1}, \dots , E_{k} \,|\, \rho ) $.) 

\end{definition}

Note in the Schr\"odinger and Heisenberg models 
that the time evolution extends naturally to 
$ \mathcal{L} ( \mathcal{H} ) $. 
Consequently, given a Hamiltonian 
these integrals are also 
time dependent. 
This definition is quite different 
from the partial trace, which is a non-commutative
sort of conditional expectation in the specific 
context of tensor product Hilbert spaces.

\section{\bf Born's rule redux}
\markboth{Born's rule redux}{Born's rule redux}

Until the end of the last section, 
I had been rather cavalier in using the term 
{\em Born's rule}. 
What I mean by it is any formula in quantum theory that
is a special case of \eqref{generalized-borns-rule}. 
Since M.~Born was the first to give such a formula 
in quantum theory, 
I have decided to credit him 
by calling \eqref{generalized-borns-rule} 
and its immediate consequences 
the 
{\em Generalized Born's Rule}. 
Actually, I have not yet presented Born's rule 
in its original form. 
It might be instructive for the reader to see 
this in detail. 
To do this I shall dive ever so shallowly into 
historical waters.  

The time independent version of Schr\"odinger's equation 
$ H \psi = E \psi $ for a self-adjoint 
partial differential operator $ H = H^{*} $ 
is an eigenvalue problem with 
two unknowns for which it must be solved:
the eigenvalue $ E \in \mathbb{R}$ 
and its corresponding non-zero eigenvector $ \psi $. 
Already in Schr\"odinger's first paper
\cite{schrodinger} on the subject it was  
realized that $ E $ represents 
an energy, but the physical significance 
of $ \psi $ was left unresolved in that paper. 
However, it seemed reasonable that the solution $ \psi $ 
should also have some physical significance. 
And a similar concern arises with 
the solution $ \psi_{t} $ of the time dependent 
Schr\"odinger equation.

In modern terminology 
M.~Born addressed this 
in the specific case that 
$ \psi \in L^{2} (\mathbb{R}^{3}) $. 
However, to avoid a lot of sub-indices 
let's consider the case 
$ \psi \in L^{2} (\mathbb{R}) $, since 
the same same ideas apply. 
So, $ \mathcal{H} =  L^{2} (\mathbb{R}) $ is 
the Hilbert space for this situation. 
The basic assumption is that the position 
of the system is a relevant observable, 
that is, the values of its pvm lie in the 
von Neumann algebra of the system. 
Here the self-adjoint position
operator $ Q : D (Q) \to L^{2} (\mathbb{R}) $
is defined on the dense subspace
$$
 D (Q) := \{ \psi \in  L^{2} (\mathbb{R}) \,|\,
 x \psi (x) \in  L^{2} (\mathbb{R}) \}
$$
by the formula $ Q \psi (x) := x \psi (x) $. 
But more importantly, the pvm of $ Q $ is 
$
    P_{Q} (B) \, \phi = \chi_{B} \, \phi 
$
for all Borel subsets $ B $ of $ \mathbb{R} $ 
and all $ \phi \in  L^{2} (\mathbb{R}) $. 
Here $ \chi_{B} $ is the characteristic function 
of the Borel set $ B $, defined for all 
$ x \in \mathbb{R} $ as
$$
    \chi_{B} (x) := 
    \left\{ 
    	\begin{array}{cc}
    	1 & \mathrm{if~} x \in B, 
    	\\
    	0 & \mathrm{if~} x \notin B. 
    	\end{array}
    \right.  
$$
All of these results about $ Q $ 
come from functional analysis. 
We continue by using Born's rule as given in 
Axiom~4 to calculate the probability 
that the position of the system is in 
a Borel subset $ B $ of $ \mathbb{R} $ 
given a unit vector $ \psi \in  L^{2} (\mathbb{R}) $ 
as follows: 
\begin{align*}
P ( Q \in B \,|\, \psi ) &=
\langle \psi , P_{Q} (B) \psi \rangle = 
\int_{\mathbb{R}} dx \, \psi(x)^{*} 
\big( P_{Q} (B) \psi (x) \big) = 
\\
&= 
\int_{\mathbb{R}} dx \, \psi(x)^{*} \chi_{B} (x) \psi (x) = 
\int_{B} dx \, \psi(x)^{*} \psi (x) 
\\
&= 
\int_{B} dx \, |\psi(x)|^{2}. 
\end{align*}
The last expression on the right here is the formula given 
by Born for the probability that the position of the system 
is in $ B $ given the unit vector $ \psi $. 
Of course, Born came to this conclusion without using 
all the tools of 
quantum probability, which came later. 
Actually, Born initiated the field of 
quantum probability by indicating 
the physical significance of this expression. 
This formula for the probability is one 
way of viewing the physical significance of the solution 
$ \psi \in  L^{2} (\mathbb{R})$ of the
eigenvalue problem $ H \psi = E \psi $. 
It we consider $ P ( T \in B \,|\, \psi) $ for 
some other self-adjoint operator $ T $, we can give 
$ \psi $ some other physical significance. 
I do not wish to elevate this last comment to the level 
of a general principle of complementarity; 
it is merely another application of Born's rule 
as given in Axiom~4. 

Concerning a solution $ \psi_{t} \in  L^{2} (\mathbb{R})$ 
of the time dependent 
Schr\"odinger's equation, Born's rule asserts that 
\begin{equation}
\label{prob-of-Q-dependent-on-t}
P ( Q \in B \,|\, \psi_{t} ) = 
\int_{B} dx \, |\psi_{t}(x)|^{2}
\end{equation}
is the probability at every time 
$ t \in \mathbb{R} $ that the
position of the system is in the Borel set $ B $. 
So we see time dependent probability in quantum 
theory in this simple example. 
However, in this treatise we take equation 
\eqref{prob-of-Q-dependent-on-t} to be the 
fundamental time evolution equation of the position
observable~$ Q $. 
Moreover, we see in 
this example that the time dependent 
Schr\"odinger's equation plays a secondary role 
in understanding the physical 
significance of $ \psi_{t} $.

\section{\bf Comparison with Classical Probability}
\markboth{Comparison with Classical Probability}{Comparison with Classical Probability}
\label{comparison=section}

This section, just as the rest of this treatise, 
does not 
address the issues of comparing quantum theory 
with classical physical theory. 
Rather, it is a comparison of Kolmogorov's 
formulation in \cite{kolmogorov} 
of classical probability in terms of
measure theory with the quantum theory of 
probability that has been presented here. 

A key difference is the mathematical 
structure of the set of events. 
In classical probability the events are 
elements of a {\em $ \sigma $-algebra} 
$ \mathcal{F} $ whose elements are subsets 
of a non-empty {\em sample space} $ \Omega $. 
In particular this is a Boolean algebra. 
This means the various rules of Boolean 
algebra hold including the 
de~Morgan identities. 
Another way of saying this is that the events
obey the rules of classical logic that go back at least
to the works of Aristotle. 
Thinking that events tell us that nature has 
certain properties, this means that those 
properties also satisfy the rules classical logic. 

In quantum theory the events are the closed subspaces
($ \equiv $ projections) of a {\em complex}
Hilbert space $ \mathcal{H} $. 
(Note that the role of the complex numbers 
seems to be essential.)  
These form a complete orthomodular lattice 
for which the de~Morgan identities fail if 
$ \dim \mathcal{H} \ge 2 $. 
Consequently, if one assigns `properties'  
to these events, then classical logic 
will not apply to them. 
So we must think differently about quantum events. 
Of course, we can always say that the event itself 
is a property, but this is quite distinct from how 
classical probability is structured. 

Quantum events when viewed as projections lie 
inside a larger structure, 
namely the complex vector space  
$ \mathcal{L}(\mathcal{H}) $. 
To draw comparisons it is 
convenient to embed classical events
in a larger structure, namely the complex vector space 
$\mathcal{M} (\Omega) := \{ X : \Omega \to \mathbb{C} \,|\, X \mathrm{~is~measurable}  \}$. 
Then a classical event $ A \in \mathcal{F} $ is 
associated with its characteristic 
function $ \chi_{A} \in \mathcal{M} (\Omega) $. 
One has $ \chi_{A}^{2} = \chi_{A} = \chi_{A}^{*} $. 
Conversely, for every 
$ X \in \mathcal{M} (\Omega) $ satisfying 
$ X^{2} = X = X^{*} $, there 
exists a unique set $ A \in \mathcal{F} $ 
such that $ X = \chi_{A} $. 
So, we can define classical events equivalently 
as those $ \chi \in \mathcal{M} (\Omega) $ satisfying 
$ \chi^{2} = \chi = \chi^{*} $. 
This compares favorably with the definition of 
a quantum event as those 
$ E \in \mathcal{L}(\mathcal{H}) $
satisfying $ E^{2} = E = E^{*} $. 

In classical probability the observables are 
called {\em random variables} and are defined 
as those elements of $ \mathcal{M} (\Omega) $ 
that are real valued, that is, 
those $ X \in \mathcal{M} (\Omega) $ satisfying 
$ X = X^{*} $. 
The {\em essential range} of any  $ X \in \mathcal{M} (\Omega) $ 
is the {\em spectrum} of $ X $, 
denoted $ \mathrm{Spec} (X) $. 
(See a text on measure theory for definitions.)
If $ X $ is a random variable, then 
the elements of $ \mathrm{Spec} (X) $ form a 
non-empty subset
of $ \mathbb{R} $, and its elements 
correspond to the values 
seen in the observations associated to $ X $. 
For each Borel subset $ B $ of $ \mathbb{R} $ there is 
an event in $ \mathcal{F} $ that is denoted as 
$ X \in B $
and is called 
the event that $ X $ is observed to have a 
value in $ B $. 
It is defined as
$ X \in B := X^{-1} (B) $. 
If a probability measure $ P $ on $ \mathcal{F} $ 
is given, 
then the {\em probability that $ X $ has a value in 
the Borel set $ B $} is defined to be 
$ P (X \in B ) = P ( X^{-1} (B) ) $. 
For $ X $ and $ P $ given, the map 
$ B \mapsto P (X \in B ) =: \mu_{X} (B) $ is a probability 
measure on $ \mathbb{R} $ which is called 
the {\em distribution of $ X $ (with respect to $ P $)}. 
The phrase in parentheses is often omitted since 
$ P $ is implicit in a many contexts. 
The probability measure $ P $, being a measure, 
has a theory of integration that comes with it for free. 
So, integrals 
$\mathcal{E} (X):= \int_{\Omega} dP(\omega) \, X(\omega) $
are defined for a wide class of 
$X \in \mathcal{M} ( \Omega )$, including 
all bounded, Borel measurable functions. 
We say that $ \mathcal{E} (X) $ is the 
{\em expected value} of $ X $ (with respect to $ P $).

An important identity for the expected value is
$$
\mathcal{E} (X)= \int_{\Omega} d P(\omega) \, X(\omega) = 
\int_{\mathbb{R}} d \mu_{X} (\lambda) \, \lambda  
$$
in the sense that if one of these two integrals exists, 
then so does the other and the equality holds. 
This expresses the expected value as the first moment of 
a probability measure on $ \mathbb{R} $. 
More generally, for any bounded, Borel 
function $ f : \mathbb{R} \to \mathbb{C} $ and 
all $ \omega \in \Omega $ 
we define $ f (X) (\omega) := f ( X ( \omega) ) $, which 
is itself bounded and Borel measurable. 
Then we have that the probability measure $ \mu_{X} $ 
satisfies 
$$
\mathcal{E} ( f(X) )= 
\int_{\Omega} dP(\omega) \, f (X) (\omega) = 
\int_{\mathbb{R}} d \mu_{X} (\lambda) \, f (\lambda).   
$$

The observables in quantum theory are 
self-adjoint operators $ T = T^{*} $, but 
the condition $ T \in \mathcal{L}(\mathcal{H}) $
is not required though it may hold. 
The spectrum of any self-adjoint operator
(bounded or unbounded) is a non-empty, closed 
subset of $ \mathbb{R} $, and its elements 
correspond to the values 
seen in the observations associated to $ T $. 
For each Borel subset $ B $ of $ \mathbb{R} $ there is 
an event in $ \mathcal{L} ( {\mathcal{H}} )$ 
that is denoted as 
$ T \in B $
and is called 
the event that $ T $ is observed to have a 
value in $ B $. 
It is defined as 
$ T \in B := P_{T} (B)$, where 
$ P_{T} $ is the pvm associated to $ T $
by spectral theory. 
If a density matrix $ \rho $ is given, 
then the probability that $ T $ has a value 
in the Borel subset $ B $ of $ \mathbb{R} $ 
is defined by Born's rule to be 
$$ 
P ( T \in B \,|\, \rho) = Tr (  P_{T} (B) \, \rho ).  
$$
In quantum theory, only rarely would $ \rho $ be  
omitted from the notation on the left side. 
Probability theory in quantum theory exists 
even prior to choosing a state $ \rho $, since 
a self-adjoint operator $ T $ has its unique associated
pvm $ P_{T} $. 
This has properties similar to those of 
a probability measure, except that 
it takes values that are quantum events. 
So a pvm has a {\em codomain} consisting of quantum events. 
On the other hand a classical probability measure  
has classical events in its {\em domain}. 
The integral 
$ \int_{\mathbb{R}} d P_{T} \, (\lambda) \, f (\lambda) $ 
exists in $\mathcal{L} ( {\mathcal{H}} )$
for a wide class of Borel measurable functions 
$ f : \mathbb{R} \to \mathbb{C} $, including all 
bounded functions. 
The `expected value' of this pvm gives
$$
   T = \int_{\mathbb{R}} d P_{T} \, (\lambda) \, \lambda
$$
by the spectral theorem. 
Maybe you did not expect this result. 
Actually, 
$$
    f (T) := 
    \int_{\mathbb{R}} d P_{T} \, (\lambda) \, f (\lambda) 
$$
defines a functional calculus for all bounded, 
Borel functions 
$ f : \mathbb{R} \to \mathbb{C} $. 

Another curious point is that a classical probability 
measure $ P $ satisfies $ 0 \le P (A) \le 1 $ for 
every event $ A $. 
This is an inequality of real numbers. 
On the other hand, a pvm $ P $ on $\mathbb{R}$ satisfies
$ 0 \le P(B) \le I $ for every Borel subset of 
$\mathbb{R}$.
This is an inequality of self-adjoint operators.  
So, the linearly ordered interval of real numbers 
$ [0,1] $ for 
probabilities in the classical case is 
replaced by the lattice of projections in the
`interval' $ [0, I] $ of self-adjoint operators. 
An even more curious point is that the interval 
$ [0,1] $ only contains real numbers, while the 
`interval' $ [0,I] $ contains self-adjoint operators
that are not projections. 
This opens the door to considering positive operator
valued measures, which will be discussed later. 

Yet another way of relating quantum probability to 
classical probability is by restricting a pvm to 
a sub-lattice $ \mathcal{E}^{\prime} $ 
of the lattice events of $ \mathcal{E} $
such that $ \mathcal{E}^{\prime} $ is a $ \sigma $-algebra. 
Then one can put any classical probability measure 
whatsoever on $ \mathcal{E}^{\prime} $. 
This probability measure need not be the 
restriction of a spectral measure associated to a pvm
defined $ \mathcal{E} $, in which case one is 
considering a structure unrelated to quantum theory. 
However, if one starts with a spectral measure on 
$ \mathcal{E} $, one can restrict it to many such 
$ \sigma $-algebra sub-lattices in order to `view' 
the pvm in a variety of classical ways. 
This could be what some would call complementarity, 
though such a specific description is not usually 
given. 
Conversely, one could have a classical probability 
measure on $ \mathcal{E}^{\prime} $ and ask whether 
this is the restriction of a spectral measure 
on $ \mathcal{E}$. 
And if it is, then whether that spectral measure is unique.
This is close to the setting of the Kadison-Singer 
conjecture (see \cite{kadison}), 
which is now a proved theorem (see \cite{marcus}). 
In that context one has a {\em maximal 
commutative} sub-$ * $-algebra
$ \mathcal{A} $ of a $ C^{*} $-algebra $ \mathcal{C} $
and a state 
$ \phi : \mathcal{A} \to \mathbb{C}$. 
Then the theorem says that there exists 
a unique extension 
$ \tilde{\phi} : \mathcal{C} \to \mathbb{C}$
of $ \phi $ which is a state. 
Colloquially, under these hypotheses 
one commutative `snapshot' of a state suffices 
to characterize it. 

Another notable 
difference is that the probability of a sequence of
events in quantum probability does not reduce in general 
to the probability of a single event, as happens in
classical probability theory. 
This entails a separate definition of quantum probability 
for sequences of events. 
The properties of there multi-event probabilities 
include new features absent in classical probability 
theory, such as interference terms and dependence 
on the order of events. 

Also notice that classical probability measures 
on $ \mathbb{R} $ arise naturally 
in quantum probability, but 
classical probability does not involve quantum 
probability. 
And finally in classical probability theory, 
there is no basic time evolution equation, 
although time dependent 
stochastic processes are a part of that theory. 
However, quantum probability is intrinsically 
a part of quantum theory, which has time evolution
as a major facet of the theory. 
In fact, the generalized Born's rule 
\eqref{generalized-borns-rule} is the 
fundamental time evolution equation 
of quantum theory.

\section{\bf Expected Value}
\markboth{Expected Value}{Expected Value}

Expected value is a mostly unremarkable, quite 
secondary aspect 
of quantum probability. 
However, many times I have heard colleagues speak 
about it incorrectly.  
Unfortunately, this misunderstanding can be found
in print, too. 
So I think that it is necessary 
to give a clarification of this topic. 

First, let's repeat what quantum theory 
says about probability, namely Born's rule 
for $ \psi $ a unit vector, $ A = A^{*} $ 
and $ B $ a Borel subset of $ \mathbb{R} $, which 
with a bit of new notation is 
$$
\mu (B) :=  P ( A \in B \,|\, \psi) =  
  \langle \psi, P_{A} (B) \psi \rangle. 
$$
As a function of the Borel subset $ B $ of $ \mathbb{R} $ 
with both $ A  $ and $ \psi $ fixed, 
$ \mu $ is a probability measure 
on the real line $ \mathbb{R} $. 
We have seen this basic point already many times. 
One can now apply concepts from standard probability 
theory to this probability measure $ \mu $. 
And this is indeed done. 
For example, one can consider the {\em moments} of 
any probability measure. 
So, for every integer $ k \ge 1 $ we define the 
{\em $ k $th moment} of $ \mu $ as
$$
  m_{k} := \int_{\mathbb{R}} d \mu (\lambda) \, \lambda^{k}, 
$$
provided that this integral converges absolutely. 
The {\em expected value} of $ \mu $ is defined to 
be $ m_{1} $, the first moment, provided again that 
the integral converges absolutely. 
The idea is that $ m_{1} $ is a {\em statistical estimator} 
given by probability theory 
of the empirically observed {\em sample average} 
$$
 \overline{m}:= 
 \dfrac{\lambda_{1} + \cdots + \lambda_{n}}{n}
$$
of $ n \ge 1 $ measured values (or {\em sample}) 
$ \lambda_{1} , \dots , \lambda_{n} $. 
The values measured are typically not all the same 
and, indeed, this is what motivates one to 
turn to probability 
theory in order to understand them. 
It is quite common that the expected value is not 
going to be equal to any of the measured values, since 
this is a common property of the sample average. 
For example, the sample average of the number of children
per family (in a sample of families in a city, say)
could be 2.3, which is not an integer. 

As a brief aside, let me note that 
one can continue by defining the {\em central moments} for 
every integer $ k \ge 2 $  by
\begin{equation}
\label{define-sigma-k}
\sigma_{k} :=  
\int_{\mathbb{R}} d \mu(\lambda)\,(\lambda - m_{1})^{k}
\end{equation}
provided that the 
expected value $ m_{1} $ exists and that the integral 
in \eqref{define-sigma-k}  
also is absolutely convergent. 
A frequently used central moment is
$ \sigma_{2} $, which is also called the {\em variance}.
And the {\em standard deviation} of $ \mu $ 
is defined by $ \sigma := \sqrt{\sigma_{2}} $. 
Standard deviations (or equivalently variances)
enter into the Roberts inequality that expresses 
mathematically the Heisenberg Uncertainty Principle. 

It is well known that the moments (or equivalently
the central moments) 
do not uniquely determine $ \mu $ in all cases. 
Even in the 
most favorable case when these moments exist 
for all $ k $ this 
moment problem might not have a solution $ \mu $ and, 
if it does, that solution might not be unique. 
Colloquially, one can say that the moments carry some information about the probability measure $ \mu $, 
but not in general all the possible information. 

We have enough context now for 
discussing a common misunderstanding. 
The false assertion is often made that the 
only information that quantum theory 
provides about a system is the expected 
value of observables in a given state. 
Actually, Born's rule provides the 
probability measure for any observable 
in a given state, which is a lot more 
than just the first moment of that 
probability measure. 
This aspect of quantum theory dates back 
at least to 1932 in the seminal book 
\cite{von-neumann} of von Neumann 
and so should be well known in the physics 
community. 
My experience is that it is known by some, 
but not by others. 
An example of this confusion is the `proof' that 
Schr\"odinger's model is equivalent to the 
Heisenberg model by showing that the expected value
is the same in both models. 
While this is a necessary condition, by itself it 
is not sufficient. 

To complete this section here is the derivation 
of the usual formula for calculating the expected value 
of $ A = A^{*} $ given a pure state 
$ \psi $ in the dense domain of $ A $. 
$$
m_{1} = 
\int_{\mathbb{R}} 
\langle \psi, dP_{A} (\lambda) \, \psi \rangle 
\, \lambda 
= 
\big\langle 
   \psi, 
   \big( 
   \!\int_{\mathbb{R}} 
      dP_{A} (\lambda) \, \lambda \, 
   \big)  
   \, \psi 
\big\rangle 
=
\langle 
\psi, A \psi \rangle .
$$

\section{\bf Dynamics: The Generalized Born's Rule
	\\
	The Final Version}
\markboth{Dynamics}{Dynamics}

The {\em dynamics}, also known as the {\em time evolution}, 
of a physical system is given in quantum theory 
by a further generalization of the 
time dependent generalized Born's rule 
\eqref{generalized-borns-rule}. 
Contrary to the confused opinion of many
authors (including me in \cite{path}), there
is only one fundamental time evolution equation
in quantum theory. 
It is the same equation in all models. 
It is the only equation which we subject to
experimental verification. 
But \eqref{generalized-borns-rule} is still 
not adequate for all purposes, since it assumes that
there is no time evolution of the quantum system 
between events. 
So now we suppose that a system is given with 
state $ \psi $ at time $ t_{0} $ and that 
we are interested in the events $ E_{1}, \dots , E_{n} $
at the times $ t_{1}, t_{2}, \dots , t_{n} $, where
$ t_{0} < t_{1} < \cdots < t_{n} $.
The probability of this sequence of events at these times is 
$$
P ( E_{1}, \dots , E_{n}, t_{0}, t_{1}, \dots , t_{n} 
| \psi ) \!=\! || E_{n} U(t_{n}, t_{n-1}) 
\cdots 
E_{2} U (t_{2}, t_{1}) E_{1} 
U (t_{1}, t_{0}) \psi ||^{2},
$$
where $ U (s , t) $ is the time evolution operator of 
the system for the times $ s < t  $.  
In the Schr\"odinger model 
$ U (s , t) = \exp ( -i (t-s) H / \hbar ) $, where 
$ H $ is the Hamiltonian of the quantum system. 
I imagine that this formula is obvious to those who
think in the Schr\"odinger model. 
But whether this formula is obvious or not is 
an independent consideration.  
It is an axiom. 
It is the final version of the 
{\em Generalized Born's Rule}. 

\vskip 0.2cm \noindent
{\bf Axiom 5 Updated:} 
Let $ \rho $ be a density matrix. 
Suppose that $ E_{1}, \dots , E_{n} $ is a 
sequence of events and that 
$ t_{0} < t_{1} < \cdots < t_{n} $ is a sequence 
of times. 
Suppose that we have  {\em time evolution operators} 
 $ U (s , t) \in \mathcal{L} ( \mathcal{H} )$ 
 for all $ s < t $. 
We define the {\em consecutive probability 
	with time evolution} 
$ P ( E_{1}, \dots , E_{n}, t_{0}, t_{1}, \dots , t_{n} 
| \rho ) $ 
to be
\begin{equation}
\label{dynamics}
   Tr ( E_{n} U(t_{n}, t_{n-1}) 
   \cdots 
   E_{2} U (t_{2}, t_{1}) E_{1} 
   U (t_{1}, t_{0}) \rho ). 
   \qquad 
\end{equation}
The {\em conditional probability with time evolution} 
reads much like 
\eqref{generalized-borns-rule} with the time 
evolution operators appropriately interspersed 
between the event operators. 
The exact, general formula becomes unwieldy to write down. 
$ \blacksquare $

\vskip 0.2cm 
This axiom holds in all models if we extend the 
group $ E_{t} $ to act on all bounded operators and so,
in particular, on the operators $ U (s , t) $. 
In the standard models (Schr\"odinger, Heisenberg, 
interaction) the action of $ E_{t} $ is via 
conjugation by operators in a unitary group. 
This conjugation acts on all of 
$ \mathcal{L} ( \mathcal{H} ) $ as well. 
As we have come to expect, the value of the probability
\eqref{dynamics} is invariant under isomorphisms of models. 
It might appear strange that the time evolution 
operators depend on the model, rather than being the same 
in all models as in the familiar Schr\"odinger model. 
But this is exactly one of the things that happens 
in the interaction model, which also should be familiar 
for the reader. 
Recall that in the interaction model one writes the
Schr\"odinger Hamiltonian as $ H = H_{free} + H_{int} $, 
the sum of a `free' term and of an `interacting' term. 
Then the operators in the Schr\"odinger model, 
including the time evolution operators, are transformed to 
operators in the interaction model by conjugating them 
with $ \exp ( -i t H_{free}/\hbar) $. 
The splitting of $ H $ into two terms is chosen to facilitate 
subsequent calculations, not to change the results. 
And that is exactly the point of having different 
isomorphic models of quantum theory. 

Again, the manner for computing this time 
dependent probability 
does depend on the model, which 
leads to much confusion. 
In the Schr\"odinger model, which is the most 
familiar and most widely used model, all the events are
time independent and only the state could possibly change 
with time. 
As is well known 
the time evolution of the state in this model 
is given by a family of equations known collectively 
as Schr\"odinger's equation, which all have the 
same form
$ i \hbar \, \psi^{\prime} (t) = H \psi (t)$,
where $ H $ is a self-adjoint 
operator, known as the Hamiltonian, 
acting in the Hilbert space. 
The multitude of physical systems covered by this
approach is due to the fact that physicists are very 
adept at finding the appropriate 
Hamiltonian for many systems. 
However, the time dependent state which solves 
Schr\"odinger's equation has no 
physical significance. 
It is an artifact of the model and nothing else. 
But it is one of the ingredients that 
for the case $ \rho = | \psi \rangle \langle \psi | $ 
in \eqref{dynamics} goes into 
computing the time dependent probability, which indeed 
does have physical significance. 
In other models one must calculate the time 
dependence of the elements in 
\eqref{generalized-borns-rule} using
other auxiliary equations. 
But again these auxiliary 
 time dependent elements do not have 
any physical significance. 

It will surely be taken as heresy on my part to say 
that Schr\"odinger's equation is without physical 
significance. 
However, 
if that is the fate of its solutions, then that must 
be the fate of Schr\"odinger's equation itself. 
It is a stepping stone, a useful tool. 
This idea 
flies in the face of long-standing traditions 
in physics, especially those that favor 
differential equations as the most fundamental 
elements of a physical theory. 
There is an expectation that the time evolution 
of a physical system should be expressed as a 
differential equation involving time as one of
the variables. 
But \eqref{dynamics} does not have that form. 
Taking the derivative of \eqref{dynamics} in order 
to find a differential equation that it must solve 
leads to a relation of the time derivative of 
the probability with the time evolutions of the events
and of the state. 
But both of those time evolutions are model 
dependent. 

This preference for differential equations 
manifests itself in the way 
the equation of motion is written in the 
Heisenberg model. 
Typically this is presented as an ill-defined 
differential equation whose so-called `solution' 
is then given. 
It is that `solution' \eqref{Heisenberg-eqn} 
which is the actual 
time evolution equation, in spite of the 
fact that it is not a differential equation.  

Another apparent problem in taking the generalized Born's 
rule \eqref{dynamics} 
as the fundamental time evolution equation of
quantum theory is that this ignores the history 
of its discovery. 
The physics community is fascinated with 
the story of how these ideas emerged and who 
gets the credit for each of them. 
And I am complicit in this tradition to the extent that 
I do use the names of scientists when discussing 
their discoveries. 
At one point while writing this treatise
I thought of removing `Lie' from Lie group  
to give one example of how one might eliminate history 
from this narrative.   
Born's rule in the traditional 
narrative is seen as a later 
add-on, an embellishment of ideas that already
`worked' but somehow seemed incomplete. 
And besides that there already was a known equation, 
namely Schr\"odinger's equation, 
that was recognized by one and all
as the time evolution {\em differential}
 equation of quantum theory.  
And this history then continues with Born's rule being 
severely criticized by some and rejected by others. 
This paragraph is one of my few excursions into the 
history of quantum theory, and the point behind it
is that the historical sequence of discoveries is not  
the logical structure of quantum theory. 
The generalized Born's rule \eqref{dynamics} 
is the fundamental time evolution equation of 
quantum theory. 

If you think that my purpose here is to remove 
Schr\"odinger's equation from its central 
position in quantum theory, 
then you are reading correctly. 
For example, on Wikipedia the various topics on 
quantum theory (which there is called quantum mechanics)
are presented in a box with 
Schr\"odinger's equation at its head. 
This should be replaced with the generalized 
Born's rule according to my thesis. 

It is not well understood that one is actually only 
using the conditional probability 
\eqref{generalized-borns-rule} 
when analyzing many physical 
phenomena. 
Let's recall how that works in the case of two events. 
Given an event $ E_{1} $ and a unit vector $ \psi $, 
the conditional probability of a 
subsequent event $ E_{2} $ is 
$$
P ( E_{2} \,|\, E_{1}, \psi ) =
\dfrac{|| E_{2} E_{1} \psi  ||^{2}}{|| E_{1} \psi ||^{2}} 
= || E_{2} \tilde{\psi} ||^{2} = 
P ( E_{2} \,|\,  \tilde{\psi} ), 
$$
where $  \tilde{\psi} := E_{1} \psi / || E_{1} \psi || $ 
is the `collapse' of the original state $ \psi $. 
Of course, this is just reading backwards 
the motivating argument for the definition of 
conditional probability. 
The point now is that the right side of this 
equation can always be translated into the 
conditional probability on the left side. 
Another common way of speaking of this situation 
is that the first event $ E_{1} $ `prepares' a  
state $  \tilde{\psi} $, which the event 
$ E_{2} $ then `measures'. 
However, it all reduces to quantum conditional 
probability of quantum events. 

Another curious aspect of quantum 
theory is that the one-event Born's rule 
in Axiom~4 is rarely ever 
used to understand observations. 
A careful analysis almost always reveals that 
one is actually considering a conditional probability 
of two events.

\section{\bf Quantum Information}
\markboth{Quantum Information}{Quantum Information}

As with any other aspect of basic quantum theory, 
information must be defined in a way that is 
invariant under isomorphisms of models. 
Since the only structure left invariant is 
the Generalized Born's Rule, information must 
be defined in terms of it. 
Most studies of quantum information are done in
the setting of a tensor product of a finite 
number of finite dimensional Hilbert spaces, 
where more tools such as partial trace are 
available. 
But that restriction will not be made here. 
Following the usual convention, I will formulate
this topic in terms of {\em entropy}, 
which is the negative of information. 

Suppose that $ T = T^{*} $ is a self-adjoint operator 
and that $ \rho $ is a density matrix. 
Then, as we have seen, 
for $ B $ a Borel subset of $ \mathbb{R} $  
the map
$$
   B \mapsto P ( T \in B \,|\, \rho ) = 
   Tr ( P_{T} (B) \, \rho)  
$$ 
is a probability measure on $ \mathbb{R} $.
We suppose that this measure is absolutely continuous 
with respect to Lebesgue measure. 
Let $ f_{T,\rho} : \mathbb{R} \to [0,\infty) $ 
be a density function for $ P ( T \in B \,|\, \rho ) $, 
that is
$$
   P ( T \in B \,|\, \rho ) = 
   \int_{B} d \lambda \, f_{T,\rho} (\lambda). 
$$
This condition uniquely determines $ f_{T,\rho} $ 
except on a subset of $ \mathbb{R} $ of 
Lebesgue measure zero. 
One can make $ f_{T,\rho} $ unique by 
requiring that it satisfy the 
{\em cadlag condition}. 

This is a model independent property in the sense 
that if $ T^{\prime} $ and $ \rho^{\prime} $ 
correspond respectively in an isomorphic model to 
$ T $ and $ \rho $, then 
$  f_{T^{\prime},\rho^{\prime}} =  f_{T,\rho} $ 
almost everywhere with respect to Lebesgue measure. 
Then we define the {\em entropy of $ T $ given $ \rho $} as 
$$
    H ( T \,|\, \rho ) := 
    - \int_{\mathbb{R}} d \lambda \, f_{T,\rho} (\lambda) 
    \log ( f_{T,\rho} (\lambda)  )
$$
where the usual definition $ 0 \log 0 := 0 $ 
is being used. 
The symbol $ \log $ can mean the logarithm to the 
base $ e $ or any  other base $ b > 1 $.  
Once a base is chosen, 
that fixes a normalization condition 
on the entropy. 
Continuing to use the notation established above, 
we have 
$H ( T \,|\, \rho ) = H ( T^{\prime} \,|\, \rho^{\prime} )$. 

Similarly, the {\em conditional entropy of $ T $ 
given an event $ S \in B $ and a density matrix 
$ \rho $} can be defined if 
$ P ( T \in C \,|\, S \in B , \rho ) $ is 
absolutely continuous with respect to 
Lebesgue measure on $ \mathbb{R} $ 
with density function 
$ f_{S,B,\rho} : \mathbb{R} \to [0,\infty)$ by
$$
 H ( T \,|\, S \in B, \rho ) := 
 - \int_{\mathbb{R}} d \lambda \, f_{S, B, \rho} (\lambda) 
 \log ( f_{S, B, \rho} (\lambda)  ). 
$$

Since there is no joint probability measure 
on $ \mathbb{R}^{2} $ in 
general for a pair of self-adjoint operators 
$ S $ and $ T $, it seems to not be possible to
define in general a conditional probability of $ T $, 
given $ S $ and $ \rho $. 
Of course, if $ S $ and $ T $ commute, then 
this can be done. 
In general the probabilities 
$ P (  S \in B,  T \in C ) $ and 
$ P (  S \in B,  T \in C \,|\, \rho) $,
where $ B,C $ are Borel subsets of $ \mathbb{R} $ 
and $ \rho $ is a density matrix, 
give numbers that do carry 
information about the {\em ordered} pair $ S,T $.

\section{\bf Is Standard Quantum Theory \\
Being Changed?}
\markboth{Is Standard Quantum Theory 
Being Changed?}{Is Standard Quantum Theory 
Being Changed?}

No, it certainly is not. 
All of the calculations and predictions 
of probabilities given by standard 
quantum theory remain the same, provided that 
one uses the Schr\"odinger model or any 
model isomorphic to it. 
In the Schr\"odinger model one has available 
the standard mathemtical methods 
just as before, namely 
Schr\"odinger's equation and the collapse of
the state. 
What is different is that it would be better 
to speak about 
these methods in a new way for 
what they are: mathematical methods and 
not physical processes. 

In my introductory text 
\cite{path} I used the expression
{\em Schr\"odinger's razor} to describe the 
desire by Schr\"odinger to find a 
partial differential equation which would
describe quantum phenomena just via its solutions
without any extra,  
auxiliary conditions. 
For a beginning student I think 
it is appropriate to present 
Schr\"odinger's equation as that 
differential equation. 
In the Hilbert space context 
there is then an 
elegant existence and uniqueness 
theorem of solutions 
for self-adjoint Hamiltonians. 
And this is rather gratifying both physically and
mathematically. 
But extra rules must be given, 
namely Born's rule in order to compute 
one-event probabilities 
and collapse of the
state in order to compute conditional 
probabilities. 
While this spoils Schr\"odinger's 
original intention, it is essential in 
making the 
Schr\"odinger model 
work and in also making it 
isomorphic to 
the Heisenberg model.
It turns out 
Schr\"odinger's equation comes 
very close
to fulfilling the desire of having a 
fundamental quantum time evolution 
equation that is also a differential equation.
The utility of 
Schr\"odinger's equation as a practical 
tool remains unchanged, and 
this is why the philosophy of ``shut up 
and compute'' works quite well. 
Still, the fundamental time evolution 
equation of quantum theory,
the Generalized Born's Rule, is not 
a differential equation. 
And that is not a bad thing. 

While I might be criticized for throwing 
differential equations to the wolves, 
I wish to emphasize their important, 
continuing role in science and, 
in particular, in physics. 
But we all have become so accustomed to them 
that we, just like Schr\"odinger, find it 
difficult to imagine a viable alternative. 

Yet, Maxwell's equations are instructive 
in this regard. 
These are usually presented as 
differential equations, although I first
learned about them as integral equations. 
For me 
the transition to the differential formulation  
was neither intuitive nor motivated. 
Besides those who have to teach what 
the differential Maxwell equations `mean' often
translate them into their integral form as a 
pedagogical aid. 
After all, how else can a student understand 
what $ \nabla \cdot \mathbf{B} = 0 $ is saying? 
Or $ \nabla \cdot \mathbf{E} = 4 \pi \rho $? 
Or what their difference is all about?  

In conclusion 
I would prefer to avoid speaking 
at all of the 
collapse of the state and 
to replace that language with 
references to quantum 
conditional probability. 
However, there is no harm done if one wishes to 
continue to 
speak of collapse as long as one realizes that
this is nothing more than code language 
for a mathematical operation used 
as a step in 
computing quantum conditional probability. 

One should be cognizant that 
a mathematical concept need not have any
physical meaning or `interpretation'. 
As an example in the context of Hilbert
space theory, the equation 
$ \psi + \psi = 2 \, \psi $ makes mathematical 
sense and is correct. 
Yet, as far as I am aware, 
it has no meaning in quantum theory. 
Another basic operation in Hilbert space theory 
is the inner product.
Again, this has no physical `interpretation'. 
The collapse of the state and, sadly, 
Schr\"odinger's equation  do not have either one in and of 
themselves any physical meaning. 
However, by combining them they do allow us to
calculate time dependent probabilities using the
Generalized Born's Rule, which does have 
a physical meaning. 
And in this sense standard quantum theory 
remains unchanged.

\chapter{Entanglement}
\markboth{Entanglement}{Entanglement}
\label{entanglement-chapter}
\label{Entanglement-Chapter}

\hfill 
I would not call that property\footnote{The property 
referred to is entanglement. See Ref~\cite{schrodinger2}.}
 {\bf one} 
but rather 

\hfill 
{\bf the} characteristic trait of quantum mechanics,

\hfill  
the one that enforces its entire departure from

\hfill 
classical 
lines of thought. 
(emphasis in original).

\vskip 0.2cm 

\hfill Erwin Schr\"odinger

\vskip 0.4 cm 
Entanglement has been one of the most puzzling topics 
in quantum theory. 
This is largely due to a detailed discussion about 
two events that presumably `know' nothing of 
each other, but nonetheless show an uncanny 
deterministic relation between them.
One wonders how a probabilistic theory could 
ever explain this determinism. 
Worse yet, these two events can be space-like with 
respect to each other and so relativity theory 
forbids information from passing from 
one to the other. 
Yet in some strange form this seems to be what 
is happening. 
There are many experiments that agree with 
entanglement, 
including some carefully constructed experiments where 
the two events are indeed space-like. 
And no experiment, as far as I am aware, 
contradicts entanglement. 
How can we deal with this conundrum? 
Must quantum theory be modified or abandoned? 
Is there some spooky action at a distance that 
accounts for entanglement? 
Is physics non-local? 
How can the collapse of a state 
by one detector possibly affect the other 
detector? 

Entanglement can be understood using just 
basic quantum theory:~events, states and 
probability. 
The point is that in any probability theory 
some of the calculated probabilities will 
turn out to be $ 1 $. 
Experimental confirmation of entanglement should not  
produce existential {\em angst} , much less to 
appeals for new scientific principles to 
`explain' the `mystery' of this confirmation. 
So what is going on here in terms of 
basic quantum theory? 
Quite simply put, entanglement always involves 
{\em three} events, never just two. 
(Or possibly more than three!) 
There always is an event that precedes in time the 
two other events. 
Of course, those two later events `know' about that
earlier event. 
When the calculated conditional quantum 
probability of these three (or more) events 
shows that they are not independent, 
this lack of 
independence is said to be {\em entanglement} 
of the later events. 
And this agrees with experiment. 

Here is some notation for 
a simple example of entanglement. 
We have an event $ E_{0} $, which we will call the 
initial event,  
and then 
two events $ E_{1} $ and $ E_{2} $, each of which occurs
after $ E_{0} $. 
The time relation between $ E_{1} $ and $ E_{2} $ 
is irrelevant to this analysis.
This example captures the essence of the 
EPR paper and is discussed further in 
Chapter~\ref{EPR-chapter}. 
We could assume that $ E_{1} $ and $ E_{2} $ commute, 
but that assumption is not used 
until the end of this chapter. 
Notice that space-like separated events {\em must} commute
according to special relativity, 
while {\em some} time-like (or light-like) 
separated events 
{\em can} commute. 

We start by taking 
$ \mathcal{H} = \mathbb{C}^{2} \otimes \mathbb{C}^{2}$. 
We let $ S_{1} $, $ S_{2} $ and $ S_{3} $ denote the 
standard $ 2 \times 2 $ spin matrices. 
In the sequel we will only use 
$$ 
S_{3} = \dfrac{1}{2}
\left( 
\begin{array}{cc}
1 & 0 
\\
0 & -1 
\end{array}
\right) 
$$
for convenience in doing calculations. 
Similar results hold for the other spin matrices. 
Then we next consider these self-adjoint 
operators acting on $ \mathcal{H} $:
$$
    S^{2}, 
    \qquad
    S_{r} := S_{3} \otimes I, \qquad 
    S_{l} := I \otimes S_{3},
$$
where
$$
 S^{2} = ( S_{1} \otimes I + I \otimes S_{1} )^{2}
+ ( S_{2} \otimes I + I \otimes S_{2} )^{2}
+ ( S_{3} \otimes I + I \otimes S_{3} )^{2}
$$ 
is the standard total spin operator. 
The three events to be considered are defined as 
$$
    E_{0}:= P_{ S^{2} }  ( \{ 0 \} ), \qquad
    E_{1}:= P_{  S_{r}} (\{ 1/2 \}), \qquad
    E_{2}:= P_{  S_{l}} (\{ -1/2 \}).
$$

The way one verbalizes this can be misleading, 
but I will venture to do it anyway. 
The event $ E_{0} $ says that initially the system 
has total spin $ 0 $. 
The event $ E_{1} $ says that a detector to the right
of the place where $ E_{0} $ occurred measures 
the $ z $ component of the spin to be $ 1/2 $. 
The event $ E_{2} $ says that a detector to the left
of the place where $ E_{0} $ occurred measures  
the $ z $ component of the spin to be $ - 1/2 $. 
The words `right' and `left' are not really important.  
They merely give a visual rendition to the formalism. 

Notice that in this example the self-adjoint operators
$ S_{r} $ and $ S_{l} $ commute. 
Consequently, the events $ E_{1} $ and $ E_{2} $ also
commute. 

Given these definitions it is an exercise, 
which we now do, to compute the appropriate conditional 
probability for these $ 3 $ events. 
The reader 
should be aware that   
the intermediate steps in this calculation are 
without physical significance. 
The only physically significant part of this 
calculation is the last step where the 
value of the conditional probability 
is given. 

Let $ \varepsilon_{1}, \varepsilon_{2} $ be the 
standard orthonormal basis of $ \mathbb{C}^{2} $, namely
\begin{equation*}
\varepsilon_{1} = 
\left( 
\begin{array}{c}
1 \\ 0
\end{array}
\right) 
\qquad \mathrm{and} \qquad 
\varepsilon_{2} = 
\left( 
\begin{array}{c}
0 \\ 1
\end{array}
\right).  
\end{equation*}
We first identify the pertinent quantum events 
in terms of this basis using Dirac notation:
$$ 
E_{1} = P_{S_{r}} (1/2) = 
| \varepsilon_{1} \rangle \langle \varepsilon_{1} | 
\otimes I_{2} \quad 
\mathrm{and} \quad 
E_{2} = P_{S_{l}} (1/2) = I_{2} \otimes  
| \varepsilon_{2} \rangle \langle \varepsilon_{2} |,  
$$
where $ I_{2} $ is the identity operator acting
on $ \mathbb{C}^{2} $. 
Notice that these are rank~$ 2 $ projections. 
The operator $ S^{2} $ is not central in 
$ \mathcal{L} (\mathcal{H})$. 
It is well known that its spectrum is
$\{ 0 , 2 \} $. 
The eigenspace for the eigenvalue $ 2 $ has 
dimension $ 3 $ (a triplet for spin $ 1 $),
while the eigenspace for the eigenvalue $ 0 $ 
has dimension~$ 1 $ (a singlet for spin $ 0 $), 
and this latter eigenspace has an orthonormal basis
consisting of the one vector
\begin{equation}
\label{entangled-state}
   \psi_{0} := \dfrac{1}{\sqrt{2}} 
   \big( \varepsilon_{1} \otimes \varepsilon_{2} -
   \varepsilon_{2} \otimes \varepsilon_{1} \big). 
\end{equation}
So we have the quantum event 
$$
E_{0} = P_{S^{2}} (\{ 0 \})
= | \psi_{0} \rangle \langle \psi_{0} |,
$$
which is a projection with rank $ 1 $. 
We now compute the quantum conditional probability 
\begin{equation}
\label{compute-entangle}
P (E_{1} \,|\,  E_{2}, E_{0} , \psi ) = 
\dfrac{|| E_{1} E_{2} E_{0} \, \psi ||^{2}}
{|| E_{2} E_{0} \, \psi ||^{2}}
\end{equation}
provided that $ || E_{2} E_{0} \psi || \ne 0 $,
(When not stated otherwise,  
$ \psi \in \mathcal{H} $ is a unit vector.) 
Intuitively, 
the expression \eqref{compute-entangle}
gives the conditional 
probability that the detector on the right 
measures the $ z $-component of spin to be $ 1/2 $
given that the detector on the left measures 
the $ z $-component of spin to be $ -1/2 $ 
and that (previously) the initial state has been 
prepared to have spin~$ 0 $. 
These words in everyday language 
only serve to reassure the 
dubious that this formalism has its 
motivation. 
We first have for all  
$ \psi \in \mathcal{H} $ that 
$$  
E_{0} \, \psi = 
| \psi_{0} \rangle \langle \psi_{0} \, | \psi 
= \langle \psi_{0}, \psi \rangle \psi_{0}. 
$$
Continuing we have for all 
$ \psi \in \mathcal{H} $ that
\begin{align*}
E_{2} E_{0} \psi  &= 
(I_{2} \otimes 
| \varepsilon_{2} \rangle \langle \varepsilon_{2} | )
\langle \psi_{0}, \psi \rangle \, \psi_{0}
\\
&= 
2^{-1/2} \langle \psi_{0}, \psi \rangle 
(I_{2} \otimes 
| \varepsilon_{2} \rangle \langle \varepsilon_{2} | )
\big( \varepsilon_{1} \otimes \varepsilon_{2} -
\varepsilon_{2} \otimes \varepsilon_{1} \big)
\\
&=
 2^{-1/2} \langle \psi_{0}, \psi \rangle 
 \big( \varepsilon_{1} \otimes 
 | \varepsilon_{2} \rangle \langle \varepsilon_{2} |\varepsilon_{2} -
 \varepsilon_{2} \otimes | \varepsilon_{2} \rangle \langle \varepsilon_{2} |
 \varepsilon_{1} \big)
 \\\
&=  2^{-1/2} \langle \psi_{0}, \psi \rangle 
\big( \varepsilon_{1} \otimes \varepsilon_{2} \big)
\\
&=  2^{-1/2} 
| \varepsilon_{1} \otimes \varepsilon_{2} \rangle 
\langle \psi_{0} | \psi. 
\end{align*}
This is non-zero if 
$ \langle \psi_{0}, \psi \rangle \ne 0$, which we assume from now on. 
We conclude 
\begin{equation}
\label{E2-E0-cond-prob}
E_{2} E_{0} =  2^{-1/2} 
| \varepsilon_{1} \otimes \varepsilon_{2} \rangle 
\langle \psi_{0} |. 
\end{equation}
Clearly, $ E_{2} E_{0} $ is a non-zero, rank $ 1 $ 
operator that is not a projection. 
Finally, for all $ \psi \in \mathcal{H} $ 
we evaluate again that 
\begin{align*}
E_{1} E_{2} E_{0} \psi &= 
\big( | \varepsilon_{1} \rangle \langle \varepsilon_{1} | 
\otimes I_{2} \big)
 2^{-1/2} \langle \psi_{0}, \psi \rangle 
 \big( \varepsilon_{1} \otimes \varepsilon_{2} \big)
 \\
 &= 
 2^{-1/2} \langle \psi_{0}, \psi \rangle 
 \big( | \varepsilon_{1} \rangle \langle \varepsilon_{1} | 
 \otimes I_{2} \big)
 \big( \varepsilon_{1} \otimes \varepsilon_{2} \big)
 \\
 &= 
 2^{-1/2} \langle \psi_{0}, \psi \rangle 
 \big( \varepsilon_{1} \otimes \varepsilon_{2} \big)
\\
 &=
  2^{-1/2} 
  | \varepsilon_{1} \otimes \varepsilon_{2} \rangle 
  \langle \psi_{0} | \psi 
\\
&= E_{2} E_{0} \psi. 
\end{align*}
This implies that 
we have two equal {\em operators} 
(not projections)
\begin{equation}
\label{two-equal-operators}
 E_{1} E_{2} E_{0} =  E_{2} E_{0}.
\end{equation}
This is more than we need to conclude that the 
conditional probability 
\eqref{compute-entangle} is equal to $ 1 $, 
namely
$$
P (E_{1} \,|\,  E_{2}, E_{0} , \psi ) = 
\dfrac{|| E_{1} E_{2} E_{0} \, \psi ||^{2}}
{|| E_{2} E_{0} \, \psi ||^{2}} = 1. 
$$
The equality \eqref{two-equal-operators} 
of these two operators is an intermediate 
step which has no physical significance. 
For example, the operator $  E_{2} E_{0} $ is {\em not}
an event since $ E_{0}  $ and $ E_{2} $ 
do not commute. 
Note that $ E_{0}  $ and $ E_{1} $ also  
do not commute. 
This lack of commutativity is good news. 
It means that we are dealing with a truly 
quantum situation. 

Similarly, one can calculate that 
$$
P (E_{2} \,|\,  E_{1}, E_{0} , \psi ) = 
\dfrac{|| E_{2} E_{1} E_{0} \, \psi ||^{2}}
{|| E_{1} E_{0} \, \psi ||^{2}} =1 
$$
provided that $ || E_{1} E_{0} \, \psi || \ne 0 $.  
The details are left to the reader. 
Using this straightforward method for
$ 3 $ events with the same 
time order, one can evaluate the conditional 
probability for other combinations of components of spin. 
Typically those probabilities will lie strictly 
between $ 0 $ and $ 1 $, but in the specific example
given here a value of $ 1 $ is the result. 
The fact that a certain probability is $ 1 $ 
does not invalidate the use of probability 
theory, but is merely a special case of it. 
In particular, probability $ 1 $ does not
imply that this is a deterministic situation. 
In other specific cases probability $ 0 $ will occur.
For example, we have 
$$
P (E_{1}^{c} \,|\,  E_{2}, E_{0} , \psi ) = 
\dfrac{|| (I - E_{1}) E_{2} E_{0} \, \psi ||^{2}}
{|| E_{2} E_{0} \, \psi ||^{2}} = 0, 
$$
where $ E_{1}^{c} = I - E_{1} $ is the event 
complementary to $ E_{1} $. 
One of the intermediate steps used to show this 
(using a formula above)
is $ (I - E_{1}) E_{2} E_{0}  =0 $.  

In any probabilistic theory, some of the calculated 
probabilities will turn out to be $ 1 $. 
I do not see anything `spooky' about that. 
Nor do I see anything `non-local' in this 
analysis. 
Nor do I see that an element of `reality' 
is required to `explain' a probability one situation. 
If you are seeking `reality' here, why not say that 
{\em all} probabilities are real? 
See Chapter~\ref{EPR-chapter} for more on this. 

Of course, we can discard the information about 
the event $ E_{2} $ and ask only for the conditional
probability of $ E_{1} $ given $ E_{0} $. 
Intuitively, this is the point of view of the 
experimenter on the right. 
So we want to compute
$$ 
P ( E_{1} \,|\, E_{0} , \psi ) =
\dfrac{|| E_{1} E_{0} \, \psi ||^{2}}
{ ||  E_{0} \, \psi ||^{2} }. 
$$
We first note that we have  
$  E_{0} \, \psi = 
\langle \psi_{0} , \psi \rangle \, \psi_{0}$ 
for all $ \psi \in \mathcal{H} $ as before. 
Next, we calculate for all 
$ \psi \in \mathcal{H} $ that
\begin{align*}
 E_{1} E_{0} \psi &= 
 \big(| \varepsilon_{1} \rangle \langle \varepsilon_{1} | 
 \otimes I_{2} \big)
 \big( \langle \psi_{0} , \psi \rangle \psi_{0} \big)
 \\
 &=
 2^{-1/2} \langle \psi_{0} , \psi \rangle 
 \big( | \varepsilon_{1} \rangle \langle \varepsilon_{1} | 
 \otimes I_{2} \big)
 \big(  \varepsilon_{1} \otimes \varepsilon_{2} -
 \varepsilon_{2} \otimes \varepsilon_{1} \big)
 \\
 &=
 2^{-1/2} \langle \psi_{0} , \psi \rangle 
 \big( \varepsilon_{1} \otimes \varepsilon_{2} \big )
 \\
 &= 2^{-1/2} |  
 \varepsilon_{1} \otimes \varepsilon_{2}\rangle 
\langle \psi_{0} | \psi.  
\end{align*}
Consequently, 
$ E_{1} E_{0} = 2^{-1/2} |  
\varepsilon_{1} \otimes \varepsilon_{2}\rangle 
\langle \psi_{0} | $ and therefore 
$$ 
P ( E_{1} \,|\, E_{0} , \psi ) =
\dfrac{ || E_{1} E_{0} \, \psi ||^{2}}
{|| E_{0} \, \psi ||^{2}}
= \dfrac{||  2^{-1/2} 
| \varepsilon_{1} \otimes \varepsilon_{2}\rangle 
\langle \psi_{0} | \psi \, ||^{2}}
{|| \langle \psi_{0} , \psi \rangle \, \psi_{0} ||^{2}} 
= 1/2, 
$$
since $ || \, | \varepsilon_{1} \otimes \varepsilon_{2}\rangle 
\langle \psi_{0} | \psi \, || =
| \langle \psi_{0} , \psi \rangle | $. 
In other words, the axiom 
for calculating the conditional 
probability gives the expected result. 
Similarly, one can show that 
$ P ( E_{2} \,|\, E_{0} , \psi ) = 1/2 $. 

We can also compute the conditional probability that 
$ E_{1} $ and  $ E_{2} $ occur given $ E_{0} $. 
This makes sense if we assume that 
the events $ E_{1} $ and $ E_{2} $
commute. 
The conditional probability for this situation is
$$
  P  ( E_{1}, E_{2} \,|\, E_{0}, \psi) =
  \dfrac{|| E_{1} E_{2} E_{0} \, \psi ||^{2}}
  {||  E_{0} \, \psi ||^{2}} 
$$
provided $ ||  E_{0} \, \psi || \ne 0 $. 
Recall  
\eqref{E2-E0-cond-prob} and
\eqref{two-equal-operators}: 
$E_{1 }E_{2} E_{0} 
=
2^{-1/2} | \psi_{0} \rangle 
\langle \varepsilon_{1} \otimes \varepsilon_{2} |$. 
This equality of {\em operators},
not of events, has 
no physical significance, but does allow us to 
proceed to the next part of the calculation. 
So here it is: 
$$
P  ( E_{1}, E_{2} \,|\, E_{0} , \psi) =
\dfrac{|| E_{1} E_{2} E_{0} \, \psi ||^{2}}
{|| E_{0} \, \psi ||^{2}} = 
\dfrac{|| \, 2^{-1/2} | \psi_{0} \rangle 
\langle \varepsilon_{1} \otimes \varepsilon_{2} | \psi \, ||^{2}}
{|| \langle \psi_{0} , \psi \rangle \, \psi_{0} ||^{2}}
= 1/2. 
$$

So, we have calculated a 
different conditional probability, and we get a 
probability that is neither $ 0 $ nor $ 1 $. 
Notice that this result agrees with what little 
intuition we can muster concerning quantum theory. 
In possibly misleading ordinary language it says that
after measuring the system to have spin $ 0 $ the 
probability is $ 1/2 $ that the detector on 
the right measures the $ z $-component of spin 
to be $ +1/2 $ and the detector on the left  
measures the $ z $-component of spin 
to be $ -1/2 $. 

What about the rest of the probability? 
Simple, similar calculations which are 
left to the reader show that 
$P( E_{1}^{c}, E_{2}^{c} \,|\, E_{0}, \psi) = 1/2 $, 
where the {\em complementary event} of an event $ E $ 
is defined as $ E^{c} := I - E $. 
One also easily verify that
$
P  ( E_{1}^{c}, E_{2} \,|\, E_{0}, \psi ) = 0
$  
and 
$   
P  ( E_{1}, E_{2}^{c} \,|\, E_{0}, \psi ) = 0. 
$
All of these probabilistic statements can be expressed 
in ordinary language, and they all agree with 
commonly held `intuition'.

One mathematical fact behind this example is that
both events $ E_{1} $ and $ E_{2} $ are products of the form
$ F_{1} \otimes F_{2} $, where each factor is an 
event in $ \mathcal{L} ( \mathbb{C}^{2} ) $, 
but the event $ E_{0} $ is not of that form.
Here the factorizations are with respect to the
given definition 
$ \mathcal{H} = \mathbb{C}^{2} \otimes \mathbb{C}^{2} $. 
This fact is not invariant in general 
under a unitary transformation 
to another isomorphic Hilbert space with its own given 
factorization as a product of two Hilbert spaces 
of dimension~$ 2 $. 
Nonetheless, it does 
make sense to speak of 
entangled events.  
It does not make sense to speak of 
entangled particles. 
A good rule of thumb is that when someone speaks of 
{\em two} entangled particles, it is best to rephrase 
that as a clearer 
statement about {\em three} entangled events.
Entanglement is a property of events. 
Of course, you may object that 
I never defined entanglement. 

But first I summarize the situation in 
our example. 
Using \eqref{E-psi-psi-is-E-psi}, 
the fact that $ P ( E_{0}) =1 $
and previously calculated 
probabilities, we see that 
\begin{align*}
P ( E_{1} \,|\, E_{0} ) &=
P ( E_{1} \,|\, \psi_{0} ) = 1/2,
\\
P ( E_{2} \,|\, E_{0} ) &=
P ( E_{2} \,|\, \psi_{0} ) = 1/2, 
\\
P ( E_{1} E_{2} \,|\, E_{0} ) = 
P ( E_{2} E_{1} \,|\, E_{0} ) &=
P ( E_{1} E_{2} \,|\, \psi_{0} ) = 1/2. 
\end{align*}
This implies that the later,
commuting  events $ E_{1} $ and 
$ E_{2} $ are not independent with 
respect to the earlier event $ E_{0} $ 
or, equivalently, with respect to the 
`initial' state $ \psi_{0} $. 
One might be misled into thinking that 
commuting events should always 
be independent,
but that need not be so, since that 
depends on other prior circumstances. 
This motivates the following  definition. 
\begin{definition}
\label{define-entangle}
    Given three time ordered events 
    $ E_{0} , E_{1}, E_{2} $ with 
    $ E_{0} \ne 0 $ being the earliest, 
    we say that these events are 
    {\em tangled} if the later two 
    (but not necessarily commuting)
    events $ E_{1}, E_{2} $ 
    are not independent 
    in some order when their 
    probabilities are 
    conditioned on 
    the first event $ E_{0} $. 
    Equivalently, 
    \begin{align}
    either \quad 
    &\Delta (E_{1}, E_{2} \,|\, E_{0}):= 
    P ( E_{1} E_{2} \,|\, E_{0} ) 
    - P ( E_{1}  \,|\, E_{0} )
      P (  E_{2} \,|\, E_{0} ) \ne 0 
      \nonumber 
    \\
    or \quad  
    &\Delta (E_{2}, E_{1} \,|\, E_{0})=
    P ( E_{2} E_{1} \,|\, E_{0} ) 
    - P ( E_{1}  \,|\, E_{0} )
    P (  E_{2} \,|\, E_{0} ) \ne 0,  
    \label{degree-of-entanglement}
    \end{align}
    
    In general a
     sequence of time ordered events is 
	{\em entangled} if some 
	subset of later events is not independent with respect to 
	their probabilities conditioned 
	on some of the earlier events.
	The noun describing such entangled events
	is
	{\em entanglement}.
	
	The numbers $ \Delta (E_{1}, E_{2} \,|\, E_{0})$ and 
	$ \Delta (E_{2}, E_{1} \,|\, E_{0})$ 
	in the interval $ [ -1, 1] $ 
	are quantitative measures 
	of the {\em degree} of entanglement. 
\end{definition}

According to this definition, entanglement is 
a probabilistic property of quantum events, namely 
a certain lack of independence among them.  

Notice that in this definition the Hilbert space 
need not be represented as a tensor product. 
Typically, I would expect that under some 
reasonable hypotheses 
three entangled events are such 
that in some unitarily 
equivalent tensor product 
Hilbert space the two 
later events are factorizable 
as a tensor product of events, 
while the earliest one of them is not. 

Another mathematical fact of this example is that 
the first event $ E_{0} $ has rank $ 1 $. 
Some might object that this event only serves to `prepare'
the `real' initial state 
$ \psi_{0} = E_{0} \psi / || E_{0} \psi ||$, 
where $ \psi \in \mathcal{H} $ is some 
`pre-initial' state satisfying $ E_{0} \psi \ne 0$. 
You could think of it that way, but by doing so you
are missing the point. 
States do not change in this analysis. 
Everything follows from  
only events and their conditional probabilities. 

Of course, anyone familiar with entanglement 
realizes immediately 
that \eqref{entangled-state} is an 
{\em entangled state}.  
How does that terminology fit in with 
the analysis in terms of the conditional 
probabilities of three events? 
First of all, to say that 
\eqref{entangled-state} is an 
entangled state means by the standard 
definition that it can not be written as 
a factorized vector, namely as 
$ \phi_{1} \otimes \phi_{2} $ where 
$ \phi_{1} $ and $ \phi_{2} $ are unit vectors 
in $ \mathbb{C}^{2}$. 
(This is a standard exercise.) 
So the vector $ \psi_{0} $ is an entangled 
space not in and of itself, but only 
relative to a given tensor product 
representation of the Hilbert space of 
the system. 

So let's see how such a tensor product Hilbert 
space 
$\mathcal{H} = 
\mathcal{H}_{1} \otimes \mathcal{H}_{2} $ 
gives a special case of entanglement. 
As usual, we avoid trivial cases by 
requiring that $ \dim \mathcal{H}_{j} \ge 2 $
for $ j=1,2 $. 
We consider any unit vector 
$ \psi_{0} \in \mathcal{H} $, 
whether it is entangled 
or not, and from it construct the non-zero event
$ E_{0}:= | \psi_{0} \rangle \langle \psi_{0} | $. 
Next we use the tensor product structure 
to define events 
$ E_{1}:= F_{1} \otimes I_{2} $ 
and $ E_{2}:= I_{1} \otimes F_{2} $, where 
for $ j=1,2 $ we let $ F_{j} $ be any event 
in $ \mathcal{L}  (\mathcal{H}_{j}) $ and 
$ I_{j} $ denotes the identity operator 
acting on $ \mathcal{H}_{j} $. 
We require $ E_{0} $ to be the 
earliest event. 
The relative time order of $ E_{1} $
and $ E_{2} $ is irrelevant. 
Clearly, $ E_{1} $ and $ E_{2} $ are 
commuting non-zero  
events with dimension $ \ge 2 $. 
In particular neither $ E_{1} $ nor $ E_{2} $ 
is a state, even though $ F_{1} $ or $ F_{2} $ 
could be a state in its respective Hilbert space. 
Using \eqref{E-psi-psi-is-E-psi} and 
$ P ( E_{0} ) =1  $ we have for any $ E $
(which is an event or a product of events)
 that 
$$
P ( E \,|\, E_{0} ) = P ( E \, E_{0} ) =
P ( E \, | \psi_{0} \rangle \langle \psi_{0} | ) = 
P ( E \,|\, \psi_{0} ). 
$$
Consequently, the conditions in 
\eqref{degree-of-entanglement} become 
respectively 
 \begin{align*}
 \mathrm{either} \quad 
 &\Delta (E_{1}, E_{2} \,|\, \psi_{0}):= 
 P ( E_{1} E_{2} \,|\, \psi_{0} ) 
 - P ( E_{1}  \,|\, \psi_{0} )
 P (  E_{2} \,|\, \psi_{0} ) \ne 0 
 \\
 \mathrm{or} \quad  
 &\Delta (E_{2}, E_{1} \,|\, \psi_{0})=
 P ( E_{2} E_{1} \,|\, \psi_{0} ) 
 - P ( E_{1}  \,|\, \psi_{0} )
 P (  E_{2} \,|\, \psi_{0} ) \ne 0,  
 \end{align*}
If one of these two conditions holds, we say 
that $ E_{1}, E_{2} $ are {\em entangled} with respect to the {\em initial state} $ \psi_{0} $. 
With this notation we  
have the next result. 

\begin{prop} 
The following statements are equivalent. 
\begin{enumerate}
	
\item $ \psi_{0} $ is an entangled state. 

\item For some choice of events $ F_{1} $ and 
$ F_{2} $, 
the two events $ E_{1} $ and $ E_{2} $ 
are entangled with respect to $ \psi_{0} $. 

\item For some choice of events $ F_{1} $ and 
$ F_{2} $, 
the three events $ E_{0}, E_{1}, E_{2} $ 
are entangled. 

\end{enumerate} 
\end{prop}
\textbf{Proof:}
The second and third statements are equivalent 
by the remarks above. 
We continue with other cases. 

Not $  1 \implies $ Not $ 2 $: 
This was Example~\ref{example-independence}. 
 
$  1 \implies  2 $: 
Suppose that $ \psi_{0} $ is entangled. 
The Schmidt decomposition   
applies to Hilbert spaces 
of all dimensions, whether finite or infinite. 
(See~\cite{busch}.)
This gives us a  
non-unique, norm convergent 
representation of $ \psi_{0} $, 
$$
    \psi_{0} = \sum_{j=1}^{n} \lambda_{j} \, 
   \phi_{j}  \otimes  \psi_{j}   
   \in \mathcal{H}_{1} \otimes  \mathcal{H}_{2}, 
$$
where for each $ j $ we have 
$ \lambda_{j} > 0 $ and $ \phi_{j} $ 
(resp., $ \psi_{j} $) is an orthonormal 
{\em set}, but
not necessarily a {\em basis}, 
of $ \mathcal{H}_{1} $ 
(resp., $ \mathcal{H}_{2} $). 
The $ \lambda_{j} $'s need not be distinct. 
However, the  upper index 
$ n \in \mathbb{N}^{+} \cup \{ +\infty \} $ 
is unique. 
Also, $ \psi_{0} $ is entangled 
if and only if $ n \ge 2 $. 
Then we define 
$ F_{1}:=  | \phi_{1} \rangle \langle \phi_{1} |  $
and 
$ F_{2}:=  | \psi_{2} \rangle \langle \psi_{2} |  $.
It then is easy to calculate that
$ P ( E_{1} \,|\, \psi_{0} ) = 
\lambda_{1}^{2} \ne 0$
and
$ P ( E_{2} \,|\, \psi_{0} ) = 
\lambda_{2}^{2} \ne 0$.
And one also readily sees that 
$ P ( E_{1} E_{2} \,|\, \psi_{0} ) =0 $ 
and therefore for these choices of $ F_{1} $ 
and $ F_{2} $ we have that $ E_{1} $ and $ E_{2} $ 
are entangled with respect to $ \psi_{0} $. 
$ \quad \blacksquare $

\vskip 0.2cm 
The point of this proposition is that 
this standard entanglement situation {\em can} 
be understood in terms of nothing other 
than probabilities and events, but
I do not mean to say that one {\em must} do 
it this way. 
As in all other aspects, this treatise does not 
discard anything in standard quantum theory. 
Rather what I am simply trying to do is provide a 
language that is both inclusive and logical. 
The mathematics of the entanglement properties of 
quantum systems with three or more entities 
gets amazingly complicated. 
For example, see \cite{bengtsson}. 
There is no reason to stop using 
standard techniques for dealing with that. 
For example, partial trace is an available 
mathematical tool, which can enter 
the theory in Axiom~3. 
But\textit{} I conjecture that 
all studies of entangled states can be 
understood using only  
events and probabilities, since the 
entangled state itself is simply an initial 
event which conditions the probabilities 
of subsequent events.  
If that turns out to be incorrect, then 
substantive changes must be made to the 
axioms proposed in 
Chapter~\ref{axiom-chapter}. 
And that would be quite interesting, since 
that should then be checked experimentally. 

The analysis of entanglement given in this chapter
does not depend 
in any way on Schr\"odinger's equation. 
Both the states and the events 
have been taken to be 
time independent. 
So this example is in both the Schr\"odinger model 
and in the Heisenberg model. 
Typically in other presentations, 
the time dependence is imposed on the 
spatial structure rather than on the spin structure 
in a way consistent with the presentation here. 
Most importantly we learn 
that Schr\"odinger's equation is not 
a basic structure in quantum theory, while 
quantum probability theory is basic. 

The critical reader may object that I have done 
nothing other than just repeat the standard analysis 
for an entanglement experiment in a disguised form. 
But that ignores certain key aspects 
of this presentation.
Firstly, it is an analysis of three events, 
not of two particles.  
And secondly, 
there is no reference to collapse,
which is only a way of describing some intermediate steps 
in other, more common ways of dealing with entanglement. 

And collapse does not require 
an `explanation' any more than
any other mathematical algorithm 
that gives the right results. 
The terminology for this particular step is 
unfortunate, since it leads one to think that 
some sort of `understanding' of a physical process 
is needed. 
Calculations are meant to help us understand physical 
phenomena, or as it has been said
since Antiquity: to save the phenomena. 
(See \cite{russo} for the original meaning of this phrase.)

But for those who cling to the collapse language, 
let me note that collapse also 
occurs in the context
of classical physics, despite the often made 
claim that entanglement is a purely quantum effect. 
One can easily produce two space-like events
which are highly correlated and are described  
classically. 
As in the quantum case, one simply allows an 
event in the common past of these two events 
to have an impact on them. 
As in the quantum case, the two later events 
would be binary: one of two related possibilities. 
The mythical determinism of classical mechanics
has nothing to do with this, since that is after 
all a chimera. 
I will come back to this later. 

Also, I have done an analysis here in terms of 
the conditional probability of events, 
which is never done when 
analyzing entanglement as far as I am aware. 
Rather the standard discussion uses these words:
particle, state, measurement and collapse. 
It turns out that the conditional 
probability in the particular case 
\eqref{compute-entangle} just happens to be $ 1 $. 
Again I want to emphasize strongly that 
the intermediate steps in the calculation  
have absolutely no physical significance. 
Only the calculated probability matters. 
The rules of quantum probability say that this 
particular combination of events occurs in $ 100\% $ of
repetitions of the experiment. 
This deserves to be checked by experiment. 
And it has been. 
The conclusion is that quantum theory is verified. 
This seems to be widely accepted in the physics community. 

Let me make clear that entanglement is a 
property of events and not of anything else. 
It makes no sense to entangle particles 
because particles are not events. 
It makes no sense to entangle the states 
(note the plural!) of two particles, because 
referring back to Axiom~2 
only the entire quantum system itself has 
associated states, and not its constituent 
particles. 
It makes no sense to entangle a system with its 
environment, since neither a system nor its 
environment is an event. 
It makes no sense to entangle an observer of a 
quantum system with that system, and so on. 
These are important and useful 
concepts when used correctly, 
but one needs only the basic quantum 
theory of events, states and probability 
to understand entanglement. 

For those accustomed to thinking in terms of 
the Schr\"odinger model it might be difficult 
to analyze phenomena using a sequence of
quantum events instead of 
a sequence of quantum states. 
Of course, 
I am claiming that all entanglement phenomena can be
analyzed as is done here using a sequence of
quantum events and their conditional probability. 
This is a sweeping claim, although it depends 
on the rigorous definition I have given of 
entanglement. 
More specifically, I am claiming that situations that 
are traditionally called entanglement in the literature 
are described by the definition \ref{define-entangle}. 
Clearly, this claim will and should be challenged. 
Let me note that 
it should be kept in mind when evaluating this claim 
that quantum states are themselves rank~$ 1 $ 
quantum events. 
But do note that in the example of this chapter 
there are also quantum events of rank $ 2 $. 

Now if you insist on using the word 
`collapse', that could be acceptable as long as you
do not assign a physical significance to it,
as long as you do not treat it as a 
physical process. 
However, in practice that does not happen. 
Many physicists go down a rabbit hole by
saying that  
collapse is a 
profound topic which requires 
some deeper `explanation' or `interpretation'. 
Then it 
becomes a source of thousands of pages of 
no significance at all.  
Many are so used to working only with the Schr\"odinger 
model that they are unaware that some aspects of that 
model, such as collapse,  are not present in other 
equivalent models, such as the Heisenberg model.  
And it is only the model independent probabilistic 
aspects of quantum theory that have a physical 
significance. 
Therefore I advocate for 
discarding ``collapse'' from the quantum 
vocabulary much as ``equant'' has been discarded from
astronomical vocabulary. 
It is not needed. 
It is best to realize that there is no there there. 

Analogies are difficult to find, since other 
branches of science do not rely on two distinct 
ways of speaking about time evolution. 
However, biologists sometimes speak of 
the evolution of species 
in teleological terms, which they take to be 
incorrect, yet useful, shortcuts for describing 
Darwinian evolution. 
They might say that a certain 
species of animals evolved to have 
a thicker subcutaneous layer of fat and white fur 
{\em in order} 
to survive to a colder, snowy climate. 
This can be easily misinterpreted by those who have 
not studied biology. 
However, biologists realize that this is just a manner
of speech. 
My point is that the two ways of speaking about
time evolution in quantum theory (Schr\"odinger's 
equation and collapse of the state) are at best a 
short cut, a manner of speech, that takes the place 
of the Generalized Born's Rule. 

There is a sort of critique of formal rules and 
axioms that objects to the lack of prior 
justification of the rules.  
In short, where do the rules come from? 
That they work well for all observations is not 
accepted as a satisfactory answer, especially 
if the rule clashes with an `intuition' of some 
sort or other. 
Of course, any proposed explanation of `why' the 
rule is correct is itself subject to the same 
criticism. 
And so on {\em ad infinitum}. 
Such criticism could be applied to the 
Generalized Born's Rule 
for computing quantum probabilities, to which 
I respond: {\em Hypotheses non fingo}.

\chapter{Schr\"odinger's Cat}
\markboth{Schr\"odinger's Cat}{Schr\"odinger's Cat}

\hfill Nothing divided people more deeply

\hfill   than how they felt about cats. 

\vskip 0.2cm

\hfill Kingsley Amis

\vskip 0.4cm \noindent 
But there is another problem with  entanglement.  
It has to do with how it is verified. 
The two events $ E_{1} $ and $ E_{2} $ can occur 
with space-like separation. 
In fact, this has been done precisely to 
see if quantum theory holds in that case. 
But how is the comparison of results from a series
of measurements made? 
After all, at the time of the measurement neither
experimenter can possibly know what the other has
measured. 
So they record their measurements and get together 
at a later time (in the intersection of their
respective future light cones) and compare data. 
Of course, such persistent data are not exactly events. 
This will lead us to consider the `paradox' of 
Schr\"odinger's cat. 

Think of special relativity theory where events in space-time
are described as points in a Minkowski space. 
Or perhaps, since points are an idealization, events
are small regions in space-time, all of whose points
are very close to each other. 
One of the conceptual difficulties with relativity theory
is that one discusses objects as if they were events,
but somehow persisting in time. 
These are called {\em world-lines}. 
For example, the physics professor holds up his piece of 
chalk for all the class to see. 
He asks how long it is. 
Everyone agrees that it is about $2$cm long. 
But no! 
That is only is one inertial frame of reference.
The endpoints of the chalk are actually world-lines
of events in Minkowski space and the spatial 
distance between these endpoints at the same
time (relative to 
another frame of reference) is the length of the
chalk (in that other frame of reference). 
The length in another appropriate inertial frame is
$ 10^{-9} $m, a nanometer. 
Students are not so easily convinced and rightly so.
After all, the concept of event has been changed
behind their backs. 
And what color is the chalk? 
Well, the white light of the class room reflects 
mostly the blue end of the spectrum. 
All agree that it is a piece of blue chalk. 
But in another inertial frame the frequency is shifted, 
and the reflected light from the chalk is red!
Even classical physics can be very non-intuitive. 

Returning to quantum theory, what is persistent data?
I doubt that it is an event which lingers for an 
indefinitely long time. 
This seems to be changing concepts, much as in the 
case of relativity theory.  
Actually, I have not given any localization information 
in speaking about quantum events; they are just 
(time) ordered projections in a Hilbert space. 
Let me remedy that by 
using ideas of Haag from his version 
of local quantum field theory in \cite{haag}. 
As usual everything is within the context of one 
given Hilbert space $ \mathcal{H} $. 
One thinks of open regions in Minkowski space. 
For any such region $ \Omega $ one has an associated 
von Neumann algebra $ \mathcal{V}(\Omega) $ in
$ \mathcal{L} ( \mathcal{H} )$. 
It is well known that a von Neumann algebra has `lots' 
(in a sense that can be made 
precise\footnote{For example, the smallest von Neumann algebra in $ \mathcal{L}( \mathcal{H} )$
containing the events of a given von Neumann algebra
$ \mathcal{V} \subset \mathcal{L}( \mathcal{H} )$ is $ \mathcal{V} $ itself.}) 
of projections, that is, of quantum events. 
One says that any projection in $ \mathcal{V}(\Omega) $ is
an event {\em localized} to the set $ \Omega $. 
Intuitively, one is thinking that it is 
something that happened within that region of space-time. 
These quantum events are the only ones which
correspond to something happening in $ \Omega $;
all other quantum events in 
$ \mathcal{L} ( \mathcal{H} )$ 
are not part of the quantum theory in $ \Omega $. 

Suppose that $ \Omega_{1} $ and $ \Omega_{2} $ 
are space-like separated regions, meaning that for 
all $ \omega_{1} \in \Omega_{1} $ and 
all $ \omega_{2} \in \Omega_{2} $ we have that 
$ \omega_{1} $ and $ \omega_{2} $ are space-like 
events. 
Under these hypotheses 
we assume, following Haag, that for all quantum events 
$ E_{1} \in \mathcal{V} ( \Omega_{1} ) $ and 
$ E_{2} \in \mathcal{V} ( \Omega_{2} ) $, we have that 
$ E_{1} $ and $ E_{2} $ commute. 
Of course, the `small' open sets are the ones 
that matter, especially those that are small 
in the temporal direction. 
I do not know how to incorporate a tubular 
neighborhood around a very long, especially 
infinitely long, world-line, into quantum theory in an intuitive way. 
But events associated with these sorts of regions 
are what one would use in quantum theory to 
describe the data recorded from scientific 
experiments. 
Or to describe any sort of information that 
is held indefinitely and can be shared. 
Of course, the entire scientific enterprise is 
based on sharing data.
And this is what happens also with every 
entanglement experiment. 
But in that context it is not considered to
be a problem.
It seems to be `natural' that the experimenters 
share their data at some future time. 
And everyone is in agreement that entanglement 
has been experimentally verified this way. 
This is the way one verifies all quantum experiments, 
actually all experiments in all the sciences! 
And this is the basis of the often heard statement 
that quantum theory has been remarkably 
successful. 

But wait!
Enter Schr\"odinger's cat, and 
the persistence of data from a quantum event 
becomes a problem. 
Let's take the innocent cat out of harm's way by
changing the set-up. 
As in the original version, there is a 
radioactive source which has a probability 
of $ 1/2 $ of decaying within a certain period 
of time. 
A diabolical  
machine is in place that can detect 
this decay always and is screened from any possible 
background noise. 
(We will see later why it is just as diabolical  
as Schr\"odinger's machine.)
If the decay is detected, then the machine 
prints YES on a sheet of paper that is blank
except for the experiment run number. 
Otherwise, it does nothing during the time
of the experiment.
When the time period ends, the machine ejects the
sheet of paper. 
To check that everything is working well, 
the experimenter repeats this many times, 
counts the number of times that YES appears
and then divides by the total number of experimental runs. 
Within statistical precision the relative 
frequency of YES agrees with $ 1/2 $.
But are these sheets of paper quantum events? 
Is there a self-adjoint operator $ P $ whose spectrum 
is $ \{ 0,1 \} $ with the quantum event $ P = 1 $
corresponding to YES appearing and $ P = 0 $ otherwise? 
If so, then $ P $ is a quantum event.  
And even if this is true, how do 
the results $ 1 $ and $ 0$ persist
in time? 
After all, the data from these events has to be published, lest the experimenter 
perish. 
And the sheets of paper can be preserved for 
future reference an indeterminable number of 
times for a more or less indefinite future. 

Here is my best attempt at dealing with 
these questions. 
I think it makes sense to use the projection 
$ P $, which in the cat context tells us whether
the cat is dead or alive. 
This operator generates a commutative 
von Neumann algebra of 
dimension $ 2 $, namely 
$
\mathcal{V} = 
\{ \alpha I + \beta P \,|\, \alpha, \beta \in \mathbb{C} \} 
$ 
in $ \mathcal{L} (\mathcal{H}) $, where the dimension of
$ \mathcal{H} $ could be quite large, even infinite. 
Now following Haag 
the only quantum events available in
this quantum theory are $ 0, \, P, \, I-P, $ and $ I $. 
Notice that these are $ 4 $ distinct events, since
$ P \ne 0 $ and $ P \ne I $ by the 
construction of the experiment.
And only two of these events are non-trivial. 
Let me note that the event $ 0 $ corresponds to
starting the experiment with a dead cat, while the 
event $ I $ corresponds to using an immortal cat. 
 
So what happens now if we take normalized eigenvectors 
$ \psi_{1}, \psi_{2} $ of
$ P $, one for the eigenvalue $ 1 $ and the other
for the eigenvalue $ 0 $?
We can form the normalized eigenvector (pure state)
$ \phi := 2^{-1/2} (\psi_{1}  +  \psi_{2}) $
in the Hilbert space $ \mathcal{H} $. 
But this state is not available in this quantum 
theory, because the corresponding quantum event 
$ | \phi \rangle \langle \phi |$ 
is not in $ \mathcal{V} $. 
In other words, the `superposition' state of a 
live cat and a dead cat is not in this quantum theory.
It is important to note that this argument does 
not use a {\em superselection rule}.
Rather it is a mistake to take 
$ \mathcal{L} ( \mathcal{H} )$ as 
the von Neumann algebra for this situation. 
The error lies in thinking that the only 
von Neumann algebras suitable for doing 
quantum theory are isomorphic to 
$ \mathcal{L} (\mathcal{K})  $ for some 
Hilbert space $ \mathcal{K} $. 
The von Neumann algebra $ \mathcal{V} $ 
is not of this type, since $ \dim \mathcal{V} =2 $. 
Rather we have 
$ \mathcal{V} \cong L^{\infty} (\{ 0,1 \}  ) $. 
Worse yet, 
$ \mathcal{V} $ is commutative, 
thereby violating the 
non-commutativity as required 
in Axiom~1. 
And any added event would have to 
correspond to some physical aspect of the `quantum cat'.  
If we only observe the results `live cat' or
`dead cat', then $ \mathcal{V} $ suffices to save 
the phenomena, but it is not a quantum theory.  
We do not need a theory that includes events 
for results 
that are not observed. 
And even more importantly, 
we do not--actually can not--give
meaning to quantum events that do not 
correspond to physical results, since it is just such a
correspondence which {\em is} the meaning of 
a quantum event.
Also, $ \mathcal{V} $ is inadequate for describing 
the radioactive nucleus as a {\em quantum} system, 

However, it seems to be folklore that ``one can not 
write something down on paper in quantum theory''. 
This is related to the no-clone theorem 
of quantum theory. 
(See \cite{park}.)
But my diabolical machine does write something down 
on paper. 
And this is why it is diabolical. 
It defies quantum theory. 
What are we to make of that? 
Some physicists hold to the view that all 
physical phenomena are described by quantum 
theory. 
So they face the challenge of describing  
in the context of quantum theory how  
data can be recorded and shared. 
It is not permitted to appeal to classical physics
in such an argument. 
It has to be a demonstration totally 
within quantum theory itself. 
This is essentially what is known as the 
{\em measurement problem} of quantum theory.
Other physicists assert that the diabolical machine
as well as all measuring devices are classical, not
quantum, systems. 
But it is still not easy to see how
classical mechanics with conservative forces can  
explain how records can be made and preserved. 
A more plausible way would be a classical system 
with dissipative forces that has at least two 
stable sinks in phase space. 
(If there are exactly two stable sinks, this 
classical system is 
called a {\em bit}. 
And bits do seem to exist.)
That would not be a Hamiltonian system, and so the 
usual methods of quantization do not apply. 
In other words. it would be difficult, though maybe 
possible, to describe the recording device using 
quantum theory. 
But somehow or another a classical system has to be 
coupled to a quantum system in such a way that 
information flows from the quantum system to the classical
system, where that information is preserved. 
It seems to be folklore that this is not possible, 
but I will not express an opinion.  
I do not propose any solution. 

Nonetheless, I wish to note that there do seem to be 
recording devices in nature that are usually 
considered to be quantum systems. 
Let us think of a nucleus that is unstable via
beta decay. 
The nucleus changes from one element in the 
periodic table to another when it decays. 
This is recorded in many ways; for example the
electric charge of the nucleus changes. 
So the electric charge serves as a way of recording
whether the nucleus has decayed or not. 
And if the daughter nucleus is stable, then this data 
is recorded for all posterity. 
Nuclei are certainly not classical systems. 
And most physicists would agree that they are 
quantum systems. 
So this is an example of how some data
(though possibly not all data) 
can be recorded in a quantum system. 
The wheel has come full circle. 
It was a radioactive decay that initiated  
the events of the diabolical machine. 
In short, forget about the diabolical machine, 
the box, and the cat inside the box. 
None of that needs any more explaining than 
beta decay itself does. 
So we find the same puzzle at the end 
of the story as at the beginning, but now 
in a quantum system. 

The same problem is present in classical physics, 
though it seems to never be discussed. 
Again, it is not simply a question of a physical 
system and some observation of it. 
Rather, the information from that situation is 
recorded and shared. 
The question is: 
``Can one  
write something down on paper in classical theory''. 
The answer may be yes, but that is not so obvious. 
Classical mechanics describes motion as a 
trajectory in phase space. 
What does that theory have to do with preserving 
information? 
A piece of paper with something written on it 
is a system with many degrees of freedom, that 
is to say, its phase space has an enormously 
high dimension. 
And what does a trajectory in that phase space 
have to do with the transmission of information 
in the scientific community? 
These are not rhetorical questions. 
My point is that the physics community is not 
agitated over the fact that these serious questions 
are not addressed when discussing classical mechanics,
say in a university course or in a popular 
exposition for a general public. 
Typically, one hears that there is no measurement 
problem in classical physics, but that there is 
such a problem in quantum physics. 
What's sauce for the goose is sauce for the gander.

\chapter{The Measurement Problem}
\markboth{The Measurement Problem}{The Measurement Problem}

\hfill Truth is truth

\hfill To the end of reckoning

\vskip 0.2cm

\hfill William Shakespeare

\vskip 0.4cm \noindent 
The measurement problem is not
a problem that one can address, let alone solve, 
in the context of an axiomatic theory. 
Let me illustrate this with an example which 
I hope is not very 
controversial: Euclidean geometry. 
In the full axiomatization of this theory 
(which eluded Euclid's efforts) 
certain statements are proved about the 
size of angles and the areas of geometric 
figures, such as squares in the Pythagorean 
theorem. 
These statements are in contradiction with 
statements in non-Euclidean geometries about
the `same things'. 
It is generally accepted that a way to test if 
Euclidean geometry holds is to measure the three physical  
angles of physical triangles and see whether their sum 
is indeed $ \pi $ radians (in modern units). 
It does not pertain to Euclidean geometry nor 
to any of its competitors to explain how to measure
angles or areas. 
That is not understood to be part of the task 
of the theory itself. 
This question is not considered to be one which 
the theory has to answer: 
How does one measure an angle? 
In fact, one assumes that physical triangles 
with three physical angles exist and, moreover, 
that all of these incompatible geometries are
speaking of these same physical objects. 
There are no Euclidean angles, no projective geometry 
angles, and so on. 
But there are measured angles. 
Similarly, there is no classical energy, no 
quantum energy and so on. 
But there is measured energy. 
For an experimental scientist measurement 
is an activity, not a problem. 

And 
I see the axiomatization of quantum theory 
in the same way. 
As presented in this treatise quantum theory is 
an axiomatic theory of states, events and probabilities. 
How these are measured is not addressed in the 
mathematical theory. 
(But I do come out in favor of relative frequency
as the way to measure probability.) 
One should think of `state, event, probability' 
as three basic concepts much like
`point, line, angle' in geometry. 
Of course, the latter geometric words do 
get a significance in terms of observations. 
And so do the former three concepts of quantum theory. 
But those are scientific questions outside the 
scope of the axiomatic theory. 

In this treatise  the question is how to treat 
measurement  
in the context of quantum theory based on 
states, events and probability. 
Appeals to classical physics, or any other theory,
are beside the point. 
One is looking for a consistent, but not circular, 
exposition of the topic. 
This will not solve the measurement problem, 
but rather explain what are the issues that 
a scientific approach must address. 
In fact, I am not sure what the 
measurement problem actually 
is, and so I 
am quite unprepared for solving it. 
I am only trying to put measurement 
into the context of this treatise 
as a certain, but not clearly defined, type of event. 

To start off I would like to introduce the idea 
that measurement is done by devices (or physical 
systems, if you wish) 
that implement sequences of two events in 
such a way that the conditional  
probability of the second event, given the first 
event (and maybe a state) is $ 1 $. 
It is definitely not assumed that such a 
device always has the same sequence of two events. 
Such a device is essentially a function that 
has a certain set of events as the possible first  
event (input) of the sequence and then has the 
second event (output) in that sequence. 
It should further be required that this function 
is one-to-one, so that the second event in the sequence
uniquely specifies the corresponding first event. 
Colloquially, the second event is characterized
by the 
first event, and it in turn uniquely 
characterizes the first event. 

Let me emphasize this one more time.
Sometimes the claim is heard that a measuring
device must be classical since its results are 
`determined'. 
The assertion is made that this is a deterministic process,  
and hence the device can not be a quantum 
system but must be classical system. 
But this logic is fallacious. 
Probability theories include the special 
case of conditional probability $ 1 $, as 
we have seen in Chapter~\ref{Entanglement-Chapter} on 
Entanglement. 

Whether such devices actually exist physically 
is another question, though it is generally 
assumed that they do. 
Anyway there is nothing in quantum theory which 
precludes such devices from existing. 
So, measuring devices as defined here are 
consistent with quantum theory. 
Sequences of two events that have conditional 
probability $ 1 $ are consistent with a 
probabilistic theory; they are just a 
special case. 
How to explain particular physical devices 
using quantum theory is something that depends on 
the specific details of those devices. 
Though I doubt there is a general answer 
for all devices, I could be proved wrong. 
In short, I suspect that giving a general quantum 
theory of measuring devices is an intractable 
problem, at least with our current level of 
understanding. 
This seems to be born out by the overwhelming 
lack of progress in addressing this issue. 
But for those who think otherwise and want to 
research this problem, please remember the 
important role of approximation. 
The point is that devices which implement sequences 
of events with probability nearly equal to $ 1 $ can 
be quite accurate, though not perfect, 
measuring instruments. 

Quantum computing can be viewed in the same way. 
Though speaking of a sequence of states being 
changed by quantum operations is valid, that 
language does not survive in the equivalent 
Heisenberg model where the state is time 
independent. 
I would rather say a hard-wired 
quantum computer is a physical device that has 
a final (output) quantum event that is determined
{\em with high conditional probability} by a given 
first (input) quantum event and any state. 
Only in this case, the function being implemented 
this way need not be one-to-one; 
distinct input could result  
with high probability in the same output. 
A programmable quantum computer would do the 
same, again 
with high probability, but the function 
would depend on a program. 
To my understanding this is what `classical' computers do, 
but with a set of events that is a Boolean algebra. 
A fully quantum computer would use all quantum events. 

Let's look at Schr\"odinger's cat using these ideas. 
A radioactive nucleus with charge $ Z $ is 
under consideration. 
Let $ E_{1} $ be the (quantum) event that has value $ 1 $
if and only if beta decay occurs in the time interval. 
Let $ E_{2} $ be the (quantum!) event that 
has value $ 1 $ if and only if 
the cat dies
in the same time interval. 
The device is constructed so that 
for any state $ \psi $ we have that 
the following conditional probabilities hold: 
\begin{align*}
 &P ( E_{2} \,|\, E_{1}, \psi ) = 1 \quad 
\mathrm{decay~occurs;~cat~dies},
\\
 &P ( E_{2}^{c} \,|\, E_{1}, \psi ) = 0 \quad 
\mathrm{decay~occurs;~cat~does~not~die},   
\\
&P ( E_{2} \,|\, E_{1}^{c}, \psi ) = 0 \quad 
\mathrm{decay~does~not~occur;~cat~dies}, 
\\
 &P ( E_{2}^{c} \,|\, E_{1}^{c}, \psi ) = 1 \quad 
\mathrm{decay~does~not~occur;~cat~does~not~die}.
\end{align*}
Here $ E^{c} = I - E $ is the complementary event to
the event $ E $. 
For example, 
the event $ E_{1}^{c} $ means that the beta decay 
did not occur in the time interval. 
It is not to be confused with the expression 
`nothing happened' in ordinary language, 
which misses the point.  
No one, starting with Schr\"odinger himself, 
 ever seems to doubt the possibility of
constructing a device which has these conditional
probabilities. 

Seeing a live cat is a way of measuring the charge 
of the nucleus and getting the result 
that it is $ Z  $. 
And seeing a dead cat is a way of measuring the charge 
of the nucleus and getting the result 
that it is $ Z  +1 $. 
Of course, there are other ways of measuring 
the charge of the nucleus. 
One might say they are more `direct'
than using the poor cat as the measuring device. 
But I claim that they will also be based on sequences of  
quantum events which have conditional probabilities 
equal only to 
either $ 0 $ or $ 1 $. 
Or maybe just with conditional probabilities only 
very near $ 0 $ or very near $ 1 $.

\chapter{The EPR paper}
\markboth{EPR paper}{EPR paper}
\label{EPR-chapter}

\hfill Face au r\'eel, ce qu'on croire savoir

\hfill offusque ce qu'on devrait savoir. 

\vskip 0.2cm

\hfill Gaston Bachelard

\vskip 0.4cm

Ever since its publication in 1935 
the famous EPR paper \cite{epr} (named for its 
authors Einstein, Podolsky and Rosen), 
has generated much 
discussion and controversy, despite the 
fact that its conclusion is correct. 
It has even been referred to as 
the EPR paradox! 
The authors come in the last 
paragraph of \cite{epr} to the following 
conclusion:
\begin{center}
While we have thus shown that the wave function 
does not provide a complete description of the
physical reality, we left open the question 
whether or not such a description exists. 
We believe, however, that 
\\
such a theory is possible. 
\end{center}

The argument of this paper is an example of
entanglement, though that word is not used,  
presented in the Schr\"odinger 
picture. 
To achieve their conclusion the authors 
have to give some idea of their concept 
of `physical reality'. 
This they do by saying the following:
\begin{center}
	 
{\em If, without in any way disturbing a system, 
we can predict with certainty (i.e., with 
probability equal to unity) the value of 
a physical quantity, then there exists an 
element of physical reality corresponding to this physical quantity.}  \hskip 6.9cm 
(Italics in the original)

\end{center}
They then add how to regard this condition:  
\begin{center}

Regarded not as a necessary, but merely 
\\
as a sufficient condition of reality . . .
\end{center}
One can quibble with their condition, but for 
the moment let's accept it and try 
to understand the paper a little better. 
At the beginning of the paper the authors state:
\begin{center}
In attempting to judge the success of a physical 
theory, we may ask ourselves two questions:
(1) ``Is the theory correct?'' and (2) ``Is the
description given by the theory complete?'' . . . 
\end{center}

A little further on they continue:
\begin{center}
It is the second question that we wish 
to consider 
\\
here, as applied to quantum mechanics. 
\end{center}
This is also the question posed in the title 
of the paper. 
However a careful consideration of both their
argument and of their very words shows that 
they did not answer this second question. 
Just to be clear here again is the essential 
part of their correct conclusion:
\begin{center}
 . . . the wave function 
 does not provide a complete description of the
 physical reality . . .
\end{center}
So the implicit assumption is that 
the wave function is the only theoretical 
element that is involved 
in the description of quantum theory. 
That is how they can jump from showing 
the incompleteness of the wave function 
to the incompleteness of quantum theory 
as a whole. 

This is an error of logic, of course. 
But it is easy to see how the authors 
arrived at it. 
Apparently, 
their tacit assumption is that the Schr\"odinger
equation is the fundamental time evolution 
equation of quantum theory and that equation 
has a unique solution (the wave function), 
given an adequate initial condition.  
This is second nature
for those accustomed to thinking in terms of 
physical theories based on differential 
equations;
this is intuitive. 
Unfortunately, as we have seen, it is wrong. 
The details in the course of their argument 
about how the state changes due to measurement 
are not viewed as contradicting this 
general idea of how physical theories work. 
Of course, 
these details are nonetheless just how one
calculates conditional probability in the 
Schr\"odinger picture, though that wording 
is not used. 

However, we can suppress this tacit assumption 
and see that the actual conclusion of the paper 
is correct. 
Indeed, the wave function alone does not suffice 
to provide a complete description. 
As presented in this treatise, events and 
probabilities 
(and their possible time evolution)
are also part of basic quantum theory. 
It might well be that the concept and importance of
quantum event were not known to the authors of 
the EPR paper in 1935. 
But probability was a contemporaneously 
available concept. 
Indeed, probability one is singled out for 
special consideration in their 
sufficient condition for an element of 
physical reality. 
However, later commentators, and perhaps 
these authors as well, have turned this into a 
necessary condition if they assert that 
something with probability less that one is 
not an element of physical reality. 
While one may wish to take their condition as 
both necessary as well as sufficient, that 
is a giant step beyond what the authors of EPR 
actually assert. 

Some confusion arises since rank~$ 1 $ events 
are also states (`wave functions' for these
authors). 
But we saw that entanglement in the simplest 
example of 
Chapter~\ref{entanglement-chapter} 
involves events that are
rank~$ 2 $ operators and therefore not  
wave functions. 
Similarly, the EPR thought experiment involves 
events which are not wave functions. 
This may seem to be just a minor technical detail,
but ignoring it leads to a fatal flaw in the 
EPR argument. 
Using the notation of 
Chapter~\ref{entanglement-chapter},  
the occurrence of $ E_{1} $, 
a rank~$ 2 $ event,  
does not tell us the state
of the particle 
on the left in the Hilbert space 
$ \mathcal{H} = \mathbb{C}^{2} \otimes \mathbb{C}^{2}$, a $ 4 $-dimensional space. 
As shown in 
Chapter~\ref{entanglement-chapter} 
entanglement can easily be understood by 
calculating the conditional probability of events without assigning `meaning' to the intermediate 
steps. 
But if one prefers the language of states, 
then the 
correct statement is that the occurrence of the 
event $ E_{1} $ tells us that the state is in the 
$ 2 $-dimensional 
subspace $ \mathrm{Ran} \, E_{1} $. 
Similarly, $ E_{2} $ does not tell us the state
of the particle 
on the right. 
The implicit error here is the assumption that 
the measurements of the  
detectors are being modeled by events in 
two different $ 2 $-dimensional Hilbert spaces. 
So one is discussing this quantum system with 
three distinct Hilbert spaces, 
and therefore three different von Neumann algebras, 
This is a clear 
violation of Axiom~1. 
In short the EPR thought experiment does not 
assign, by using the authors' sufficient condition,  
an `element of reality' to the state of
either particle.

Let's examine this in a bit more detail. 
To describe the detector on the left we only 
need the von Neumann algebra 
$\mathcal{L} (\mathbb{C}^{2} )$. 
A different copy of the same von Neumann algebra 
suffices for the other detector. 
But if the space-like separated events of the 
detectors are compared and found to have some
correlations among them, then a scientific 
puzzle has been noted that requires further 
explanation. 
Of course, we already know what that explanation 
is in this case. 
And a part of that is including the two copies 
of the von Neumann algebra of the individual 
detectors as sub-algebras of 
$\mathcal{L}(\mathbb{C}^{2}\otimes\mathbb{C}^{2} )$, 
the new von Neumann algebra for the combined system. 

Let's take this one step further and suppose that 
the detector on the left not only says that 
$ E_{1} $ occurred, but the experimenter on the 
left is told that the earlier event 
$ E_{0} $ had already occurred. 
To accommodate the event $ E_{0} $ 
would require using a new model,
the simplest one of which would be the von 
Neumann algebra 
$\mathcal{L}(\mathbb{C}^{2}\otimes\mathbb{C}^{2} )$. 
Using this model the experimenter on the left 
then has sufficient information for calculating 
the {\em probability} of the event $ E_{4} $ of 
any spin $ 1/2 $ 
measurement by the detector on the right. 
As I will argue in a moment, this is sufficient
for asserting that the value 
($ \equiv $ eigenvalue) of the 
event $ E_{4} $ has an element of physical reality. 

Now I wish to 
suggest that their sufficient condition for 
element of physical reality is unnecessarily 
linked to probability one events. 
Leaving aside the vague restriction about 
not disturbing the system, I propose another 
more general, sufficient condition for an 
element of physical reality: 

\begin{center}
If we can predict with non-zero 
probability the value of 
\\
a physical quantity, then there exists an 
element of 
\\
physical reality corresponding to this physical 
quantity.
\end{center}

I might even be disposed to deleting the 
word `non-zero' from the above condition. 
After all, violation of a conservation law 
has a physical reality of a 
certain sort---the reality
of never occurring (to date). 
This condition seems to be more in the spirit of 
basic quantum
theory viewed as being a new type of 
probability theory. 
Of course, in some sense I am proposing a 
broader definition of `element of 
physical reality'. 
However, this condition is intuitive in the sense 
that people act as if it holds. 
For example, a house burning down 
and resulting in (financial) value zero 
has an 
element of physical reality 
due to its non-zero probability, even though 
it has never happened and most likely  never 
will. 
Yet this element of physical reality is sufficient 
to motivate buying fire insurance for the  
house. 

Anyway, with this enhanced sufficient condition 
physical quantities such as position or spin
component are elements of 
physical reality. 
Moreover, such physical quantities do find a 
counterpart in quantum theory. 
Actually, their pvm's and probabilities 
are counterparts. 
This would make quantum theory nearer to being 
complete than the authors might have thought 
according to their own criterion given in 
the EPR paper: 
\begin{center}
Whatever the meaning assigned to the term
{\em complete}, the following requirement for a 
complete theory seems to be a necessary one: 
{\em every element of the physical reality 
must have a counterpart in physical theory}. 
(Italics in original)
\end{center}

The authors of the EPR paper include 
as an element of 
physical reality any
probability $ 1 $ physical event. 
Its counterpart in the quantum physical theory 
is the theoretical probability as given by the 
Generalized Born's Rule, which better give the 
value $ 1 $ as well. 
The wave function alone is not sufficient  
information for calculating that theoretical 
probability. 
I am befuddled by those who speak of the 
EPR paper as a paradox. 
However, it is a poorly written paper which 
still somehow manages to arrive at a correct
conclusion. 

Clearly, quantum theory remains incomplete but
for other reasons such as its current inability to 
predict the properties of the elementary  particles, 
the values of fundamental constants, 
the nature of gravitation and the origin of the 
universe to name some examples.

\chapter{Determinism and Probability}
\markboth{Determinism and Probability}{Determinism and Probability}

\hfill The epistemological value of
  probability theory is based on

\hfill  the fact that chance 
 phenomena, considered collectively 

\hfill and on a grand scale, create non-random 
regularity. 

\vskip 0.2cm 

\hfill Andrei Kolmogorov

\vskip 0.4cm
\noindent
There is a probabilistic aspect in 
all the standard formulations of quantum theory.  
There is rarely a probabilistic aspect in 
standard formulations of classical physics 
and, when there is, it is chalked up to 
being due to a lack of complete information 
rather than a fundamental aspect of that theory. 
But it is also said that classical physics is 
deterministic. 
And this is not a logical consequence 
of what I just said. 
Rather it is known as jumping to a conclusion. 
The determinism ascribed to classical physics 
is a myth repeated over and over by so many 
experts  
that it has become heresy to question it. 
Note that a mathematical 
formulation of determinism 
is not found in the standard texts 
on classical physics. 
The best one can find in the 
scientific literature 
are various theorems in 
mathematics which state that some classes of 
time dependent  
differential equations have unique solutions provided
they are accompanied by appropriate initial 
conditions. 
But it turns out that there are other 
time dependent  
differential equations that do not have 
this property and, according to theorem, 
such equations are non-linear. 
Also there is nothing in classical Newtonian 
mechanics which tells us which sort of 
differential equations will arise in that 
theory, that is, whether they have unique 
solutions with appropriate initial conditions
or not. 
But it is known that the second law of motion of
Newtonian mechanics is non-linear in almost all 
examples. 

Since this is the relevant moment, 
let me pause from continuing this 
discussion in order to call out a confusion that
is rampant in the physics community. 
The point is that differential equations do not 
automatically imply determinism. 
To suppose otherwise is wrong-headed, to say 
the least. 
Some differential equations are consistent with 
determinism, but others are not. 
We often work with a linear approximation.
Linear differential equations with 
appropriate initial conditions do have unique 
solutions, which are 
also global in time. 
But this property of a particular approximation 
does not always hold in other non-linear 
contexts. 
So it is a {\em non sequitur} to say
that a theory based on differential 
equations is automatically deterministic.  

Let me be clear about this difference between 
classical mechanics and quantum theory. 
The theory of classical mechanics does not 
answer the question of whether determinism 
or probability applies to its subject matter. 
On the other hand quantum theory at its most basic
level makes statements  
about probability and, according to this treatise, 
about nothing else. 
One should not say that quantum theory 
eliminated determinism from physics;  
rather it put probability into physics. 
And probability is very non-intuitive. 

However, we must pause, since it appears that Schr\"odinger's equation is deterministic. 
After all, for any
self-adjoint Hamiltonian $ H $ there is a 
unique solution, global in time $ t $, 
given any initial state $ \varphi $ in the 
Hilbert space, namely, 
$ \psi(t) = e^{- i t H / \hbar } \, \varphi $ 
in the Schr\"odinger model. 
At a purely mathematical level, this is a sort
of determinism. 
But this does not give us determinism of a
quantum system, since the states $ \psi(t) $ 
by themselves only determine probabilities when 
combined with 
the events associated to a system. 
And this role of probability 
is not the lockstep relation of cause and effect as  
envisioned by the usual idea of determinism. 
The Heisenberg model also has a mathematical 
determinism, but now of the events 
as given by Heisenberg's equation 
\eqref{Heisenberg-eqn}. 
And again this is not a physical determinism, since 
in general events are not sufficient for computing 
quantum probabilities. 

And moving on now to experiment the situation is 
even shakier. 
The idea is that by knowing all about a physical 
system completely at one instant of time, the
future of that system is known or,
as one says, determined. 
How could one possibly know that? 
How could one possibly falsify that? 
If some systems behave in what seems to be a
deterministic manner (at least for some experimentally 
finite period of time), that does not mean they 
will continue to behave in that manner forever.  
And if they do not behave in  a
deterministic manner, it is unacceptable to say 
that is due to lack of knowledge about the initial 
situation. 
How could one possibly know {\em in all cases}
that there is more to know? 
It is logically possible that there is nothing more
to know. 
There is a lot of muddled thinking 
about determinism. 
Of course, there is a lot of muddled thinking 
about probability theory, too. 

Something that is truly confusing for me is 
that people, who can not look at a probability 
$ 1 $ event without running to the refuge of
deterministic language, can accept without 
any qualms any number of probability $ 0 $ 
events in quantum systems. 
Yet such events are literally duals of
each other. 
One says that a certain quantum transition is 
forbidden (i.e., has probability $ 0 $), 
since it violates some conservation law. 
However, that same conservation 
law holds with probability $ 1 $. 
And this all should be 
explained in terms of quantum theory,
not with determinism. 

From this perspective it is not so strange 
to say that quantum theory is probabilistic
and, in fact, that is thought to be one of the 
basic aspects of the theory. 
I have done just that in this treatise. 
What is strange is that a deterministic 
time dependent differential equation is claimed to be 
another basic aspect of
quantum theory. 
Of course, I refer to the time dependent 
Schr\"odinger's equation
\begin{equation}
\label{sch-eq}
   i \hbar \dfrac{d \psi}{dt} = H \psi. 
\end{equation}
Let me remind the reader what the 
mathematical theory actually says 
about this equation. 
We assume that $ H $ is a self-adjoint densely 
defined linear operator acting in a 
Hilbert space $ \mathcal{H} $. 
Then a theorem says that 
for any $ \phi  $ in the dense domain of $ H $  
there exists a unique solution $ \psi_{t} $ of 
\eqref{sch-eq} for all $ t \in \mathbb{R} $
such that $ \psi_{0} = \phi $. 
Moreover, that theorem asserts that 
$ \psi_{t} = e^{-i t H / \hbar} \phi $, where the 
globally defined unitary operators 
$ e^{-i t H / \hbar}  $ 
are defined for all $ t \in \mathbb{R} $ by 
the functional calculus of spectral theory applied
to the operator $ H $. 

In the Schr\"odinger model
one has Schr\"odinger's 
equation, a deterministic equation,  
and a probabilistic state collapse condition. 
It seems that the time evolution chugs along 
on its merry way changing the state is a continuous
deterministic way, when--poof!--somehow a 
probabilistic event occurs that changes 
the state in a discontinuous way. 
How can there be two distinct types of
time evolution? 
This perennial puzzle about quantum
theory arises as we have seen by focusing on
details in the Schr\"odinger model which 
are model dependent. 
Of course, in the equivalent Heisenberg model 
there is a deterministic time evolution of the pvm's 
while the state remains constant in time.  
In both models the (identical!) probability measures 
are continuous functions of time. 
The continuity of these probability measures 
follows from the continuity of 
$ t \mapsto e^{-it H / \hbar} $ in the
strong operator topology. 

For now I would like to discuss another sort of
probability in one aspect of quantum theory. 
But this aspect is quite different. 
Again, consider a nucleus that can undergo beta decay. 
Before the decay occurs the electron and anti-neutrino 
of the final state do not exist. 
They are created by the decay process. 
And no matter where the nucleus is in the universe, 
the electron has exactly the same characteristic 
properties as any other electron in the universe. 
Its mass, electric charge, spin and magnetic moment 
are always the same. 
In other words these are probability one properties 
of any electron. 
The same holds for the anti-neutrino; it is the same 
as any other anti-neutrino produced in beta decay. 
To make this sound more `paradoxical' 
imagine two beta decays of the same isotope nucleus,
but with a space-like separation between these 
two events.
How can one event possibly `know' about the other? 
Yet they produce the same decay products, although 
the momenta of the decay products are not always 
the same. 
Also, these decays violate parity conservation, 
but they do that in the same way always. 
On the other hand an alpha decay of a nucleus  
always preserves parity. 
These are all rightly called {\em particle properties}, 
though one could also describe them as 
{\em spooky action at a distance}.

\chapter{Interpretation}
\markboth{Interpretation}{Interpretation}

\hfill 
Interpretation is the revenge of intellect upon art.

\vskip 0.2cm
\hfill Susan Sontag 

\vskip 0.4cm
The astute reader will have noticed that I have  
only 
used the expression {\em physical significance}
instead of `interpretation' as was done 
in days of yore for describing the relation 
between theory and observation. 
This is due to the unfortunate situation that 
`interpretation' has come to denote something 
quite different from `save the phenomena', 
a phrase also from days of yore. 
(See \cite{russo}.)
Nowadays `interpretation' refers to  
extra-scientific statements that are 
intended to `save the theory'. 
They are neither verifiable nor falsifiable
and consequently have no role whatsoever in science. 
Somehow they hang in there as some sort 
of assurance that everything is `intuitive' 
after all and so serve to make some people feel 
happy. 
The theses of this treatise are scientific 
and are subject to the standard critiques 
of scientific methodology. 
Any assertion that I make may be right 
or it may be wrong. 
But in the 
contemporary sense of the word I do not give any `interpretation'  of quantum theory. 
Still, some comments are called for about 
some of the 
more common aspects of 
`interpretations' of quantum theory. 

The collapse of the wave function is quite often the 
central concern of an interpretation. 
This results in  tilting against windmills. 
There is nothing about the collapse condition 
that merits much attention once one realizes that it 
is a mathematical step used in the Schr\"odinger model 
of quantum theory, but that it has no correlate
in other isomorphic models. 
Typically, discussions that deal exclusively with the
Schr\"odinger model fall into the trap of 
treating a characteristic specific to that model 
as universally
applicable to quantum theory, when they clearly are not
due to the lack of that characteristic in 
another isomorphic model of quantum theory. 
The more-or-less standard Copenhagen Interpretation 
simply asserts that collapse is something that 
happens. 
But collapse is not a physical process, but rather a 
mathematical formula. 
Of course, the Copenhagen Interpretation 
is not just one assertion. 
And it morphs. 
So it is difficult to totally debunk it based 
on a detailed analysis of what it says. 
But it is clearly barking up the wrong tree. 

Another common goal of `interpretation' is to 
explain what probability `means' or how it arises 
in quantum theory. 
This comes from thinking that Born's 
rule only concerns expected values. 
But the quantum expected value is just the first 
moment of a classical probability measure, 
something which 
enters quantum theory naturally, 
as explained in Chapter~\ref{qprob-chapter}. 
And overlooking 
this leads to much ado about nothing. 
In this treatise I give 
the explicit 
properties of probability, namely,  
the Generalized Born's Rule.  
This can and should be checked experimentally. 
This theory is not designed 
to be intuitive or 
to make anyone happy,
though that may happen too. 
It is meant to {\em save the phenomena} of the relative 
frequencies measured in experiments. 

The consistent history approach 
(see \cite{griffiths}) is based on 
the histories of 
quantum events 
(which are called properties)
and their probabilities and therefore bears a 
superficial resemblance to the thesis of this treatise. 
The idea is to introduce lots of 
classical Kolmogorovian probability spaces 
in order to address the problem of finding 
a mathematically consistent approach for  
applying Kolmogorovian
probability theory to quantum systems. 
This leads to new constructs of which the most important 
is that of a history of (usually finitely many, though
possibly very many) time ordered quantum events. 
Each history then defines a (usually finite, though
possibly very large) classical Kolmogorovian  
sample space on which classical probabilities are 
defined and then related to Born's rule. 
The formalism depends on forming an inner product
from the trace as 
$ Tr (A^{*} B)  $ for an unspecified class of 
bounded operators $  A$ and $ B $. 
Of course, for infinite dimensional 
Hilbert spaces this is the rather small space 
of Hilbert-Schmidt operators. 
As noted in \cite{griffiths} this only works well 
for all operators in the finite dimensional case. 
While it is worthwhile to use finite dimensional 
examples to present concepts, it is not a useful
approach  when those concepts do not readily 
generalize to the infinite dimensional case. 
So the use of the trace places an enormous 
restriction on the scope of this approach. 
But next the first attempt to assign probabilities 
to the events in a history runs into an obstacle,
namely, finite additivity fails. 
To avoid this a sufficient orthogonality condition 
(see Eq.~(10.20) in \cite{griffiths}) is given. 
While this condition is not necessary, without some 
other condition being imposed  
finite additivity does fail in general. 
The author
of \cite{griffiths} does not seem to consider 
$ \sigma $-additivity, which becomes relevant 
when histories with infinitely many events are 
considered, if one wishes to use classical 
probability spaces. 
But even so, 
the restriction is then imposed to consider only
consistent (or weakly consistent) histories as having 
a physical meaning, while 
histories not satisfying this condition 
(or some possibly weaker condition) are 
considered to be neither true nor false, but
instead to be meaningless and therefore excluded 
from quantum theory. 
This is a rather strong restriction with 
nothing corresponding to it in standard quantum mechanics. 
In short, this approach adds new features to standard 
quantum theory and produces a theory which says less. 

The contrast of the current histories approach with  
the theses of this treatise should now be clear enough. 
I do not construct any new classical 
(Kolmogorovian) sample spaces, 
but rather 
use the spectral measures on $ \mathbb{R} $
associated to a self-adjoint operator and a state. 
I do not require extra special conditions to guarantee 
$ \sigma $-additivity in this case, 
since that is a consequence of spectral theory. 
In the case of multi-event probabilities $ \sigma $-additivity 
simply fails as was shown previously in detail in 
Chapter~\ref{qprob-chapter}. 
This is not surprising since the domain of definition of these 
multi-event probabilities is not even a $\sigma$-algebra. 
This is just one of many ways in which quantum probability, the central topic of this treatise,  
differs from classical probability as commented on 
extensively in 
Section~\ref{comparison=section}. 
Also, everything here 
works equally well in the infinite 
dimensional case, even though examples are 
given mainly in the finite dimensional case. 
There are also no restrictions placed on the sequences of events 
for which probabilities are defined. 
I use only absolutely standard functional analysis and 
spectral theory  
(and the measure theory behind much of it) plus a 
straightforward generalization of Born's rule to 
define quantum conditional and consecutive probabilities 
with values in the unit interval $ [0,1] $.
I consider this treatise to be a new organization of
standard quantum theory and 
in no way an interpretation or extension of it. 

I think that the entire enterprise of finding the
`correct interpretation' of quantum theory, as far 
as scientific activity is concerned, is 
quite besides the point and so is not interesting
for me at all. 
Simply put,  
I am interested in scientific 
theories of nature rather than  
who-knows-what theories of theories. 
I have not even mentioned other popular 
`interpretations' of quantum theory. 
Rather I include this chapter only to distance the 
contents of this treatise from all that.  
Interpretation is the revenge of intellect upon 
science.

\chapter{The Wave Function}
\markboth{The Wave Function}{The Wave Function}

\hfill When we want to understand something strange, 

\hfill previously unknown to anyone, we have to begin

\hfill with an entirely different set of questions. 

\hfill What is it? How does it work?

\vskip 0.2cm

\hfill Margaret Mead

\vskip 0.4cm

I do not want to discuss classical physics at all. 
But the wave function has the dubious role of being 
an aspect of quantum theory that is often 
considered in language that is classical. 
Here is some of that language. 
In classical physics
the time evolution  
is described in terms of the 
points in a particular
space,  
which is called the {\em phase space}. 
These points are the classical pure states.
The phase space typically has high dimension. 
Then the classical dynamics specifies the possible 
one-dimensional curves in phase space. 
These curves are parameterized by time. 
Each point on such a curve represents a complete 
description of the physical system at the 
corresponding time. 

All of this is temptingly analogous to the Schr\"odinger 
model of quantum theory, where 
the time dependent 
solution of Schr\"odinger's equation, which is 
inappropriately\footnote{Schr\"odinger's 
equation is not a wave equation.} 
called a {\em wave function}, gives a 
curve in a space whose points are 
called pure states. 
But analogies are not explanations. 
Unfortunately, this one is completely 
misleading, since 
in the isomorphic Heisenberg model the state
is always constant in time. 
In the classical case the dynamical curve wending its
way through a high dimensional phase space is used to 
deduce how physical objects move in three 
dimensional physical space. 
Analogously, the time dependent wave function 
is used to deduce properties that one can 
visualize in terms of the geometry of three 
dimensional physical space. 

However, the analogy fails in part because 
Schr\"odinger's equation
is not the fundamental time evolution equation of
quantum theory. 
It also fails because, although it is 
not widely recognized, 
classical mechanics also has a Heisenberg-type
model provided that the classical 
dynamical flow 
$ \phi_{t} : \mathcal{P} \to \mathcal{P}  $ 
is globally defined on the phase space 
$ \mathcal{P}  $. 
Classical observables are functions 
$ f : \mathcal{P} \to  \mathbb{R}$, and the 
classical pure states are the points in $\mathcal{P}$. 
What one observes is the time dependent quantity 
$ f (  \phi_{t} ( p_{0} ) ) $, 
where  $ p_{0} $ is the initial state. 
This is usually considered in a Schr\"odinger-type 
model as $ f (p_{t}) $ with  the observable 
$ f $ being constant in time and the initial 
state $ p_{0} $ evolving to 
$p_{t}:= \phi_{t} ( p_{0} )  $ at time $ t $. 
This easily generalizes to the time evolution 
of mixed states ($ \equiv $ probability measures on $ \mathcal{P} $) in which case this is the 
Liouville formulation of classical mechanics as  
used mostly in statistical mechanics. 
But $ f (p_{t}) = f_{t} (p_{0}) $ where we 
take the state $ p_{0} $ to
be constant in time and let the observable evolve
from its initial 
$ f $ to $ f_{t}  :=  f \circ \phi_{t}  $
at time $ t $. 
This Heisenberg-type model is called the Hamiltonian 
picture in \cite{faddeev}. 

There are also interaction-like models in 
classical mechanics, in which one takes any 
two families of functions 
$\alpha_{t}, \beta_{t} : 
\mathcal{P} \to \mathcal{P}  $ 
for $ t \in \mathbb{R} $ 
that satisfy 
$ \alpha_{t} \circ \beta_{t} = \phi_{t} $. 
Then one lets the 
states flow according to 
$ p \mapsto \beta_{t} (p) $ while 
the observables flow according to 
$f \mapsto  f \circ \alpha_{t}$. 
Even if $ \phi_{t} $ is a 
$ C^{\infty} $ function, 
its factors $ \alpha_{t} $ and $ \beta_{t} $
need not even be continuous. 
But nonetheless, such a non-intuitive
factorization could help one do calculations. 
I am not aware of such interaction-type 
models being used in classical mechanics. 

One way classical thinking can creep into 
quantum theory is when the 
Hilbert space for the quantum theory 
is $ L^{2} ( \mathbb{R}^{3} )$, 
since a normalized wave function $ \psi $ 
in that space has an associated probability 
density $ |\psi|^{2} $ on $ \mathbb{R}^{3} $. 
So, computer displays of these densities 
are made and can be seen in textbooks, 
in popular expositions and 
on the Internet. 
The graph of 
$
    |\psi|^{2} : \mathbb{R}^{3} \to [0, \infty)
$
is a subset of $ \mathbb{R}^{4} $, making direct 
visualization a bit tricky for most of us. 
But level sets in $ \mathbb{R}^{3} $ can be encoded 
in computer memory and then their $ 2 $-dimensional 
projections can be displayed on a screen or printed 
on paper. 
If one uses instead a curve of normalized
solutions of the time dependent 
Schr\"odinger's equation, then one can make a
computer video. 
Of course, it is easy to misinterpret what the 
moving blob in such a video means. 
But a more profound problem is that, except for
toy models, $ L^{2} ( \mathbb{R}^{3} )$ is not the
Hilbert space used to describe quantum systems.  
In the following examples spin and statistics are
omitted, since even without that 
the basic idea is clear. 

Consider the ever popular example of the hydrogen 
atom. 
In the texts the stationary states 
$\psi \in L^{2} ( \mathbb{R}^{3} )$ are found, 
together with their corresponding energy levels. 
Many images have been made for $ |\psi|^{2} $. 
The formulas for the stationary states as well as
the images made from them are all quite pretty. 
But all of that is misleading, since the
(electrically neutral) hydrogen 
atom is a two body problem, whose Hilbert space 
in the Schr\"odinger model is  
$ L^{2} ( \mathbb{R}^{6} )$ with Euclidean coordinates
$ x_{1}, x_{2}, x_{3} $ for the electron and
$ y_{1}, y_{2}, y_{3} $ for the nucleus.
A mathematical technique (change to center-of-mass 
coordinates) shows that the Hamiltonian of 
this system can be transformed in such a way 
as to give an equivalent problem
with two Schr\"odinger operators, each 
acting in $ L^{2} ( \mathbb{R}^{3} )$. 
But neither one of these operators 
is the Hamiltonian of the hydrogen atom 
nor of any other physical entity. 
The wave functions for each of these two 
Schr\"odinger operators can be combined and
then with an inverse change of coordinates can be
written as 
$ \psi (x_{1}, x_{2}, x_{3}, y_{1}, y_{2}, y_{3} )$ 
in $ L^{2} ( \mathbb{R}^{6} )$. 
But this last step is 
rarely done in the elementary texts. 
Rather only one of these two Schr\"odinger operators
acting in $ L^{2} ( \mathbb{R}^{3} )$ is analyzed. 
For example this is what I did in \cite{path}. 

Sometimes, instead of presenting the change of coordinates, a model is used with the nucleus 
fixed and immovable at some point in space. 
Then the electron is analyzed with a 
one-body  Schr\"odinger operator. 
Among other things the mass parameter in such 
an approach should not be the electron mass. 
In fact, the mass parameter to be used depends 
on the isotope of the nucleus. 
But justifying that is difficult, to say the least. 
So the electron mass is used, which  
introduces a small error. 
But when the electron-positron two body 
problem
is considered, the error is no longer small. 
Worse yet, this `approximation' contradicts the 
supposedly sacred Uncertainty Principle 
by fixing the values of 
both the position and the momentum of the nucleus.

The neutral hydrogen molecule is a four body 
problem and so has its wave functions in 
$L^{2} (\mathbb{R}^{12})$.
The water molecule is a 21 body problem and 
so has its wave function in 
$ L^{2} ( \mathbb{R}^{63} )$. 
And so it goes. 
But you can find images of the wave function 
of the ground state of the water molecule 
H$_{2}$O in the literature. 
This is done by 
changing to 
center-of-mass 
coordinates, 
fixing the values all of the 
variables for the 3 nuclei as well as
integrating out all 
the variables for 17 of the 18 electrons.  
The modulus squared of the
 resulting `wave function' in 
$L^{2} (\mathbb{R}^{3})$ can then be published 
in a chemistry textbook or 
on the cover of a popular science magazine. 
This sort of visualization loses a lot of 
information that is encoded in the 
correct wave function of the molecule. 
Of course, it is approximately correct information 
that remains.  
(It is only approximate since there are only approximations of the original 
wave function itself.)
But one is easily mislead into thinking that one has
visualized the spatial structure of the molecule. 

I will certainly be criticized for downplaying 
the role of Schr\"odinger's equation in quantum theory. 
But that equation stands firm on the basis of its 
many successes, which are found in
the scientific literature. 
I emphasize in \cite{path} the central role of 
Schr\"odinger's equation in the scientific 
activity of quantum physics. 
The two-slit experiment is but one grain 
of sand on the vast beach of successes of 
the Schr\"odinger model.
They all speak for themselves.\footnote{In  
poker the saying is that the cards speak 
for themselves.
But if they do not speak clearly to you, 
see for example \cite{beltrametti} 
for an explanation of the two-slit 
experiment in terms
of events, states and probabilities}
I need not repeat those details. 
I only wish to place things in the correct perspective
with respect to the Generalized Born's Rule.
Those who wish to continue working with the
Schr\"odinger model do not need permission to 
do so. 
Their results, if obtained correctly, 
will have scientific value. 
However, if one gives a physical significance to 
the model dependent aspects 
of the Schr\"odinger model, then I claim 
one is 
{\em overinterpreting} the mathematics. 

I most likely expect to be criticized for taking 
a viewpoint that is `too mathematical' in this treatise. 
But ironically it is those persons ascribing deep 
meanings to mathematical structures with 
no physical significance who are the ones that are
`too mathematical'.

This topic is related to {\em quantization}, even though
that is not a part of quantum theory but rather
a way of arriving at a quantum theory from some 
other starting point. 
Typically, one starts from classical physics, but
that is not essential. 
Actually, {\em second quantization} starts with one 
quantum theory in order to produce another. 
I do not wish to belittle research 
on quantization. 
It has its importance, but only as a way to
quantum theory. 
In this treatise I want to describe the basics of
the quantum world,
and not go into the details of the journey for 
getting there. 
I have even done research on quantization, but this 
is not the place to go into that. 
My only goal now is to comment on its lack 
of relevance to doing and 
understanding quantum theory itself. 
Even worse, quantization is used to leak ideas of 
classical 
physics into quantum theory. 
Of course, some classical ideas do transfer 
to quantum theory, but others not. 
So that issue must be faced and understood 
on a case by case basis.

\chapter{Beyond Conventional Quantum Theory}
\markboth{Beyond Conventional Quantum Theory}{Beyond Conventional Quantum Theory}

\hfill The most interesting ideas are heresies. 

\vskip 0.2cm
\hfill Susan Sontag 

\vskip 0.4cm

The main thesis of this treatise is that there is
one basic time evolution equation 
in quantum theory. 
That equation is the generalized Born's rule. 
But there is also a secondary equation in the 
Schr\"odinger model, namely Schr\"odinger's
equation. 
In the isomorphic Heisenberg model there is 
a quite different secondary time evolution equation. 
These are equations that have a unique solution
for a given initial condition. 
This looks suspiciously like determinism, but it
is only a mathematical property.
We hear it said about  Schr\"odinger's
equation that if we know the state of the system
at some initial time, then we know it at all 
future (and even past!) times. 
That is a big `if' and then some. 
In the first place, some states are not known and
are even unknowable. 
In the second place, collapse in the 
Schr\"odinger model precludes knowing both 
future and past. 

We use Schr\"odinger's
equation for the simple reason that 
it has been so successful. 
But its success has not been complete 
as is evident 
from the number of confusions, among other things,  
that it gives rise to. 
Now the central role of the
generalized Born's rule clarifies matters, 
as we have seen. 
Because of this I have been led to wonder if the 
Schr\"odinger model is the only possible 
quantum theory (up to isomorphism).
And I have come to think not. 
Still I must explain why the 
Schr\"odinger model has been so successful 
and what could be an alternative to it. 
Before getting into that let me note that 
Accardi and his collaborators have a well 
developed theory which is not an alternative
to standard quantum theory, but rather approximates 
standard Hamiltonian 
unitary flows for systems with many degrees of
freedom with singular Hamiltonian 
unitary flows in terms of fewer variables.
See \cite{accardi} and references therein for 
this important achievement, 

I shall use classical mechanics as an analogy. 
Now, classical Newtonian 
mechanics is a theory of masses,
forces and motion. 
The basic 
axioms\footnote{I believe their are at least four axioms. 
An overlooked axiom states that force is a vector quantity.}
(called 
{\em Newton's Laws of Motion}) 
of that theory give the relations among these. 
But that theory is divided into two parts. 
This division is based on the type of forces 
of a particular situation. 
These are called {\em conservative forces} and 
(rather unimaginatively) {\em non-conservative forces}. 
(I dislike the alternative phrase 
{\em dissipative force}
since it suggests that heat should be introduced  
into classical mechanics.) 
The theory for conservative forces has a lot of 
quite wonderful properties. 
This is related to 
a new form of energy, the {\em potential energy}, which 
then casts a new light on 
{\em kinetic energy} (which was originally confused with 
the expression $ m v^{2} $) to give the principle of
{\em conservation of mechanical energy}. 
And the potential energy opens the door to an elegant 
theory. 
One can then define the Hamiltonian function, Poisson 
brackets and Hamilton's 
equations of classical motion. 
Under certain hypotheses these equations 
of motion have a unique {\em local} (in time) 
solution given appropriate initial conditions. 
This seems to be determinism, but {\em caveat emptor}. 
In the first place, local solutions can and often do 
have singularities, which preclude extrapolating into 
the indefinite future. 
In the second place, 
the Hamiltonian theory does not apply 
to situations with non-conservative forces. 
In the context of conservative forces one 
arrives eventually at symplectic geometry. 
I have even heard some colleagues maintain that 
classical mechanics is a topic in symplectic 
geometry. 
But that is not so. 

Classical mechanics includes the analysis of situations 
with friction and other non-conservative forces, in which 
case there is no potential energy and no conservation 
of mechanical energy. 
The motion of a boat in water 
is a simple example of classical
mechanics with non-conservative forces. 
Other examples abound in day-to-day life. 
Actually, there is such a sparsity of examples with only 
conservative forces that the everyday `common sense' 
of most people conceals their importance. 
Ask any high school physics teacher 
how difficult it is to 
teach classical mechanics with only 
conservative forces. 
(Angular momentum is essentially impossible.)
Yet it is an error to think that 
conservative forces are basic while 
non-conservative forces are not. 
Within the context of classical mechanics 
neither is more basic than the other. 

I suppose that something analogous occurs in 
quantum theory, namely that 
the Schr\"odinger model is similar to  
conservative classical mechanics. 
That model also has a lot of 
quite wonderful properties, 
but that does not imply that they are 
necessarily properties of all quantum systems. 
I find it quite plausible that a different 
equation for the time evolution of the  
states could be posited and that (together with
time independent pvm's as just one example) would give an
acceptable quantum theory that is not isomorphic to the 
Schr\"odinger model. 
Such an alternative to the Schr\"odinger model
might not be deducible from the 
Schr\"odinger model. 
Rather it could be an independent
model, also consistent with the basic 
axioms of quantum theory. 
Let me be completely clear here.  
I am saying that Schr\"odinger's 
equation is not a necessary part of
quantum theory, just as
conservative forces are not a 
necessary part of classical mechanics. 
Of course, Schr\"odinger's equation is an important, 
interesting part of quantum theory much as 
conservative classical mechanics is an important, 
interesting part of classical mechanics. 

Another well studied extension of quantum theory is to
include {\em positive operator valued measures (povm)}, 
a generalization of projection (quantum event) valued 
measures. 
These arise so naturally that many researchers consider
this to still be quantum theory. 
It is fine to think of them that way. 
My only point is that they are not included in 
the axiomatization of Chapter~1. 
If they are included in new axioms, then the definition 
of isomorphism of quantum theories should also be 
modified. 
And the Generalized Born Rule would also have to 
be changed. 
But I would expect that 
the new Generalized Born Rule would continue to 
be the fundamental time evolution equation 
of quantum theory. 

One of the quite wonderful properties of 
the Schr\"odinger model is called {\em unitarity}. 
It says that the time evolution of a state of 
a quantum system 
is realized using the unitary group $ e^{-i t H/\hbar} $
for all $ t \in \mathbb{R} $, 
where $ H $ is a special self-adjoint operator 
associated to the system. 
It is well known that this unitarity group 
leads to a unique global (in time) solution of the appropriate 
initial value problem. 
This leads to time reversal invariance of the 
theory, but not necessarily of the physical 
quantum system.   
I find it perfectly plausible to replace the unitary 
group with a semi-group that violates unitarity as well as
time reversal invariance. 
And that semi-group would describe the model dependent
time evolution of the states. 
As always, the only time evolution relevant to 
observations, to saving the phenomena, is that of
the generalized Born's rule. 
And such a quantum theory would not be less basic
than the `friction-less' Schr\"odinger model.
Nor would it nor should it 
be derivable from the Schr\"odinger model. 
Again, let me note that a different approach 
is taken in \cite{accardi}. 

Of course, such semi-groups exist and have been 
studied extensively. 
My point is that they should not be considered as
approximations in any sense of that word to a
more `basic' Schr\"odinger model. 
Neither should they be considered as being derived 
nor even capable of being derived from the 
Schr\"odinger model. 
As with Schr\"odinger's equation itself, 
they are secondary mathematical structures 
with no physical significance in themselves, 
except that they provide a key ingredient for 
calculating the time evolution of the 
generalized Born's rule. 
That is, of course, a legitimate and important role, 
which can be recognized as such by including 
an axiom on semi-groups  into quantum theory as a 
replacement of Schr\"odinger's equation. 
The idea of using semi-groups in quantum theory 
is found in the literature 
going back to the 1970's. 
See \cite{ingarden}, \cite{kossakowski} and 
\cite{mehra}. 

My point is that semi-groups should be taken as 
seriously as Schr\"odinger's equation is. 
Any supposedly 
`deeper explanation' of the semi-group as a merely 
phenomenological and convenient approximation to 
a more `general' formulation based on the
Schr\"odinger model may be beside the point. 
Any theory should be viewed as 
phenomenological if it saves the phenomena, 
which is one goal of any theory. 
Further, any theory should be viewed as convenient 
if it aids our understanding of nature, 
which is another goal of any theory. 

A quite nice and successful family of semi-groups 
is what I wish to call the {\em Lindbladian model},
which is based on {\em quantum dynamical semi-groups}. 
This choice of name honors the author of \cite{lindblad}. 
However, as noted in \cite{lindblad} there are earlier 
articles on this topic, some of which I cited above. 
Also see \cite{davies1}, \cite{davies2} and 
\cite{davies3}.  
Unfortunately, many authors including Lindblad himself 
say that they are using a formulation in either 
the Schr\"odinger picture or the Heisenberg picture. 
I prefer to view their work as dealing with 
a new {\em model} 
not previously considered in the quantum literature. 
One of the motivations for introducing this model 
is to facilitate the description 
of irreversible quantum processes. 
This is achieved by allowing non-Hamiltonian systems that 
do not satisfy the principle of unitarity. 
However, one very nice feature of this model 
(and models isomorphic to it) is that it includes 
standard quantum theory as a special case. 
So it is also compatible with reversible quantum
processes and unitarity. 
Another nice feature is that it gives a context 
in which {\em decoherence} can be studied. 
The main difficulty with this model is that it can be 
less than obvious how to introduce a specific 
semi-group for studying a specific system. 
This difficulty makes intuition hard to come by, at 
least for me. 
In standard quantum theory this is the problem 
of {\em quantization}, which is usually resolved on a 
case-by-case basis by appealing to conservative 
classical mechanics. 

To understand better the role of unitarity in this model 
and other models of 
quantum theory we now turn to Wigner's theorem. 
Wigner's theorem is one of those wonderful 
results related to standard quantum theory. 
It is based on the idea that the Hilbert 
space expressions 
with physical significance for quantum theory 
have the form 
\begin{equation}
\label{Wigners-idea}
  | \langle \psi , \varphi \rangle |^{2} \quad 
  \mathrm{for~unit~vectors~} \psi, \varphi \in \mathcal{H}. 
\end{equation}
We have already noted that these other 
Hilbert space structures do not have any 
physical significance: 
vector sum, scalar product, inner product. 

First, let's note that \eqref{Wigners-idea} is 
a probability that comes from Born's rule.
Let $ E = | \varphi \rangle \langle \varphi | $ 
be the rank $ 1 $ event associated to $ \varphi $. 
Then we calculate 
\begin{align*}
P ( E \,|\, \psi ) &= \langle \psi , E \psi \rangle 
\\
&= 
\langle 
\psi , | \varphi \rangle \langle \varphi | \, \psi 
\rangle 
\\
&=
\langle \psi, \langle \varphi, \psi \rangle \varphi \rangle 
\\
&=
\langle \varphi, \psi \rangle \langle \psi, \varphi \rangle
\\
&= 
| \langle \psi , \varphi \rangle |^{2}. 
\end{align*}
Consequently, the expression 
in \eqref{Wigners-idea} is the probability 
that the event $ E $ associated to $ \varphi $ occurs 
given the state $ \psi $. 
Since \eqref{Wigners-idea} is invariant if we 
interchange of $ \psi $ and $ \varphi $, 
\eqref{Wigners-idea} also is the
probability 
that the event associated to $ \psi $ occurs 
given the state $ \varphi $. 
So our axiomatization of Born's rule gives the 
usual physical significance to the 
expression \eqref{Wigners-idea}. 

Moreover, as we noted before, the expression 
\eqref{Wigners-idea} is well defined for pure 
states, not just for unit vectors in the 
Hilbert space $ \mathcal{H} $. 
Specifically, this means that if we replace 
$ \psi $ with $ \lambda \psi $, where 
$ \lambda \in \mathbb{C} $ satisfies $ | \lambda | =1 $
and also $ \varphi $ with $ \mu \varphi $ 
with $ | \mu | =1 $, then the expression 
\eqref{Wigners-idea} does not change, even though 
the inner product 
$ \langle \psi , \varphi \rangle $ does change. 

Next, Wigner introduces the idea that 
a transformation of the set $ \mathcal{S} $ 
of pure states to 
itself is an isomorphism from the point of 
view of quantum theory if it is a bijection 
$ T : \mathcal{S} \to \mathcal{S} $ such that 
the expression \eqref{Wigners-idea} is preserved; 
this is a special case of {\em conservation of probability}. 
Such a bijection is called a {\em Wigner transformation}. 
Then, Wigner proves that every Wigner transformation 
is the quotient of a bijection
$ U : \mathcal{H} \to \mathcal{H} $ that is linear 
and unitary or else of a bijection
$ V : \mathcal{H} \to \mathcal{H} $ that is anti-linear 
and anti-unitary. 
This leads to the powerful corollary that every group of
Wigner transformations 
$ T_{t} : \mathcal{S} \to \mathcal{S} $ 
for $ t \in \mathbb{R} $ is the quotient of a group 
of unitary transformations 
$ U_{t} : \mathcal{H} \to \mathcal{H} $. 
With a little continuity in the group $ T_{t} $ 
and hence also in the group $ U_{t} $, we can apply 
Stone's theorem to write $ U_{t} = e^{-i t H / \hbar} $ 
for a unique densely defined self-adjoint operator $ H $ 
acting in the Hilbert space $ \mathcal{H} $. 
We have arrived at the unitary group of the Schr\"odinger
model (also of the Heisenberg model) from general 
principles. 

But if the Schr\"odinger model as well as models 
isomorphic to it are not the only possible
models of quantum theory, then one is obliged to 
conclude that 
some hypothesis used in Wigner's proof is not a 
fundamental principle of quantum theory. 
I propose that the conservation of probability may 
not be a fundamental principle of quantum theory. 
In my opinion  
it could be analogous to the 
principle of conservation of mechanical 
energy in classical mechanics. 
Conservation of {\em mechanical energy} holds in many 
interesting, important 
cases but is not a universal principle 
of classical mechanics. 
Analogously, I am proposing that conservation of 
probability--and the consequent unitarity--are not 
necessarily 
universal principles of quantum theory. 
They are fine in the `fiction-less' case
of quantum theory as seen in the Schr\"odinger model, 
but not in general.  
In this regard one has to view Wigner's theorem 
as a {\em No-Go theorem}. 
It tells us neither conservation of probability 
nor unitarity {\em go} as  
an aspect of a more general quantum theory that is not
isomorphic to the Schr\"odinger model. 
We now see that in 
the Lindbladian model 
the semi-group in general 
does not come from a Wigner transformation.
And it is precisely this that makes the 
Lindbladian model so interesting and 
meriting further research. 
I realize that this idea is controversial, but 
experiment will be the judge. 

Finally, I conclude with a rather wild speculation. 
Since events and states are all that enter the 
generalized Born's rule, together with rules 
for their time evolution, 
it seems natural to consider a theory with 
only these elements. 
So, no Hilbert space. 
Maybe even no $ \hbar $. 
There would be a set $ \mathcal{E} $ of events 
and a set $ \mathcal{S} $ of states. 
There would be a function 
$$
   P :  \mathcal{E} \times \mathcal{S} \to [0,1]. 
$$
For 
$ E \in \mathcal{E} $ and 
$ S \in \mathcal{S} $ 
we would say that $ P(E,S) $ 
is the probability of the event $ E $, 
given the state $ S $. 
Adding in two one-parameter groups of bijections
of $ \mathcal{E} $ and $ \mathcal{S} $, 
respectively, 
we would have a time dependent probability theory. 
Models of such a theory would be in abundance. 
Isomorphism would be defined in terms of 
preservation of probability. 
Some other mathematical structures would eventually 
have to be imposed on these sets in order to get 
a viable theory. 
For example, consecutive and conditional probabilities 
could be introduced. 
My main point here is that the lattice of projections 
in a von Neumann algebra might not be the best 
place to look for a model of physical events.

\chapter{A Modest Proposal}
\markboth{A Modest Proposal}{A Modest Proposal}

\hfill Bidding good-bye 
 is such sweet sorrow

\hfill that I could bid adieu
 til it be morrow

\hfill Romeo in {\em Romeo and Juliette}

\vskip 0.2cm 
\hfill William Shakespeare 

\vskip 0.4cm \noindent 
It seems to me that the expression
`quantum mechanics' does not adequately 
describe the physical theory. 
It is based on two misleading words. 
The first is `mechanics' which originally 
meant the study of machines but changed 
to mean the study of motion. 
Unfortunately, the concept of motion 
(as understood as an object moving along a
curve) is not central to the physics we 
wish to discuss at the atomic and smaller 
length scales. 

The word `quantum' is even more unfortunate. 
It was used to contrast the new theory with 
classical theory where all measured values 
could assume a continuum of possible values. 
All of a sudden there were physical quantities 
whose measured values were in a discrete set. 
These measured values were said to be `quantized'. 
While this is an important part of the story, 
it is not the whole story. 
Even in this new theory there are still some 
physical quantities with values in a 
continuum, such as the energy of a free 
particle. 
Other quantities are always `quantized', such as
the energy of a harmonic oscillator 
(realized as a diatomic molecule, say).
But there are mixed cases, such as the hydrogen atom 
(with fixed nucleus), 
which has `quantized' energy levels (associated 
with the bound states) as well as a continuum 
of energy levels (associated with scattering 
states). 

What is important here 
are the probabilities that arise from
the pvm  
of a self-adjoint operator and more generally
its spectral theory. 
So it seems to be more appropriate to say 
{\em Spectral Probability Physics} 
instead of saying 
Quantum Mechanics for the physical theory 
based on self-adjoint operators. 
Another possibility is {\em von Neumann Algebra 
Probability Physics}. 
Whatever one wishes to call it, this is an 
ongoing scientific project, whether it is done
in the language of the Schr\"odinger model 
(with Schr\"odinger's equation and collapse),
in the Heisenberg model (without those 
mathematical tools) or any other model 
satisfying the Axioms, but always including
the Generalized Born's Rule. 

With this modest proposal I bid my gentle 
reader good-bye.

\backmatter


\markboth{References}{References}
\begin{thebibliography}{99}

\bibitem{accardi}
L.~Accardi, Y.G.~Lu and I.~Volovich,
{\em Quantum Theory and Its Stochastic Limit}, 
Springer, 2002. 

\bibitem{bargmann}
V.~Bargmann, 
On a Hilbert space of analytic functions and an 
associated integral transform, Part I, 
Commun. Pure Appl. Math. {\bf 14} (1961) 187--214. 

\bibitem{beltrametti}
E.G.~Beltrametti and G.~Cassinelli, 
{\em The Logic of Quantum Mechanics}, 
Addison-Wesley, 1981. 

\bibitem{bengtsson}
I.~Bengtsson and K.~Zyczkowski, 
{\em The Geometry of Quantum States, 
An Introduction to Quantum Entanglement},
Cambridge University Press, 2006. 

\bibitem{busch}
P.~Busch et al., {\em Quantum Measurement}, 
Springer, 2016. 

\bibitem{cassinelli}
G.~Cassinelli and N.~Zanghi, 
Conditional Probabilities in Quantum 
\\
Mechanics, 
Nuovo Cimento {\bf 73} B (1983) 237--245.

\bibitem{davies1}
E.B.~Davies, 
Quantum stochastic processes, 
Commun. Math. Phys. {\bf 15} (1969) 277--304.

\bibitem{davies2}
E.B.~Davies,
Quantum stochastic processes II, 
 Commun. Math. Phys. {\bf 19} (1970) 83--105.

\bibitem{davies3}
E.B.~Davies, 
Quantum stochastic processes III, 
Commun. Math. Phys. {\bf 22} (1971) 51--70. 

\bibitem{dirac}
P.A.M.~Dirac, {\em Principles of Quantum Mechanics}, 
4th Ed., Oxford 
\\
University Press, 1958. 

\bibitem{epr}
A.~Einstein, B.~Podolsky and N.~Rosen, $ \, $
Can Quantum-Mechanical 
Description of
Physical Reality Be Considered Complete?, 
Phys. Rev. {\bf 47} (1935) 777--780. 

\bibitem{faddeev}
L.D.~Faddeev and O.A.~Yakubovskii, 
{\em Lectures on Quantum Mechanics for 
Mathematics Students}, Am. Math. Soc., 2009. 

\bibitem{gleason}
A.M.~Gleason, 
Measures on the closed subspaces of a Hilbert space, \\
J. Math. Mech. {\bf 6} (1957) 
885--893. 

\bibitem{griffiths}
R.B.~Griffiths, {\em The Consistent Quantum Theory},
Cambridge, 2002.

\bibitem{gudder}
S.P.~Gudder, {\em Quantum Probability}, 
Academic Press, 1988. 

\bibitem{haag}
R.~Haag, {\em Local Quantum Physics}, 2nd ed., 
Springer, 1996. 

\bibitem{hudson}
R.~Hudson, A short walk in quantum probability, 
Phil. Trans. R. Soc.~A {\bf 376:} 20170226. 

\bibitem{ingarden}
R.S.~Ingarden and A.~Kossakowski, 
On the connection of nonequilibrium information 
thermodynamics with non-hamiltonian quantum mechanics of open systems, 
Ann. Phys. {\bf 89} (1975) 451--485. 

\bibitem{kadison}
R.V~Kadison and I. Singer,  
Extensions of pure states. 
Am. J. Math. {\bf 81} (1959) 383--400. 

\bibitem{kossakowski}
A.~Kossakowski, 
On quantum statistical mechanics of
non-Hamiltonian systems, 
Rep. Math. Phys. {\bf 3} (1972) 247--274. 

\bibitem{kolmogorov}
A.N.~Kolmogorov, {\em Foundations of the Theory of Probability},  2nd ed., Chelsea Pub. Co., 1956. 
(English translation of:
{\em Grundbegriffe der Wahrscheinlichkeitsrechnung}, Springer, 1933.)

\bibitem{lindblad}
G.~Lindblad, 
On the generators of quantum dynamical semigroups, 
Commun. Math. Phys. {\bf 48} (1976) 119--130.

\bibitem{luders}
G.~L\"uders, 
\"Uber die Zustands\"anderung durch Messprozess, 
Ann. Phys.  (Leipzig), 
{\bf 8}  (1951) 322--328. 
English translation: 
Concerning the State-change due to 
the Measurement Process, 
Ann. Phys. (Leipzig) {\bf 15} (2006) 663--670.

\bibitem{marcus} 
A.~Marcus, D.~Spielman and N.~Srivastava,  Interlacing families II: mixed characteristic 
polynomials and the Kadison-Singer problem. 
Ann. Math. {\bf 182}  (2015) 327--350. 

\bibitem{mehra} 
J.~Mehra and E.C.G~Sudarshan, $ ~ $
Some reflections on the nature of entropy, 
irreversibility and the second law of thermodynamics, 
 Nuovo Cimento, {\bf 11B} 
(1972) 215--256.

\bibitem{von-neumann}
J.~von~Neumann, 
{\em Mathematische Grundlagen der Quantenmechanik}
Springer, 1932;
English translation: 
{\em Mathematical Foundations of Quantum Mechanics}, 
Princeton Univ. Press, 1955. 

\bibitem{park}  
J.~Park, The concept of transition in quantum mechanics, 
Found. Phys.~{\bf 1} (1970) 23--33. 

\bibitem{parthasarathy}
K.R. Parthasarathy, 
{\em An Introduction to Quantum Stochastic Calculus}, 
Monographs in Math. Vol. 85,
Birkh\"auser Verlag, 1992. 

\bibitem{russo}
L.~Russo, {\em The Forgotten Revolution}, Springer, 2004. 

\bibitem{schrodinger}
E.~Schr\"odinger, Quantisierung als Eigenwertproblem 
(Erste Mitteilung), Ann. Phys. {\bf 79} (1926) 361--376. 

\bibitem{schrodinger2}
E.~Schrödinger, 
Discussion of Probability Relations 
between Separate Systems, 
Proceedings of the Cambridge Physical Society, 
{\bf 31} (1935) 55.

\bibitem{sinha}
K.B.~Sinha and D.~Goswami, 
{\em Quantum Stochastic Processes and 
Noncommutative Geometry}, 
Cambridge Univ. Press, 2007. 

\bibitem{path}
S.B.~Sontz, {\em An Introductory Path 
to Quantum Theory}, 
Springer, 2020. 

\bibitem{woit}
P.~Woit, {\em Quantum Theory, Groups and 
	Representations}, Springer, 2017.


\end{thebibliography}
\end{document}